\documentclass[%
 aip,
 amsmath,amssymb,
 reprint,%
]{revtex4-1}

\usepackage{graphicx}
\usepackage{dcolumn}
\usepackage{bm}

\usepackage[utf8]{inputenc}
\usepackage[T1]{fontenc}
\usepackage{mathptmx}
\usepackage{etoolbox}
\usepackage{bigints}
\usepackage[dvipsnames]{xcolor}
\usepackage{comment} 
\usepackage{multirow}

\makeatletter
\def\@email#1#2{%
 \endgroup
 \patchcmd{\titleblock@produce}
  {\frontmatter@RRAPformat}
  {\frontmatter@RRAPformat{\produce@RRAP{*#1\href{mailto:#2}{#2}}}\frontmatter@RRAPformat}
  {}{}
}%
\makeatother
\begin{document}

\preprint{AIP/123-QED}

\title[Dissociation line of N$_{2}$ hydrates]{Dissociation line and driving force for nucleation of the nitrogen hydrate from computer simulation. II. Effect of multiple occupancy}
\author{Miguel J. Torrej\'on}
\affiliation{Laboratorio de Simulaci\'on Molecular y Qu\'imica Computacional, CIQSO-Centro de Investigaci\'on en Qu\'imica Sostenible and Departamento de Ciencias Integradas, Universidad de Huelva, 21006 Huelva Spain}

\author{Jes\'us Algaba}
\affiliation{Laboratorio de Simulaci\'on Molecular y Qu\'imica Computacional, CIQSO-Centro de Investigaci\'on en Qu\'imica Sostenible and Departamento de Ciencias Integradas, Universidad de Huelva, 21006 Huelva Spain}

\author{Felipe J. Blas*}
\affiliation{Laboratorio de Simulaci\'on Molecular y Qu\'imica Computacional, CIQSO-Centro de Investigaci\'on en Qu\'imica Sostenible and Departamento de Ciencias Integradas, Universidad de Huelva, 21006 Huelva Spain}
\email{felipe@uhu.es}

\begin{abstract}

In this work, we determine the dissociation line of the nitrogen (N$_{2}$) hydrate by computer simulation using the TIP4P/Ice model for water and the TraPPE force field for N$_{2}$. This work is the natural extension of our previous paper in which the dissociation temperature of the N$_2$ hydrate has been obtained at $500$, $1000$, and $1500\,\text{bar}$ [\emph{J. Chem. Phys.} \textbf{159}, 224707  (2023)] using the solubility method and assuming single occupancy. We extend our previous study and determine the dissociation temperature of the N$_2$ hydrate at different pressures, from $500$ to $4500\,\text{bar}$, taking into account single and double occupancy of the N$_{2}$ molecules in the hydrate structure. We calculate the solubility of N$_2$ in the aqueous solution, as a function of temperature, when it is in contact with a N$_{2}$-rich liquid phase and when in contact with the hydrate phase with single and double occupancy via planar interfaces. Both curves intersect at a certain temperature that determines the dissociation temperature at a given pressure. We observe a negligible effect of the occupancy on the dissociation temperature. Our findings are in very good agreement with the experimental data taken from the literature. We have also obtained the driving force for nucleation of the hydrate as a function of the temperature and occupancy at several pressures. As in the case of the dissociation line, the effect of the occupancy on the driving force for nucleation is negligible. To the best of our knowledge, this is the first time that the effect of the occupancy on the driving force for nucleation of a hydrate that exhibits sII crystallographic structure is studied from computer simulation.

\end{abstract}

\maketitle

%

\section{Introduction}

Hydrates are non-stoichiometric inclusion solid compounds in which guest molecules, such as methane (CH$_{4}$), carbon dioxide (CO$_{2}$), hydrogen (H$_{2}$), nitrogen (N$_{2}$), and tetrahydrofuran (THF), among numerous other species, are enclathrated within the voids left by a periodic arrange of water particles (host).~\cite{Sloan2008a,Ripmeester2022a} In the last decades, hydrates have been the subject of fundamental and applied research~\cite{Sloan2008a,Ripmeester2022a,Ripmeester2016a,Ratcliffe2022a} because of their promising applications in CO$_2$ capture,~\cite{ma2016review,dashti2015recent,cannone2021review,duc2007co2,choi2022effective,lee2014quantitative} H$_2$ storage and transport,~\cite{veluswamy2014hydrogen,lee2005tuning} N$_2$ recovery from industrial emissions,~\cite{Yi2019a,hassanpouryouzband2018co2} and they are also interesting from an energetic point of view since there is more CH$_4$ as hydrates in the nature than in conventional fossil fuel reservoirs.~\cite{lee2001methane,ruppel2017interaction}

Nowadays, the use of hydrates of N$_{2}$ and H$_{2}$, in combination with THF, as strategic materials for gas transport and storage, is one of the most significant and promising applications of hydrates in the environmental, energetic, and economic context. This offers an alternative to the metal hydrides that are now in use, whose development has not yet been put into practice because of the lack of information about the thermodynamics and kinetics of these new storage media.~\cite{Tsimpanogiannis2017a,Brumby2019a,Yi2019a,Michalis2022a} In this application, the use of hydrates would result in lower raw material costs while maintaining a comparable volumetric storage capacity. To achieve this, a thorough understanding of thermodynamics is required, with particular emphasis on phase equilibria and the kinetics of the formation and growth of these hydrates.

Due to the non-stoichiometric nature of hydrates, the composition of these compounds can vary as a function of the thermodynamic conditions, nature of the guest, and structure. The N$_{2}$ and H$_2$ hydrates exhibit sII structure with $136$ water molecules distributed in $16$ D (pentagonal dodecahedron or $5^{12}$) cages and $8$ H (hexakaidecahedron or $5^{12}6^{4}$) cages. When a sII hydrate is formed by small molecules such as N$_2$ and H$_2$, the structure is stabilized by the multiple occupancy of the H cages.~\cite{Sloan2008a,Ripmeester2022a} Also, it is interesting to remark that small molecules can stabilize the small D cages better than large molecules, which are in great number in sII restructure (16 D and 8 H cages per unit cell). This is the reason because N$_2$ and H$_2$ molecules form sII hydrates rather than sI. 

From the point of view of gas transport and storage, the ability of sII hydrates to encapsulate more than one molecule per cage is a very attractive property. In the case of the N$_2$ hydrate, Kuhs \emph{et al.}~\cite{Kuhs1997a} demonstrated by experiments that the small D cages of the N$_2$ hydrate are occupied by a single molecule of N$_2$ while the large H cages are doubly occupied. These results have been corroborated by additional experiments~\cite{Chazallon2002a,Sasaki2003a} as well as by simulation results.~\cite{vanKlaveren2001a,vanKlaveren2001b,VanKlaveren2002a,Tsimpanogiannis2017a} However, as far as the authors know, the effect of the multiple occupancy on the three-phases (hydrate-aqueous-N$_2$) coexistence line of the N$_2$ hydrate has not been studied yet. This information is crucial in order to exploit the capability of hydrates for gas transport and storage. It is mandatory to obtain in-depth knowledge about how the occupancy of hydrates affects the thermodynamic stability conditions of these compounds and vice versa, i.e., how the thermodynamic conditions at which hydrates are formed can affect the occupancy of these compounds. However, as Michalis \emph{et al.} claimed in a recent work~\cite{Michalis2022a} where the three-phase coexistence line of the H$_2$ hydrate was studied assuming single occupancy, it is not easy to perform molecular dynamic simulations when multiple occupancy is taken into account. Actually, the authors of that work claimed that it was not possible to stabilize the hydrate phase and they could not study the effect of the multiple occupancy on the three-phase coexistence line of the H$_2$ hydrate.

Very recently, the authors of this work have published a study\cite{algaba2023b} where the three-phase coexistence line of the N$_2$ hydrate was obtained from $500$ to $1500\,\text{bar}$ performing molecular dynamic simulation and using the solubility method.~\cite{Grabowska2022a,Algaba2023a,algaba2023b} Following this methodology, the N$_2$ solubility in an aqueous solution (L$_{\text{w}}$) when is in contact with a N$_{2}$-rich liquid phase (L$_{\text{N}_2}$) and when in contact with the N$_2$ hydrate phase (H) via planar interfaces is calculated as a function of the temperature. The temperature at which the three phases coexist is the temperature at which both curves intersect. Results from the solubility method can also be used to estimate the driving force for the nucleation of hydrates, $\Delta\mu_{\text{N}}$. This quantity is a key property for estimating homogeneous nucleation rates using different approaches.~\cite{Debenedetti1996a,Kashchiev2000a,Maeda2020a,Kashchiev2002a,Kashchiev2002b,Jacobson2010a,Jacobson2010b,Walsh2011a,Jacobson2011a,Sarupria2011a,Knott2012a,Sarupria2012a,Liang2013a,Barnes2014a,Yuhara2015a,Warrier2016a,Zhang2018a,Arjun2019a,Zhang2020a,Arjun2020a,Arjun2021a,Wang2022a,Grabowska2022a,Grabowska2023a,Algaba2023a,Wang2023a} In our previous work,~\cite{algaba2023b} the N$_2$ hydrate under consideration presented single occupancy. We also obtained the driving force for nucleation via route 1 and the dissociation route proposed by some of us,~\cite{Grabowska2022a,Algaba2023a,algaba2023b} as well as the L$_{\text{w}}$--L$_{\text{N}_{2}}$  interfacial tension. This study was carried out using the well-known TIP4P/Ice~\cite{Abascal2005b} and TraPPE\cite{Potoff2001a} models for water and N$_2$ respectively. Also, it was necessary to increase the dispersive interactions between water and N$_2$ molecules by modifying the corresponding Berthelot combining rule. The main goal of this work is to extend our previous study to higher pressures (until $4500\,\text{bar}$) and to take explicitly into account the double occupancy of H cages in the N$_2$ hydrate. We also determine the three-phase coexistence line (from $500$ to $4500\,\text{bar}$), as well as the driving force for nucleation taking explicitly into account the multiple occupancy of the N$_2$ hydrate. 

The organization of this paper is as follows: In Sec.~II, we describe the models, simulation details, and methodology used in this work. The results obtained, as well as their discussion, are described in Sec.~III. Finally, conclusions are presented in Sec.~IV.

\section{Models, simulation details, and methodology}

In this work, we use the GROMACS molecular dynamics package (2016.5 version)~\cite{VanDerSpoel2005a} to carry out all molecular dynamics simulations. N$_{2}$ molecules are modeled using the TraPPE (Transferable Potentials for Phase Equilibria) force field~\cite{Potoff2001a} and water molecules are described using the well-known TIP4P/Ice model.~\cite{Abascal2005b}As well as in our previous work,~\cite{algaba2023b} unlike dispersive interactions between the oxygen atoms of water and the nitrogen atoms of N$_{2}$ are calculated using a modified Berthelot combining rule as,

\begin{equation}
    \epsilon_{\text{ON}}=\xi_\text{{ON}}(\epsilon_{\text{OO}}\,\epsilon_\text{{NN}})^{1/2}
    \label{unlike}
\end{equation}

\noindent where $\epsilon_{\text{ON}}$ is the well-depth associated with the LJ potential for the unlike interactions between the oxygen atoms of water, O, and the nitrogen atoms of N$_2$, N, and $\epsilon_\text{{OO}}$ and $\epsilon_\text{{NN}}$ are the well-depths for the like interactions between the O and N atoms respectively. $\xi_\text{{ON}}$ is the factor that modifies the Berthelot combining rule. We perform simulations using Eq.~\eqref{unlike} with an optimized value of $\xi_\text{{ON}}$ determined in our previous work,~\cite{algaba2023b} $\xi_{\text{O}\text{N}}=1.15$.  With this value, we determine the solubility of N$_{2}$ in the aqueous phase when in contact with the N$_{2}$-rich liquid phase and the hydrate. From this information, we obtain the values of the dissociation temperature, $T_{3}$ along the whole N$_2$ hydrate dissociation line.

In this work, as in our previous study,~\cite{algaba2023b} we use a Verlet leapfrog\cite{Cuendet2007a} algorithm with a time step of $2\,\text{fs}$ to solve Newton's motion equations. We also use the Nosé-Hoover thermostat\cite{Nose1984a} and the Parrinello-Rahman barostat,\cite{Parrinello1981a} with a time constant of $2\,\text{ps}$, to ensure that simulations are performed at constant temperate and pressure. We use a cut-off of $1.0\,\text{nm}$ for the Coulombic and dispersive interactions. The Fourier term of the Ewald sums is evaluated using the particle mesh Ewald (PME) method.~\cite{Essmann1995a} The width of the mesh is $0.1\,\text{nm}$ with a relative tolerance of $10^{-5}$.

We simulate L$_{\text{w}}$--L$_{\text{N}_{2}}$ and H--L$_{\text{w}}$ two-phase equilibria to determine the three-phase coexistence line of the N$_2$ hydrate according to the solubility method following our previous works.~\cite{Grabowska2022a, Algaba2023a,algaba2023b} The solubility method is an interesting and alternative technique to the usual method of calculating coexistence temperatures, i.e., the direct coexistence technique.~\cite{Ladd1977a,Vega2008a} For the first time, Conde and Vega used this technique to determine the hydrate--water--guest three-phase coexistence or dissociation temperature of the CH$_{4}$ hydrate.{\cite{Conde2010a,Conde2013a}} After their seminal work, different authors have also determined the dissociation temperature of CH$_{4}$,~\cite{Michalis2015a} CO$_{2}$,~\cite{Miguez2015a,Costandy2015a,Perez-Rodriguez2017a} N$_{2}$,~\cite{Yi2019a} H$_{2},$\cite{Michalis2022a} and THF~\cite{Algaba2024c} hydrates and also CH$_{4}$ hydrates in salty water~\cite{Fernandez-Fernandez2019a,Blazquez2023b} and in confined conditions,~\cite{Zhang2022a,Zhang2023a,Fernandez-Fernandez2024a} among many others. It is also interesting to mention that other authors have used different methods for determining dissociation temperatures of hydrates using free energy calculations~\cite{Waage2017a} and Monte Carlo simulations to determine the chemical potentials of the mixture's components in the hydrate and fluid phases.~\cite{Jensen2010a} To calculate the driving force for nucleation of this hydrate we also perform bulk simulations at different conditions of pressure and supercooling temperatures. For systems that exhibit L$_{\text{w}}$-L$_{\text{N}_{2}}$ equilibria, the simulations are carried out in the $NP_{z}\mathcal{A}T$ ensemble, in which only $L_z$, the side of the simulation box perpendicular to the L$_{\text{w}}$--L$_{\text{N}_{2}}$ planar interface, is varied. In the case of systems that exhibit H--L$_{\text{w}}$ equilibria, we perform simulations in the anisotropic $NPT$ ensemble in which each side of the simulation box is allowed to fluctuate independently to keep the pressure constant. This ensures that the solid hydrate structure is equilibrated without stress. Finally, bulk systems of N$_2$, water, and hydrate are performed in the isotropic $NPT$ ensemble, i.e., the sides of the simulation box fluctuate isotropically to keep the pressure constant.

\begin{figure*}
\centering
         \centering
         \includegraphics[width=0.49\textwidth]{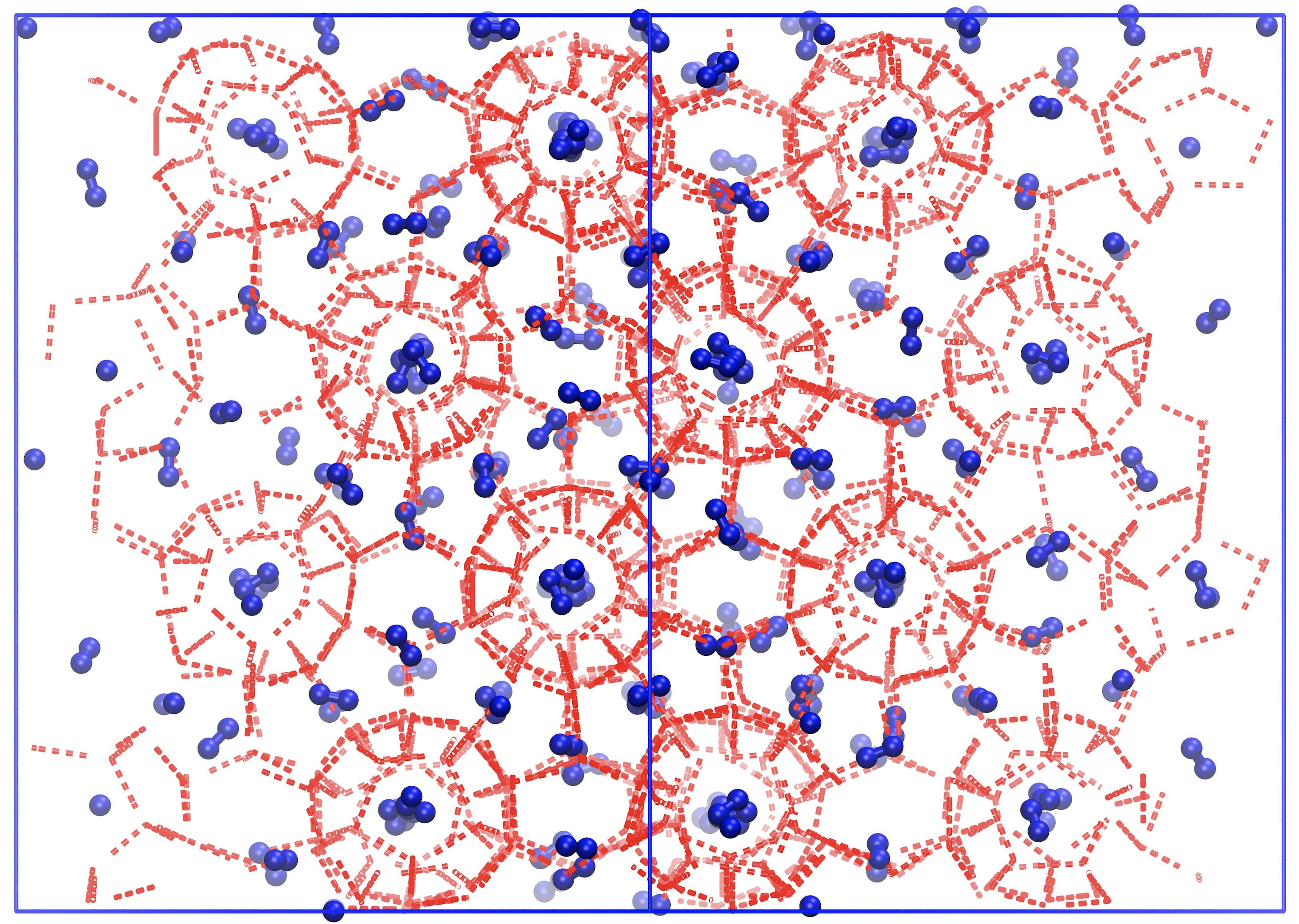}
         \includegraphics[width=0.49\textwidth]{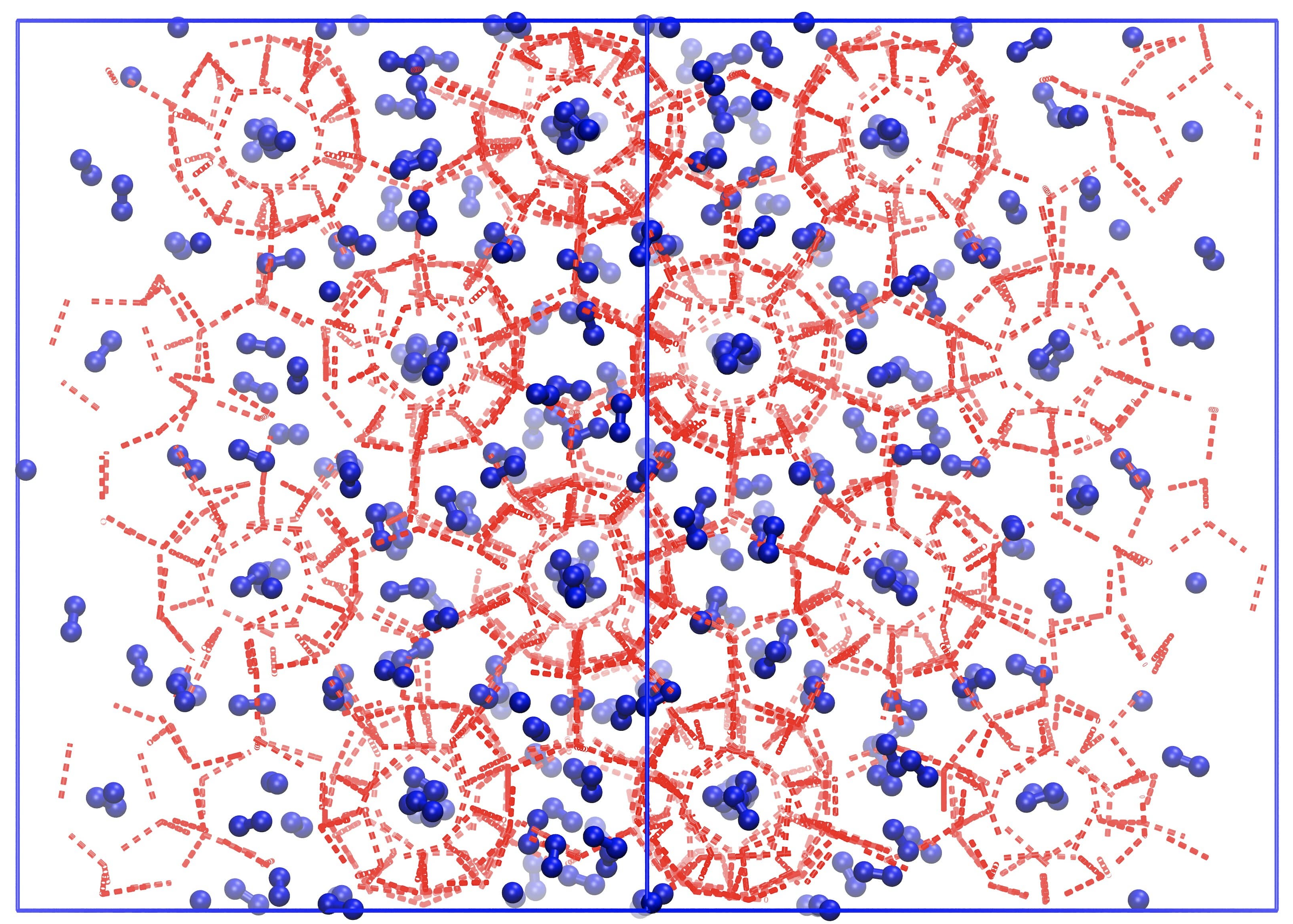}
\caption{Snapshots of the simulation box of the N$_{2}$ hydrate with single occupancy in D (small) and H (large) cages (left panel) and single occupancy in D (small) cages and double occupancy in H (large) cages (right panel) at $500\,\text{bar}$ and $240\,\text{K}$. Water molecules are represented using a red hydrogen bond representation, and N$_{2}$ molecules are represented using blue spheres for the nitrogen atoms.}
\label{figure0}
\end{figure*}

\section{Results}
In our previous work,~\cite{algaba2023b} we have studied the dissociation line of the N$_2$ hydrate from $500$ to $1500\,\text{bar}$. In this work, we extend the previous study to higher pressures. We now consider an extended range of pressures from $500$ to $4500\,\text{bar}$. Also, the effect of multiple occupancy of N$_2$ in the voids left by the hydrate structure is explicitly taken into account. Particularly, we examine the dissociation line and the driving force for nucleation when the large (H) cages of the sII hydrate are occupied by one molecule (single occupancy) and two molecules (double occupancy) of N$_2$. Fig.~\ref{figure0} shows two snapshots of the N$_{2}$ hydrate at $500\,\text{bar}$ and $240\,\text{K}$. The left panel shows the 
hydrate with single occupancy in both D (small) and H (large) cages. The right panel shows the hydrate with single occupancy in D (small) cages and double occupancy in H (large) cages. As can be seen, in both snapshots the D cages have the same number of molecules encaged. Although it is difficult to distinguish, these cages are only occupied by one N$_{2}$ molecule per cage. Note that the D cages are easily identified since one of their pentagonal faces is exposed directly to the plane of the snapshots. However, the H cages in the left panel are less occupied (just one N$_{2}$ molecule per cage), whereas in the same cages in the right panel the number of N$_{2}$ is larger since are occupied by two N$_{2}$ molecules per cage.

\subsection{L$_{\text{w}}$--L$_{\text{N}_{2}}$ equilibria: Solubility of N$_2$ in liquid water and interfacial tension}

In this Section, we focus on the study of the L$_{\text{w}}$--L$_{\text{N}_{2}}$ equilibria at high pressures (from $2500$ to $4500\,\text{bar}$) and several temperatures. This is necessary in order to obtain the solubility of N$_2$ in the aqueous solution phase, which is required to determine the dissociating line of the N$_2$ hydrate and the driving force for nucleation. We also focus on the determination of the L$_{\text{w}}$--L$_{\text{N}_{2}}$ interfacial tension. Notice that since this work is a continuation of our previous paper,~\cite{algaba2023b} we show not only the results obtained in this work (from $2500$ to $4500\,\text{bar}$) but also our previous results (from $500$ to $1500\,\text{bar}$). This allows to analyze the effect of pressure on the L$_{\text{w}}$--L$_{\text{N}_{2}}$ equilibria properties along the whole N$_2$ hydrate dissociation line.

\subsubsection{Solubility of N$_2$ in liquid water}
The solubility of N$_2$ in the aqueous solution phase when in contact via a planar interface with a rich-N$_2$ liquid phase is determined at several temperatures and pressures. Note that we have extended the range of pressures studied in our previous work\cite{algaba2023b} (from 500 to 1500\,\text{bar}) to obtain the whole N$_2$ hydrate dissociation line running from 500 up to $4500\,\text{bar}$, approximately. To this end, we have studied three additional pressures: 2500, 3500, and 4500\,\text{bar}. The initial simulation box used in this work is identical to that used in our previous study.~\cite{algaba2023b} An initial pure N$_2$ liquid phase, with 1223 N$_2$ molecules, is put in contact via a planar interface with an initial pure water phase with 2800 water molecules. We choose arbitrarily the $x$- and $y$-axes parallel to the L$_{\text{w}}$--L$_{\text{N}_{2}}$ interface, with the $z$-axis direction perpendicular to it. The total dimensions of the initial simulation box are $L_{x}=L_{y}=3.8\,\text{nm}$ and $L_{z}\simeq15\,\text{nm}$. Also, since both phases are in contact via a planar interface along the $z$-axis direction, simulations are carried out in the $NP_{z}\mathcal{A}T$ ensemble. In this ensemble, the interfacial area,
$\mathcal{A}=L_{x}\times L_{y}$, remains constant during the simulation, and only fluctuations of the simulation box along the $z$-axis direction are allowed. The normal component of the pressure tensor is fixed at the equilibrium pressure (2500, 3500, and $4500\,\text{bar}$). The simulations are run for 800 ns, with the first 200 ns used as an equilibration period and only the final 600 ns are used to average the thermodynamic properties of the system. Density profiles are obtained by dividing the simulation box into 200 slabs perpendicular to the $z$-axis direction. The position of the center of mass of the molecules is assigned to each slab and the density profiles are obtained from mass balance considerations.

\begin{figure}
\includegraphics[width=0.9\columnwidth]{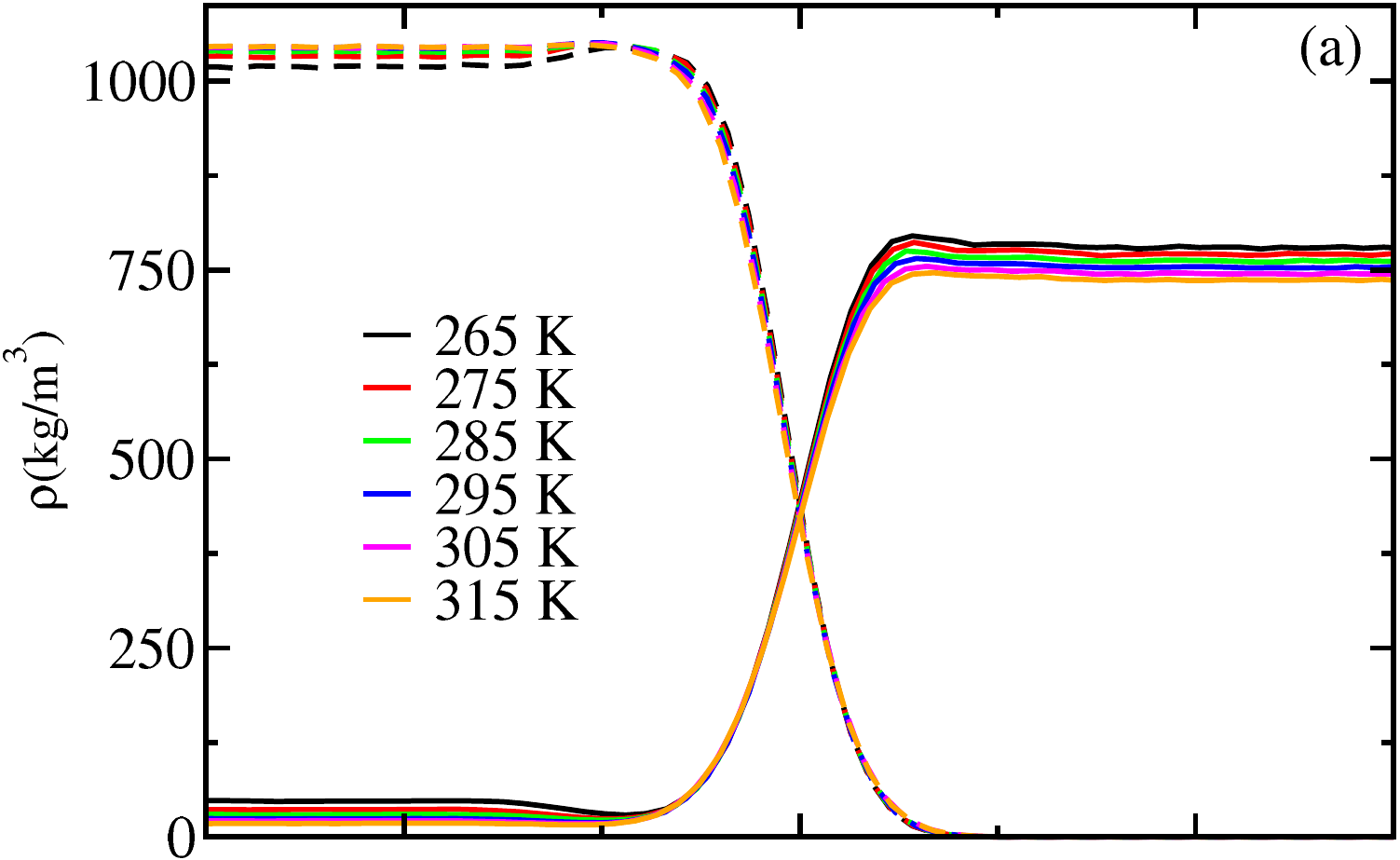}

\includegraphics[width=0.9\columnwidth]{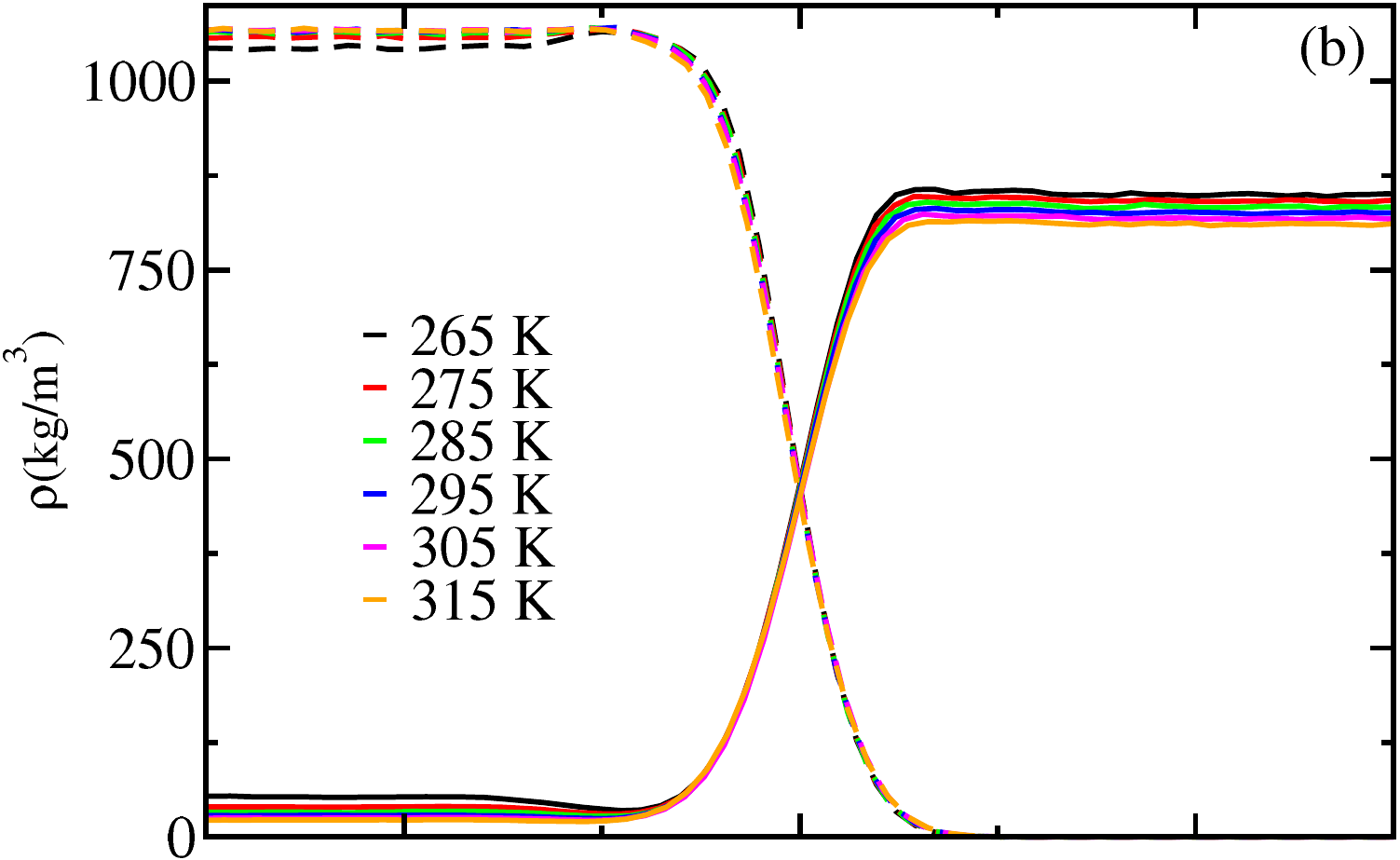}

\includegraphics[width=0.9\columnwidth]{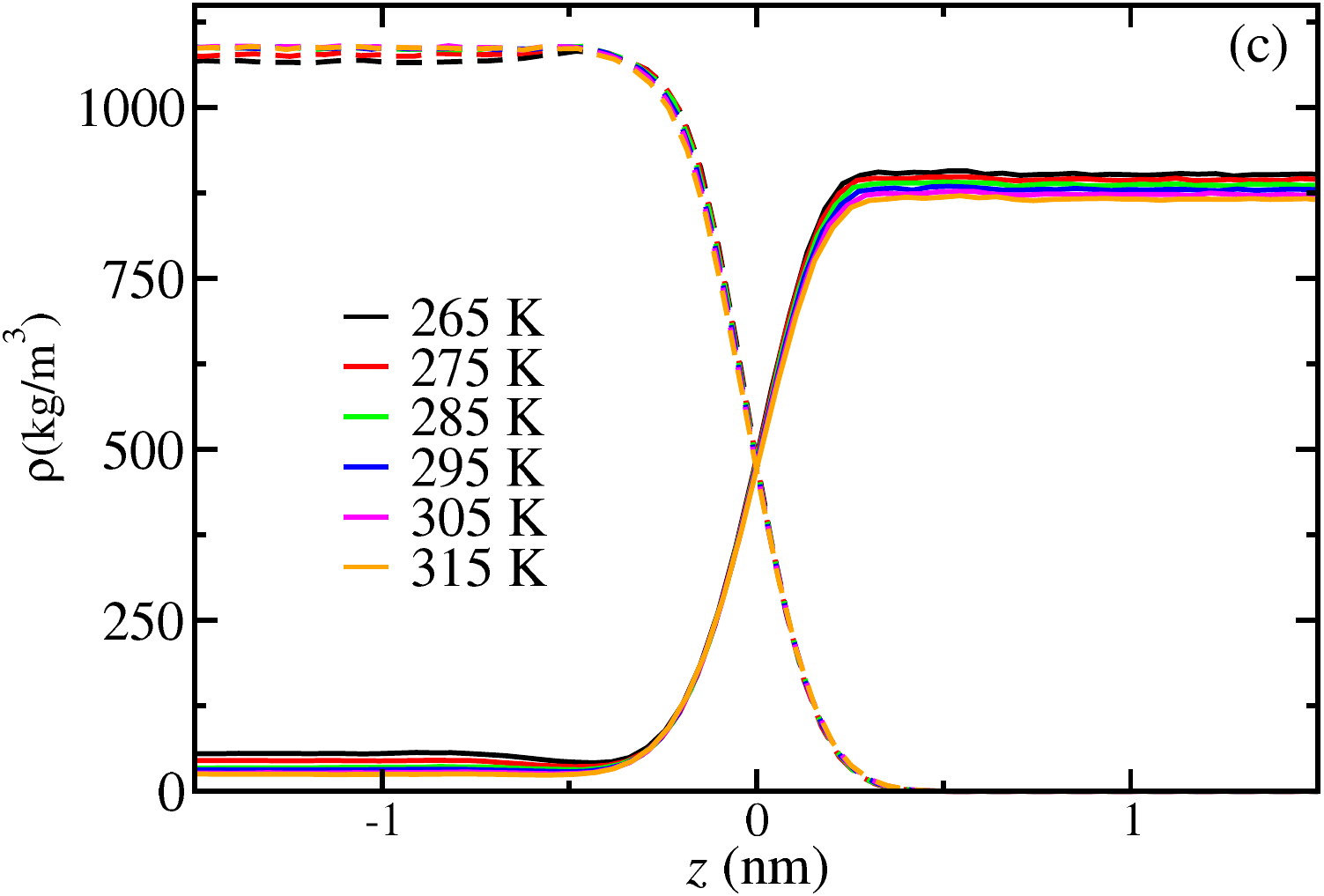}
\caption{Equilibrium L$_{\text{w}}$--L$_{\text{N}_{2}}$ density profiles, $\rho(z)$, along the interface of N$_{2}$ (continuous curves) and water (dhased curves) as obtained from MD $NP_{z}\mathcal{A}T$ simulations at $2500$ (a), $3500$ (b), and $4500\,\text{bar}$ (c) and several temperatures (see legends).}
\label{figure1}
\end{figure}

Figure~\ref{figure1} shows the L$_{\text{w}}$--L$_{\text{N}_{2}}$ equilibrium density profiles, $\rho(z)$, at 2500, 3500, and $4500\,\text{bar}$ and several temperatures. As can be observed in Fig.~\ref{figure1}, the density profiles at 2500, 3500, and $4500\,\text{bar}$ present the same qualitative behavior. We focus first on the density profiles of N$_2$. As can be seen, the density of N$_2$ in the N$_2$-rich phase (right side of the profile) decreases when the temperature is increased. Also, the amount of N$_2$ in the aqueous solution phase (left side of the profile) decreases when the temperature is increased. Notice that this is the expected behavior since the solubility of gases in water increases when the temperature is decreased. It is also interesting to observe the behavior at the interface where the density of N$_2$ shows preferential adsorption in the N$_2$-rich phase (at 2500 and 3500 bar) as well as preferential desorption in the aqueous solution phase (at 2500, 3500, and $4500\,\text{bar}$). The N$_2$ adsorption and desorption phenomena decrease when the temperature is increased. It is also interesting to point out that the adsorption of N$_{2}$ disappears at high pressures. In addition,  the desorption phenomenon decreases also with the pressure but it is still observable at $4500\,\text{bar}$.

In the case of the density profile of water, we observe that the amount of water in the N$_2$-rich phase is negligible and there is no temperature or pressure dependence. In the aqueous solution phase, the density of water increases when the temperature is increased. Also, it is possible to observe preferential adsorption of water close to the interface in the aqueous solution phase. This is correlated with the preferential desorption of N$_2$ at the same position along the interface. As well as in the case of N$_2$, the preferential adsorption of water decreases when the temperature and the pressure are increased (still observable at $4500\,\text{bar}$).

\begin{figure}
\includegraphics[width=\columnwidth]{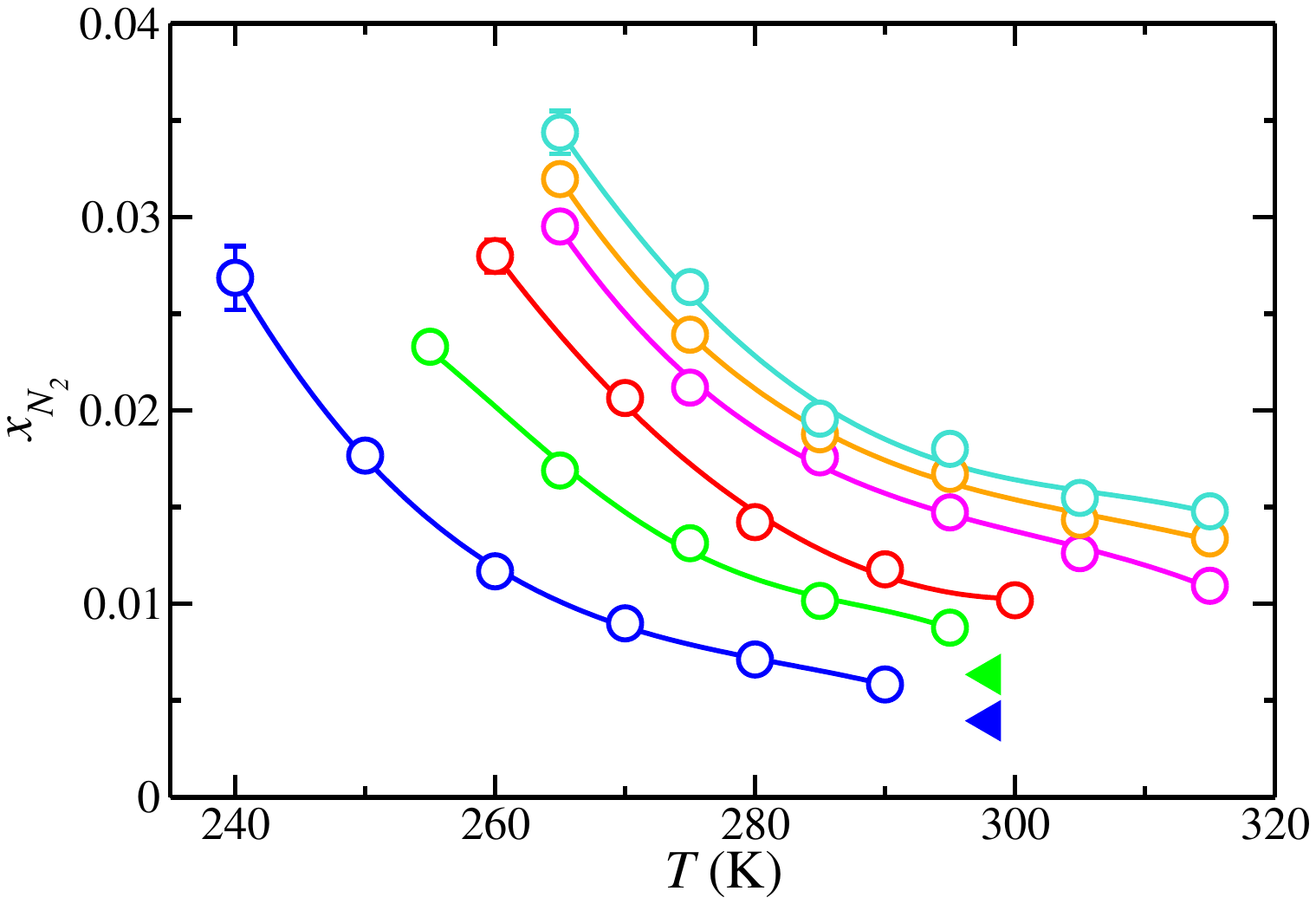}\\
\caption{Solubility of N$_2$ in the aqueous solution phase, as a function of temperature and at different pressures, from L$_{\text{w}}$--L$_{\text{N}_{2}}$ equilibria. Solubilities at $500$ (blue), $1000$ (green), and $1500\,\text{bar}$ (red) have been taken from our previous work\cite{algaba2023b} and solubilites at $2500$ (magenta), $3500$ (orange), and $4500\,\text{bar}$ (cyan) have been obtained in this work. The open circles correspond to solubility values obtained from MD $NP_{z}\mathcal{A}T$ simulations and the filled blue (500 bar) and green (1000 bar) up triangles represent the experimental data taken from the literature.~\cite{Wiebe1933} The curves are included as guides to the eyes.}
\label{figure2}
\end{figure}

Now we analyze the solubility of N$_2$ in water, as a function of the temperature, at the pressures considered in this work. From the analysis of the densities profiles it is possible to determine the solubility of N$_2$ in the aqueous solution as a function of the temperature for each pressure. The bulk densities of water and N$_2$ in the aqueous phase are calculated by averaging the values of the density profiles in each slab in the aqueous phase. Calculations are performed far enough from the interface to avoid the effect of preferential adsorption and desorption on the calculated equilibrium average densities in the bulk aqueous solution phase. Fig.~\ref{figure2} shows the solubility of N$_{2}$ in water obtained in this work at 2500, 3500, and $4500\,\text{bar}$, as well as the results of our previous work\cite{algaba2023b} at 500, 1000, and $1500\,\text{bar}$. As can be seen, the solubility decreases with temperature at a fixed pressure. This is the expected behavior of the solubility of a gas in water. In addition to this, the solubility of N$_{2}$ in water increases with the pressure at a fixed temperature. This effect is larger for low temperatures. We have also compared the predictions of the models for water and N$_{2}$ with experimental data taken from the literature.~\cite{Wiebe1933} Unfortunately, there are only two experimental points at 500 and 1000 bar and 298 K, at conditions similar to those considered in this study. In both cases, there is a good agreement between experimental data and simulation results.

\subsubsection{Interfacial tension}
From the L$_{\text{w}}$--L$_{\text{N}_{2}}$ simulations it is possible to determine the interfacial tension between the aqueous solution and the N$_2$-rich liquid phase from the diagonal components of the pressure tensor:~\cite{Rowlinson1982b,deMiguel2006c,deMiguel2006b}

\begin{equation}
\gamma=\frac{L_{z}}{2}\left[\left<P_{zz}\right>-\frac{\left<P_{xx
}\right>+\left<P_{yy}\right>}{2}\right]
\label{EQ_tension}
\end{equation}

\noindent
The additional $2$ factor in Eq.~\eqref{EQ_tension} arises because there are two L$_{\text{w}}$--L$_{\text{N}_{2}}$ interfaces. Similarly to the density profiles and solubilities, the L$_{\text{w}}$--L$_{\text{N}_{2}}$ interfacial tension has been obtained by averaging over the last 600 of the 800 ns used in the simulations. The first 200 ns are used to correctly equilibrate the system. The 600 ns of the production period are divided into 10 blocks of 60 ns each of them. The final value of the L$_{\text{w}}$--L$_{\text{N}_{2}}$ interfacial tension is obtained from the average value over the 10 blocks. The corresponding uncertainties are estimated using the standard deviation of the average.~\cite{Flyvbjerg1989a}

\begin{figure}
\includegraphics[width=\columnwidth]{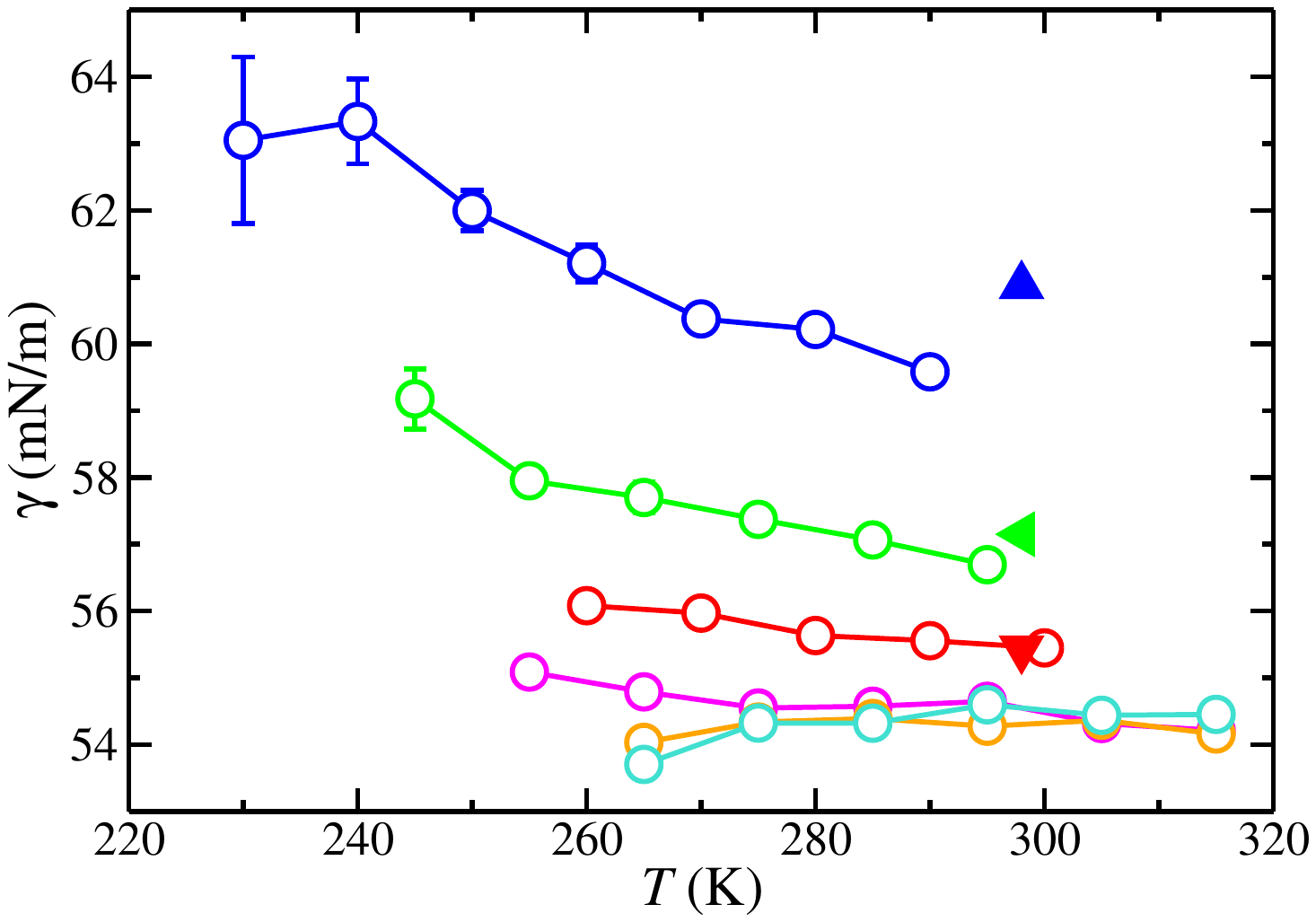}\\
\caption{L$_{\text{w}}$--L$_{\text{N}_{2}}$ interfacial tension, $\gamma$, as a function of the temperature between the aqueous solution and the N$_{2}$ liquid phase. Interfacial tension values at $500$ (blue), $1000$ (green), and $1500\,\text{bar}$ (red) have been taken from our previous work\cite{algaba2023b} and interfacial tension values at $2500$ (magenta), $3500$ (orange), and $4500\,\text{bar}$ (cyan) have been obtained in this work. The open circles correspond to solubility values obtained from MD $NP_{z}\mathcal{A}T$ simulations and the filled black triangles up ($500\,\text{bar}$), triangles left ($1000\,\text{bar}$), and triangles down ($1500\,\text{bar}$) are experimental data taken from the literature\cite{Wiegand1994}. The curves are included as guides to the eyes.}
\label{figure3}
\end{figure}

The results of the L$_{\text{w}}$--L$_{\text{N}_{2}}$ interfacial tension obtained in this work, from 2500 to $4500\,\text{bar}$, as well as the results obtained in our previous work ($500$, $1000$, and $1500\,\text{bar}$),~\cite{algaba2023b}  are shown in Fig.~\ref{figure3}. We first focus on the temperature dependence of $\gamma$ at each pressure. As can be seen, the interfacial tension decreases when the temperature is increased at the lowest pressures, $500$ and $1000\,\text{bar}$. However, at $1500\,\text{bar}$ the interfacial tension value still decreases when the temperature is increased but it is almost independent of the temperature in the range of temperatures simulated. Finally, at $2500$, $3500$, and $4500\,\text{bar}$, the L$_{\text{w}}$--L$_{\text{N}_{2}}$ interfacial tension value is completely independent of the temperature and pressure and equal to $\gamma\approx 54.2\,\text{mJ/m}^{2}$. Analyzing the behavior of $\gamma$ with the pressure is also interesting. When the temperature is fixed, the L$_{\text{w}}$--L$_{\text{N}_{2}}$ interfacial tension decreases when the pressure is increased from $500$ to $2500\,\text{bar}$. The $\gamma$ values at $3500$ and $4500\,\text{bar}$ are identical to those obtained at $2500\,\text{bar}$. In other words, the L$_{\text{w}}$--L$_{\text{N}_{2}}$ interfacial tension becomes independent of temperature and pressure at these conditions.

\subsection{H--L$_{\text{w}}$ equilibria: Solubility of N$_2$ in liquid water when in contact with N$_2$ hydrate.}

In this section, we focus on determining the solubility of N$_2$ in the aqueous solution phase when in contact via a planar interface with a N$_2$ hydrate phase. In addition, we explicitly account for the effect of the occupancy of the N$_{2}$ in the cages of the sII structure on the solubility of N$_{2}$ in the aqueous phase. Particularly, we first consider hydrate phases with only one N$_{2}$ molecule occupying each of the D (small) and H (large) hydrate cages (single occupancy). We then consider hydrate phases with one molecule of N$_2$ in each (small) D cage but two molecules of N$_2$ inside the H (large) cages of the sII structure (double occupancy).

\subsubsection{N$_2$ hydrate with single occupancy}

\begin{figure}
\includegraphics[width=0.9\columnwidth]{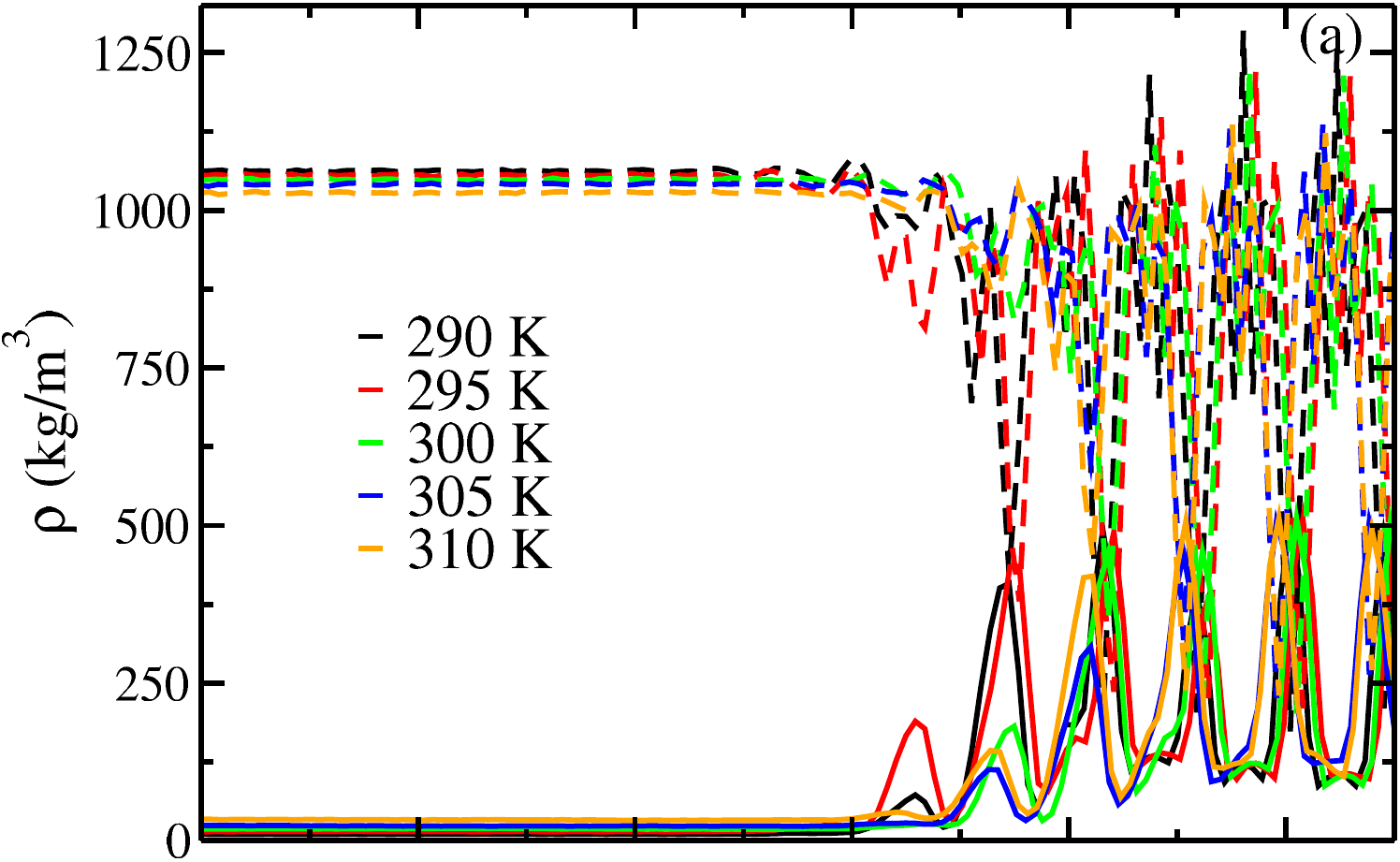}

\includegraphics[width=0.9\columnwidth]{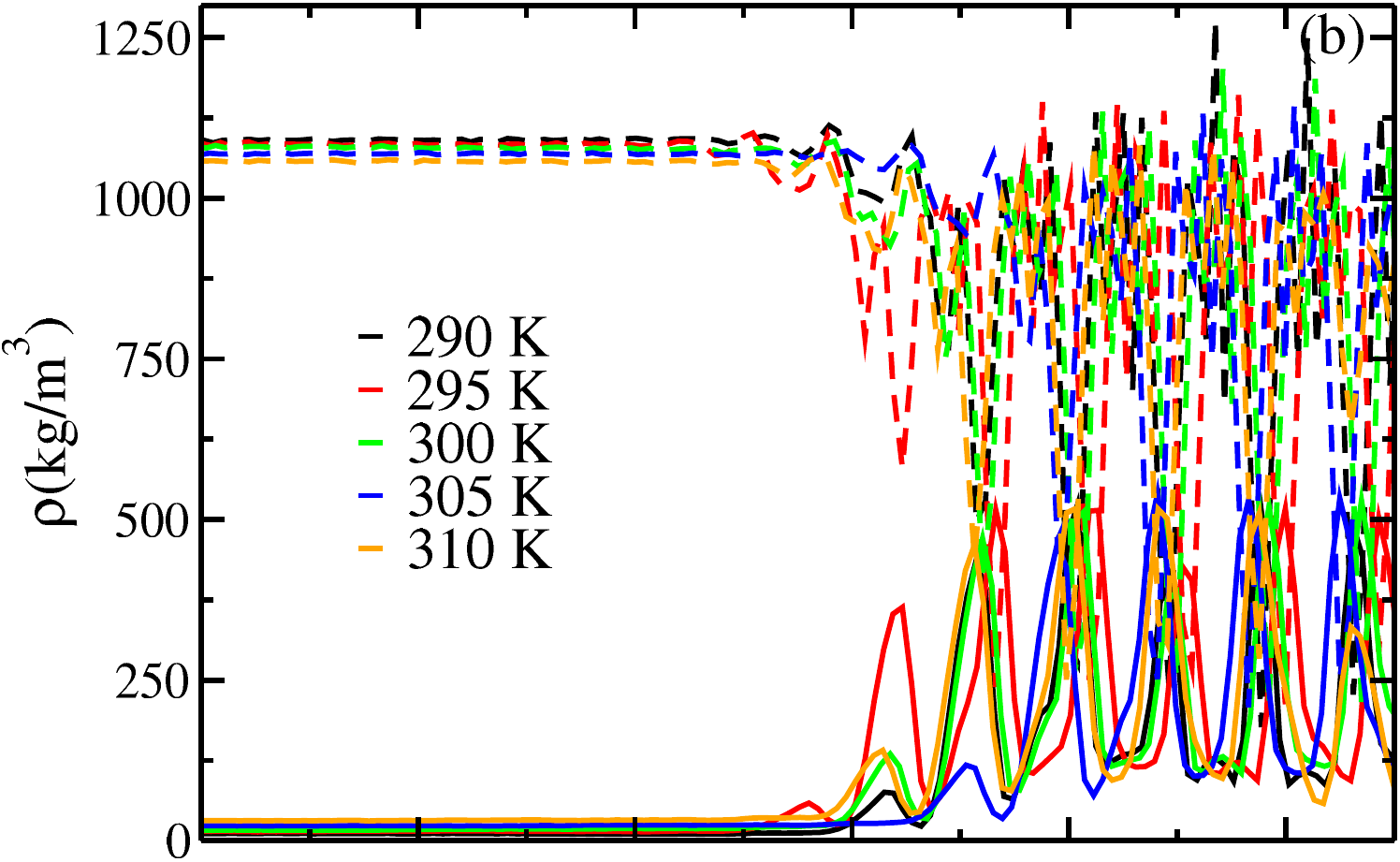}

\includegraphics[width=0.90\columnwidth]{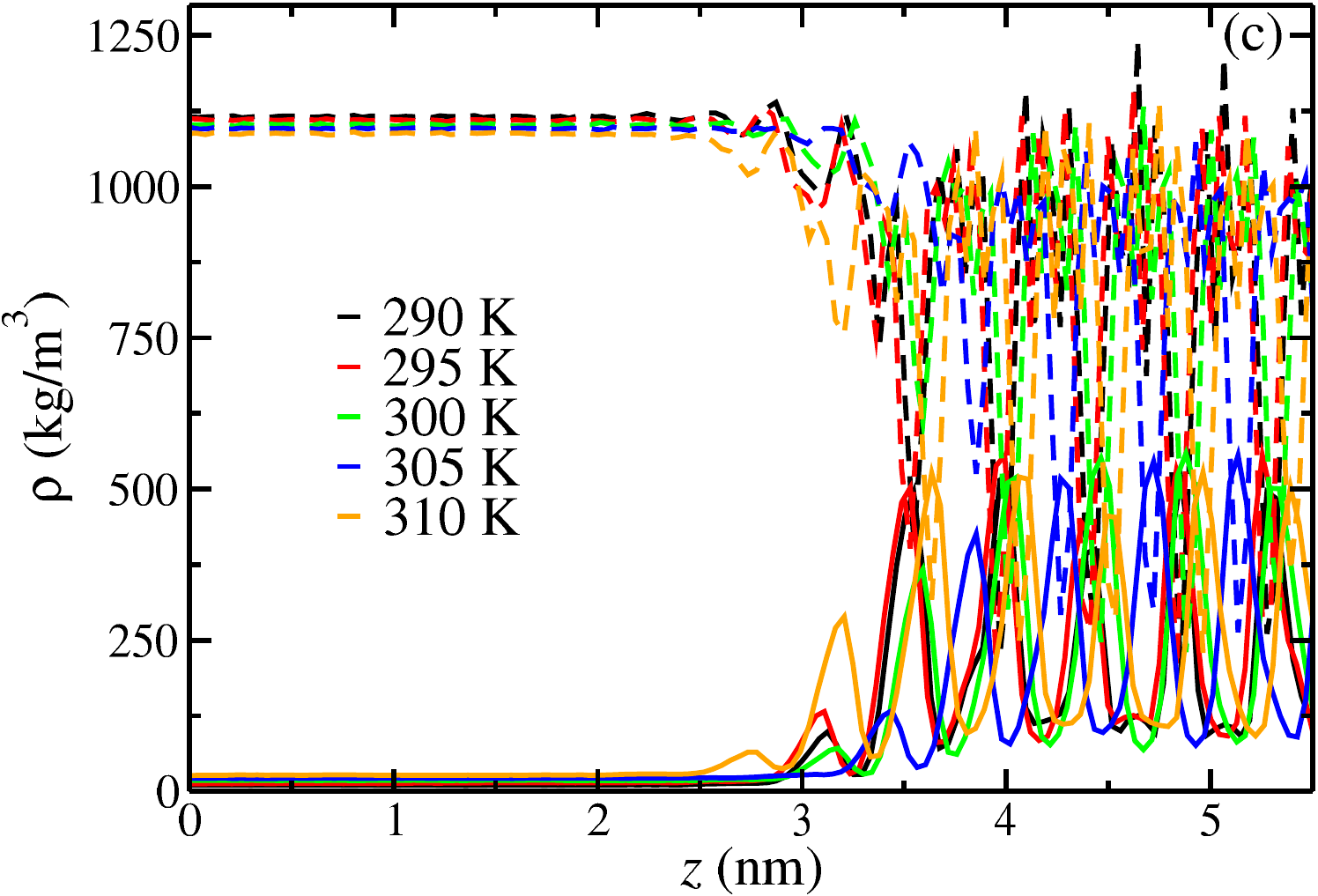}
\caption{Equilibrium H--L$_{\text{w}}$ density profiles, $\rho(z)$, along the interface of N$_2$ (continuous curves) and water (dashed curves)
as obtained from MD $NPT$ simulations at $2500$ (a), $3500$ (b), and $4500\,\text{bar}$ (c) and several temperatures (see legends). In all cases, single occupancy of the hydrate is assumed.}
\label{figure4}
\end{figure}

We extend our previous study\cite{algaba2023b} and determine the H-L$_{\text{w}}$ density profiles at $2500$, $3500$, and $4500\,\text{bar}$. This allows us to analyze the density profiles of water and N$_{2}$ molecules in a wide range of pressures, from $500$ up to $4500\,\text{bar}$. Note that this is the range at which the dissociation line of the N$_{2}$ hydrate is available from experiments.

The initial simulation box is prepared by replicating a unit cell of the sII crystallographic structure twice in each spatial direction. We take into account explicitly the proton-disordered nature of hydrates.~\cite{Buch1998a,Bernal1933a} The final $2\times 2\times 2$ hydrate phase, assuming single occupancy, is formed by $1088$ and $192$ water and N$_{2}$ molecules, respectively. The simulation box is equilibrated during $20\,\text{ns}$ in the isothermal-isobaric or $NPT$ ensemble using an anisotropic barostat to avoid stress in the crystallographic hydrate structure. Then, the $2\times 2\times 2$ hydrate phase is put in contact along the $z$-axis direction with a pure phase formed from $2176$ water molecules. Note that we use the double number of water molecules in this phase compared to the hydrate phase. The final dimensions of the initial simulation box used in this work for all the pressures are $L_x=L_y\approx 3.5$ nm and $L_z\approx 9\,\text{nm}$.

It is important to mention that the size of the system is not chosen arbitrarily. We double the number of molecules in the initial water phase for two reasons. Firstly, 
the solubility of N$_{2}$ in water is very low.  Increasing the size of the aqueous phase provokes an increase in the number of N$_{2}$ molecules in that phase, in equilibrium with the hydrate. This helps to get better statistics and more accurate solubility values. Secondly, it avoids finite-size effects on the N$_{2}$ solubility values and the dissociation line determination. Finite-size effects, as well as the effect of dispersive interactions on the dissociation temperature of hydrates, have been recently studied and quantified by some of us.~\cite{Blazquez2024a,Algaba2024a,Algaba2024b}

As in our previous work,~\cite{algaba2023b} we determine the density profiles of N$_2$ and water of the H--L$_\text{w}$ equilibrium systems (single occupancy) at $2500$, $3500$, and $4500\,\text{bar}$ and several temperatures by running simulations during $800\,\text{ns}$. Density profiles are obtained from the last $600\,\text{ns}$  by dividing the simulation box into $200$ slabs perpendicular to the $z$-axis direction. The position of the center of mass of the molecules is assigned to each slab and the density profiles are determined from mass balance considerations. Finally, only one of the interfaces is presented to avoid repetition. As can be seen in Fig.~\ref{figure4}, the density profiles show the same qualitative behavior with the temperature at each pressure. When the temperature increases, the density of water and  N$_{2}$ in the aqueous phase (left side of the profiles) decreases and increases respectively. The right side of the profiles shows the characteristic density peaks associated with the water and N$_{2}$ molecules corresponding to the crystallographic positions of both components in the hydrate phase.

\begin{figure}
\includegraphics[width=\columnwidth]{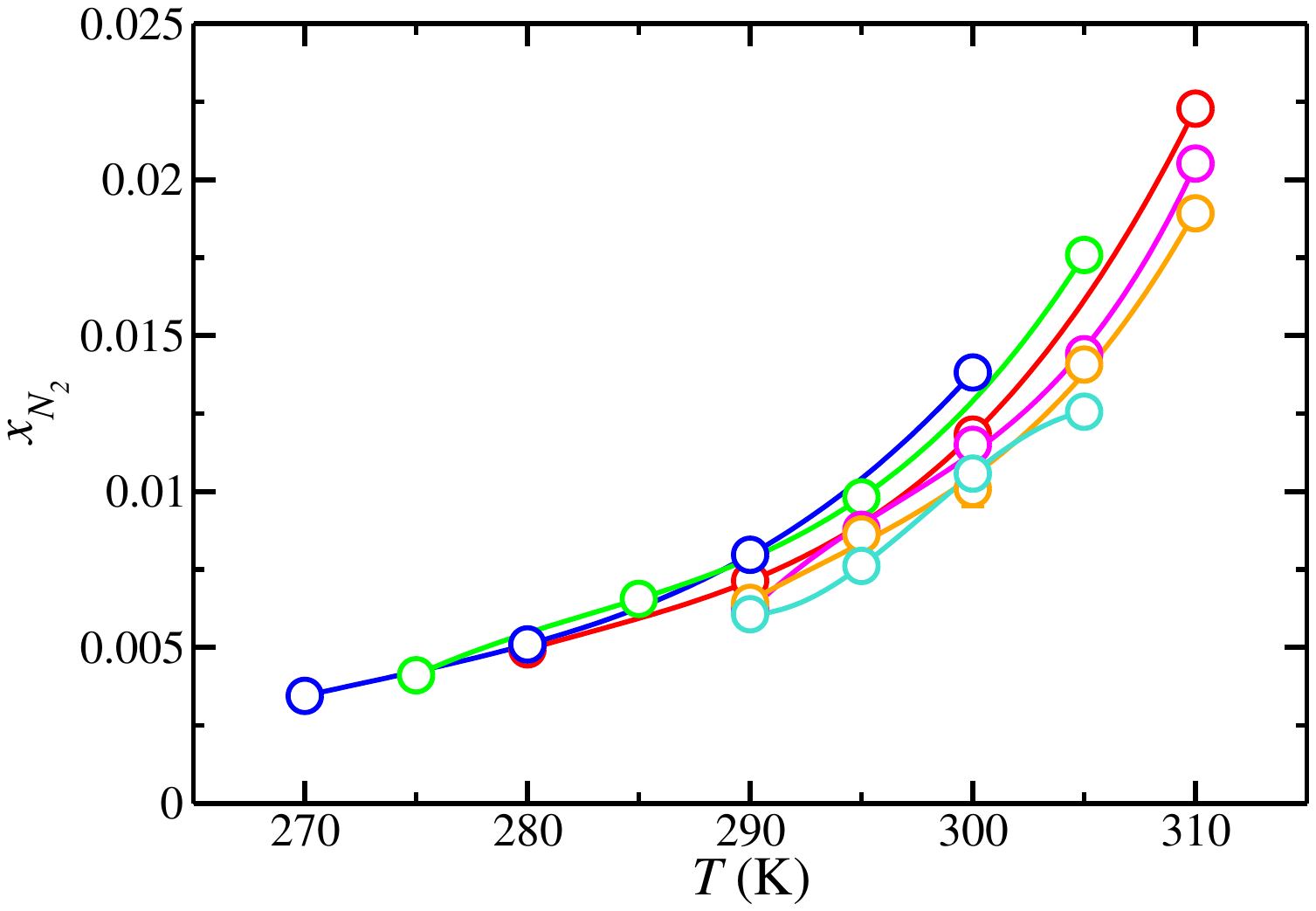}\\
\caption{Solubility of N$_2$ in the aqueous solution phase, as a function of temperature and at different pressures, from H--L$_{\text{w}}$ equilibria. Solubilities at $500$ (blue), $1000$ (green), and $1500\,\text{bar}$ (red) have been taken from our previous work \cite{algaba2023b} and solubilities at $2500$ (magenta), $3500$ (orange), and $4500\,\text{bar}$ (cyan) have been obtained in this work. The symbols correspond to solubility values obtained from MD $NPT$ simulations and the curves are included as guides to the eyes. In all cases, single occupancy in the hydrate is assumed.}
\label{figure5}
\end{figure}

We follow the same procedure used in the case of the L$_{\text{w}}$--L$_{\text{N}_{2}}$ equilibrium and determine the solubility of N$_2$ in the aqueous phase when it is contact with the single-occupied N$_2$ hydrate phase via a planar interface (from the analysis of the corresponding density profiles). The solubility of N$_{2}$ in water obtained at $2500$, $3500$, and $4500\,\text{bar}$ are shown in Fig~\ref{figure5}. Note that we have also included the results determined at $500$, $1000$, and $1500\,\text{bar}$ from our previous work.~\cite{algaba2023b} As can be seen, the solubility of N$_{2}$ in water shows the same qualitative behavior with the temperature at all the pressures: when the temperature is increased, the solubility of N$_{2}$ in water increases. This is the expected behavior since the hydrate becomes less stable when the temperature increases. According to this, part of the hydrate melts and releases N$_2$ into the aqueous solution phase. This effect is more important as the temperature increases. Here it is important to remark that the initial aqueous solution phase is pure water. Consequently, part of the hydrate has to melt to reach the equilibrium solubility value of N$_{2}$ in water. Note that this is crucial to study the effect of the occupancy on the solubility of N$_2$ in the aqueous phase and the dissociation temperature of the N$_2$ hydrate. Since the hydrate does not grow, the stoichiometric of the hydrate phase remains constant and equal to the initial one throughout the whole simulation.

\subsubsection{N$_2$ hydrate with double occupancy}

\begin{figure}
\includegraphics[width=0.9\columnwidth]{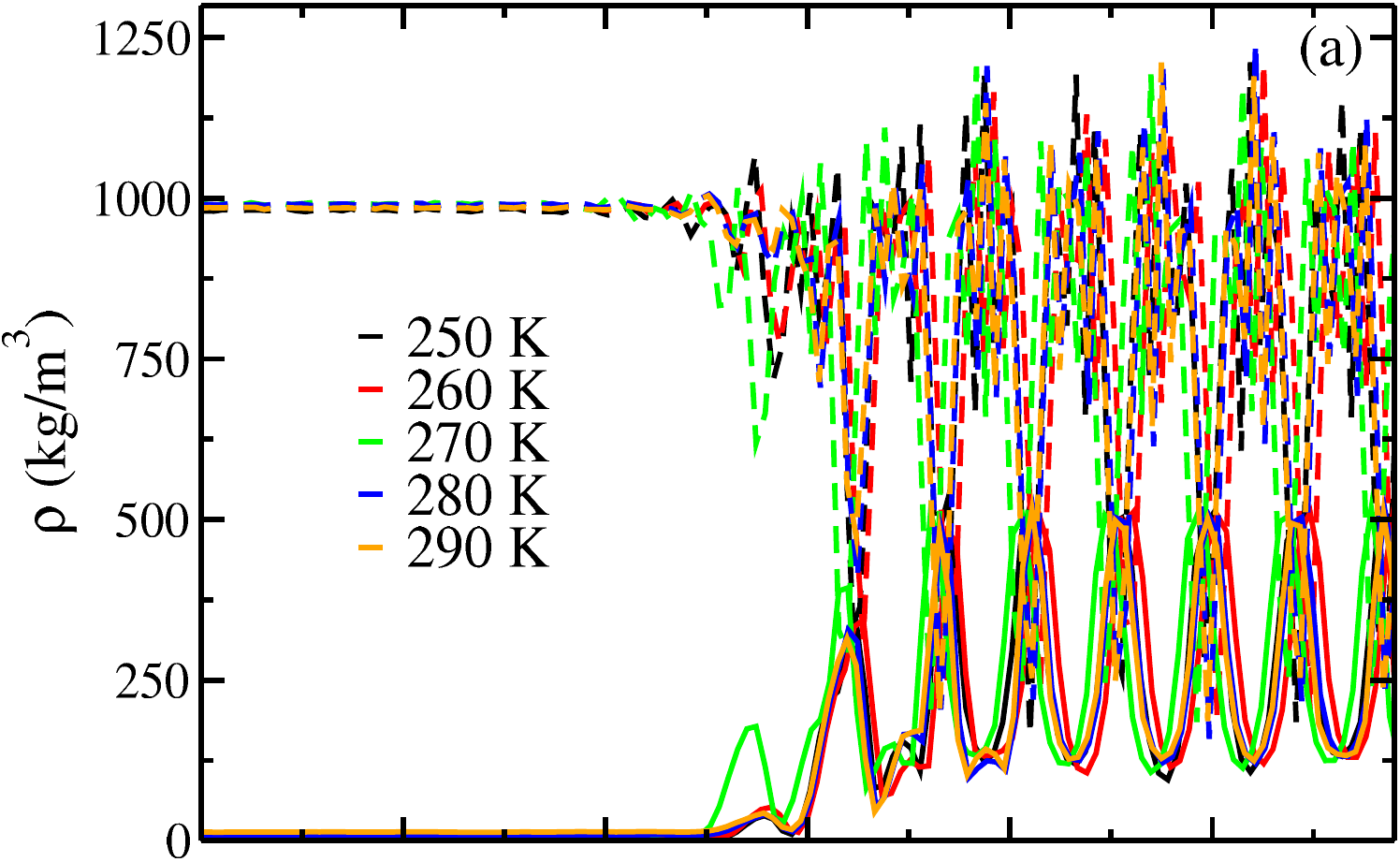}

\includegraphics[width=0.9\columnwidth]{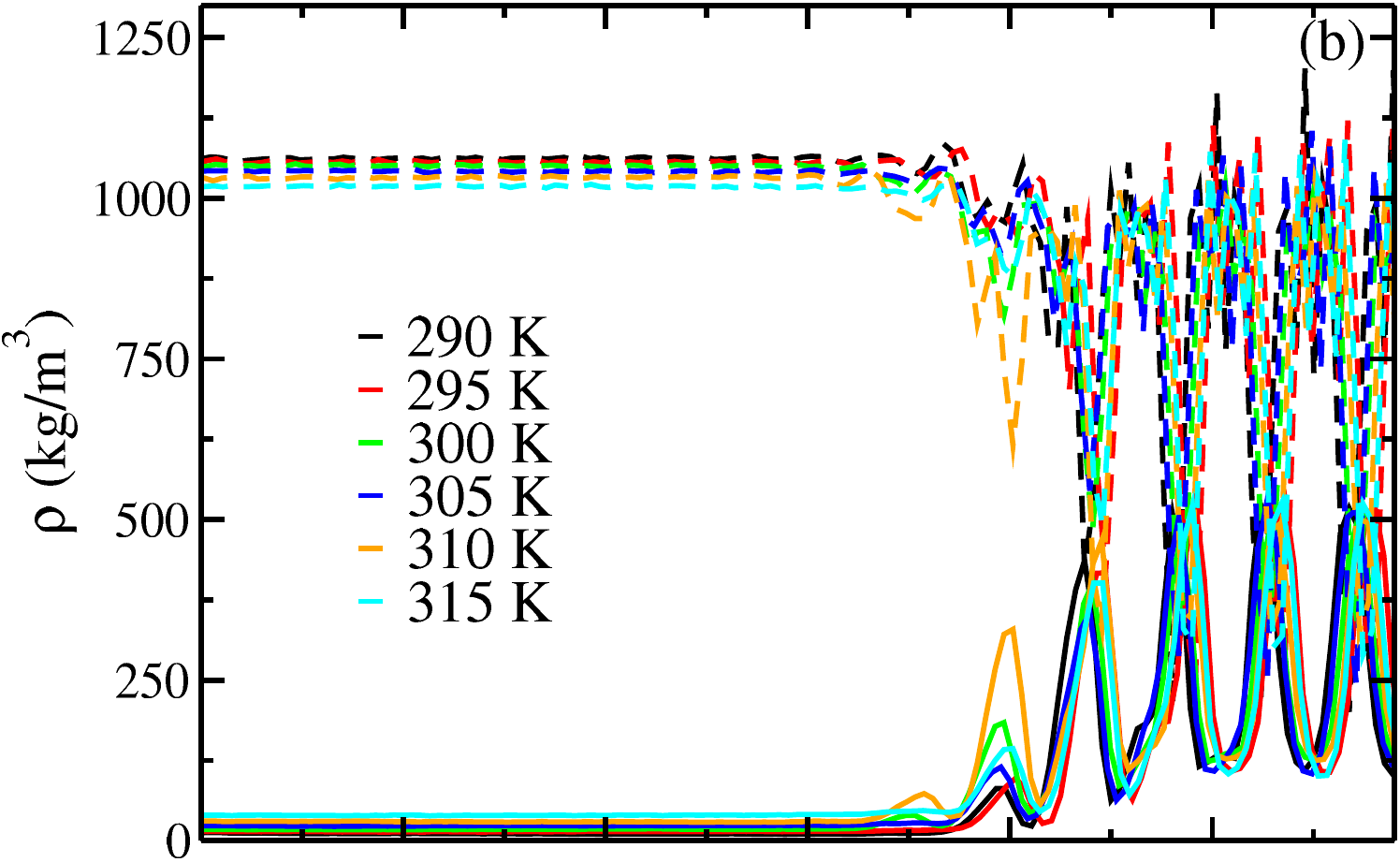}

\includegraphics[width=0.9\columnwidth]{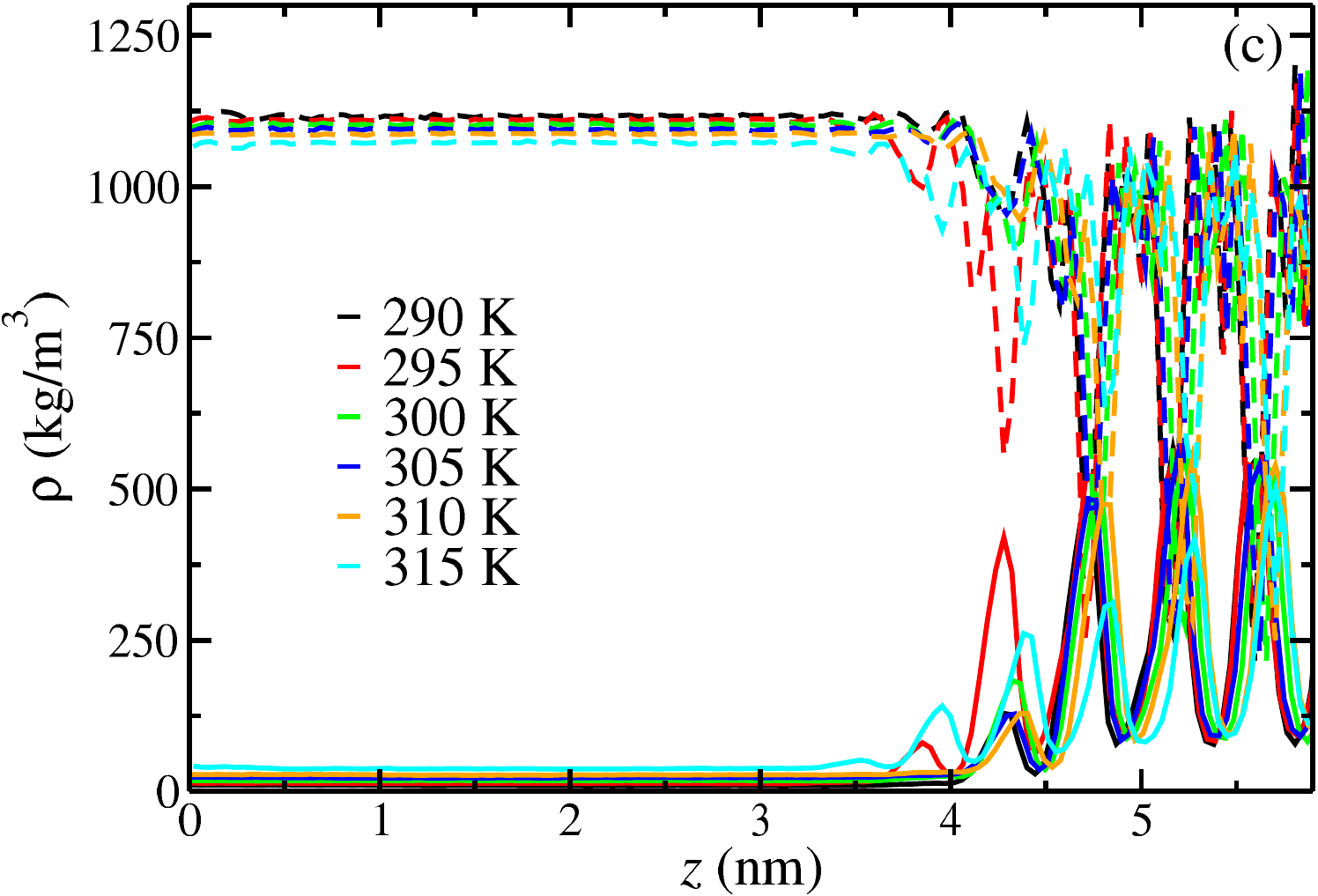}
\caption{Equilibrium H--L$_{\text{w}}$ density profiles, $\rho(z)$, along the interface of N$_2$ (continuous curves) and water (dashed curves)
as obtained from MD $NPT$ simulations at $500$ (a), $2500$ (b), and $4500\,\text{bar}$ (c) and several temperatures (see legends). In all cases, double occupancy of the hydrate is assumed.}
\label{figure6}
\end{figure}

We now consider the effect of the double occupancy on the solubility of N$_{2}$ in the aqueous phase when in contact with the hydrate. The initial simulation box is prepared in the same way as in Section III.B.1. In this case, the initial hydrate phase contains a molecule of N$_{2}$ in each small D cage (5$^{12}$) while the large H cages present double occupancy (two N$_{2}$ molecules in each 5$^{12}6^4$ H cage). The position of the molecules of N$_{2}$ in the H cages is selected to avoid repulsion or overlap caused by the double occupancy inside the cage. The final $2\times 2\times 2$ hydrate phase, assuming double occupancy, is formed from $1088$ and $256$ water and N$_{2}$ molecules, respectively. The initial hydrate phase is equilibrated during $20\,\text{ns}$ in the isothermal-isobaric or $NPT$ ensemble using an anisotropic barostat to avoid any stress from the crystallographic hydrate structure. Finally, as in the case of the single occupancy, a pure phase formed from $2176$ water molecules is put in contact with the hydrate phase along the $z$-axis direction. The final dimensions of the initial simulation box are identical to that used in the single-occupied system ($L_{x}=L_{y}\approx 3.5\,\text{nm}$ and $L_{z}\approx 9\,\text{nm}$). As in the case considered in Section III.B.1, the final size of the system is large enough to avoid any finite-size effect.~\cite{Blazquez2024a,Algaba2024a} We also ensure that the stoichiometric of the final hydrate phase in contact with the initial pure water phase is the same as the initial double-occupied hydrate phase since the hydrate phase never grows in these simulations (see Section III.B.1 for further details).

Following the same procedure as in the case of the single-occupied N$_2$ hydrate, we obtain the H~--~L$_\text{w}$ density profiles at $500$, $1000$, $1500$, $2500$, $3500$, and $4500\,\text{bar}$ and several temperatures. Note that these are the same pressures studied in the case of a single-occupied N$_2$ hydrate in this work ($2500$, $3500$, and $4500\,\text{bar}$) in combination with our previous results ($500$, $1000$, and $1500\,\text{bar}$).~\cite{algaba2023b} Fig.~\ref{figure6} shows the density profiles obtained at $500$, $2500$, and $4500\,\text{bar}$ when an aqueous solution phase is in contact with a double-occupied N$_2$ hydrate phase. The density profiles obtained at $1500$ and $3500\,\text{bar}$ show the same qualitative behavior and we have not represented them to avoid repetition. As can be observed, the qualitative behavior of the density profiles is the same at all the pressures and, also, is the same as that shown in Fig.~\ref{figure4} in the case of a single-occupied hydrate. When the temperature increases, the water/N$_2$ density decreases/increases in the aqueous phase. This is the expected behavior since the hydrate becomes less stable as the temperature is increased, and part of the hydrate melts increasing the amount of N$_{2}$ released into the aqueous phase. 

\begin{figure}
\includegraphics[width=\columnwidth]{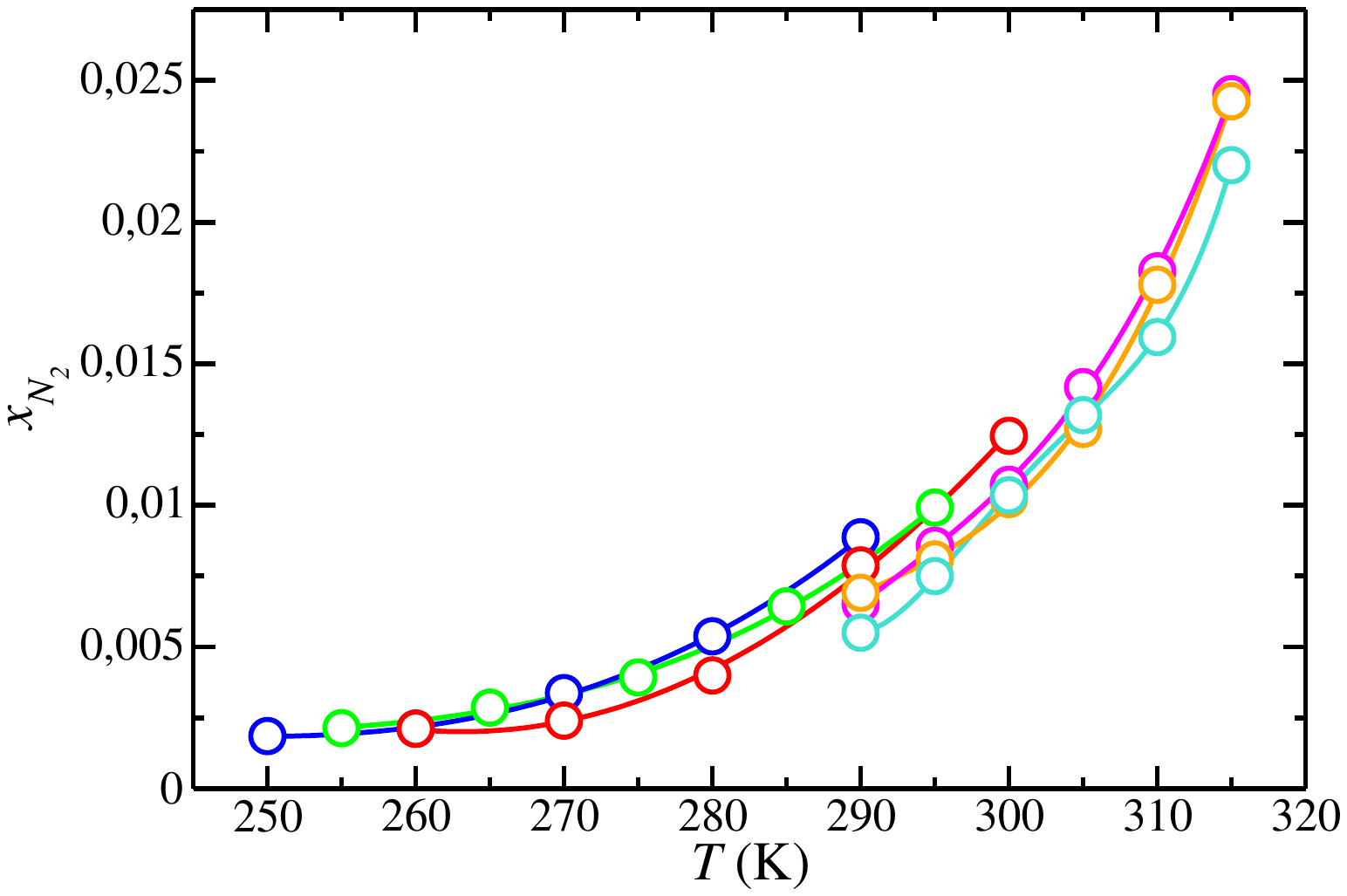}\\
\caption{Solubility of N$_2$ in the aqueous solution phase, as a function of temperature at $500$ (blue), $1000$ (green), $1500$ (red), $2500$ (magenta), $3500$ (orange), and $4500\,\text{bar}$ (cyan) when the solution is in contact with the hydrate phase via a planar interface. The symbols correspond to solubility values obtained from MD $NPT$ simulations and the curves are included as guides to the eyes. In all cases, double occupancy in the hydrate is assumed.}
\label{figure7}
\end{figure}

Following the analysis of the density profiles performed in Section III.B.1, we determine the solubility of N$_{2}$ in the aqueous phase when it is in contact with a double-occupied N$_{2}$ hydrate phase via a planar interface. As can be observed, Figs.~\ref{figure5} and \ref{figure7} are almost identical. There are no substantial differences between the solubility values of N$_{2}$ in water obtained when a single- or double-occupied hydrate is in contact with an aqueous phase. As we will see later in Sections III.C and III.D, the occupancy has a negligible effect on the dissociation temperatures of the N$_{2}$ hydrate.

\subsection{Three phase coexistence line ($T_3$) determination from solubility method}

Following the solubility method,~\cite{Grabowska2022a,Algaba2023a,algaba2023b} 
it is possible to obtain the hydrate~--~aqueous solution~--~N$_{2}$-rich three-phase or dissociation temperature, $T_{3}$, of the hydrate at a given pressure from the solubility of N$_{2}$ in water when in contact with the hydrate phase and the N$_{2}$-rich liquid phase via planar interfaces. At a given value of pressure, the temperature at which the solubility of N$_2$ in the aqueous phase reaches the same value from the L$_{\text{w}}$~--~L$_{\text{N}_{2}}$ and the H~--~L$_{\text{w}}$ equilibria is the temperature at which the three phases coexist, $T_{3}$.

\begin{figure*}
     \centering
         \centering
         \includegraphics[width=0.30\textwidth]{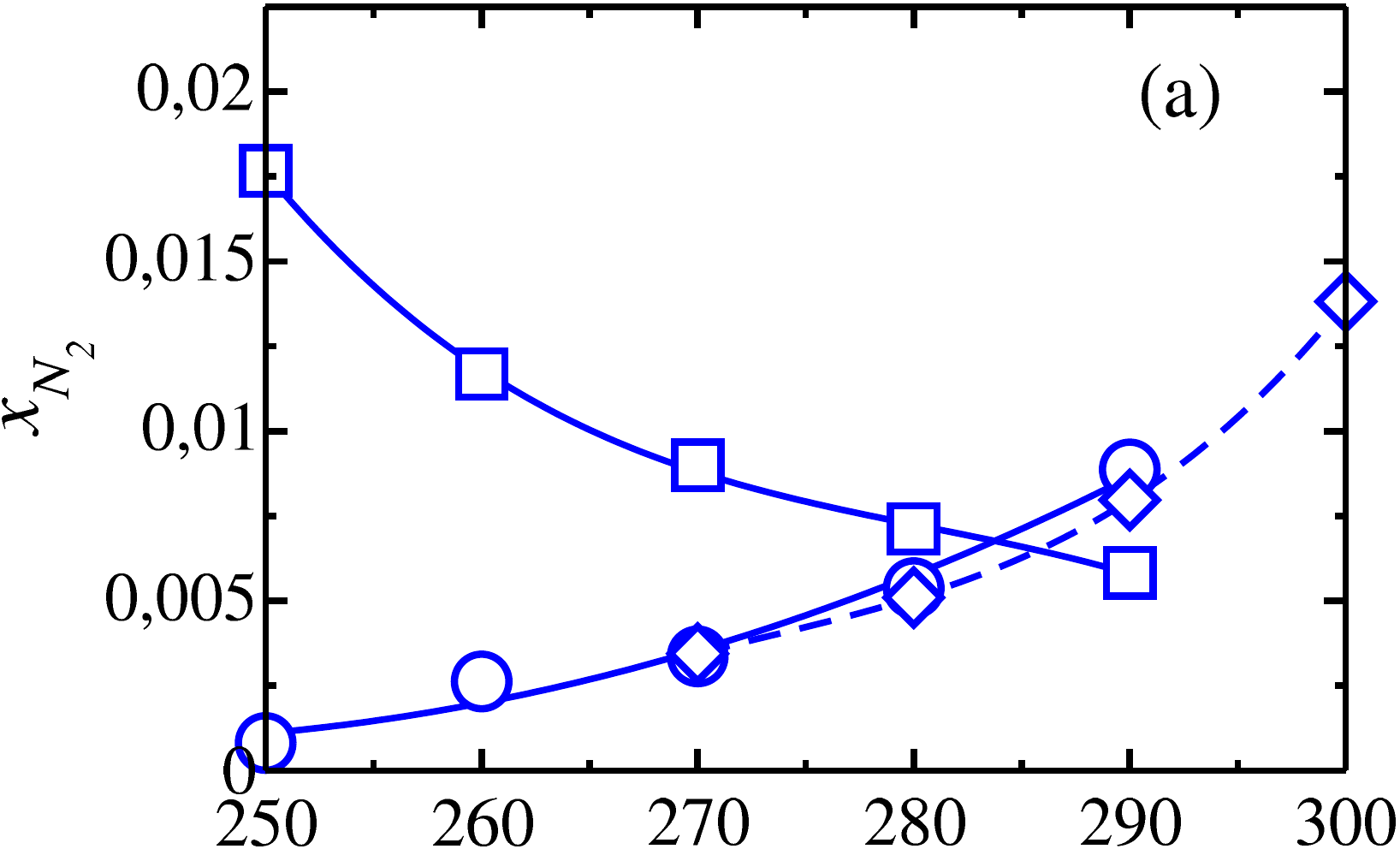}
         \includegraphics[width=0.28\textwidth]{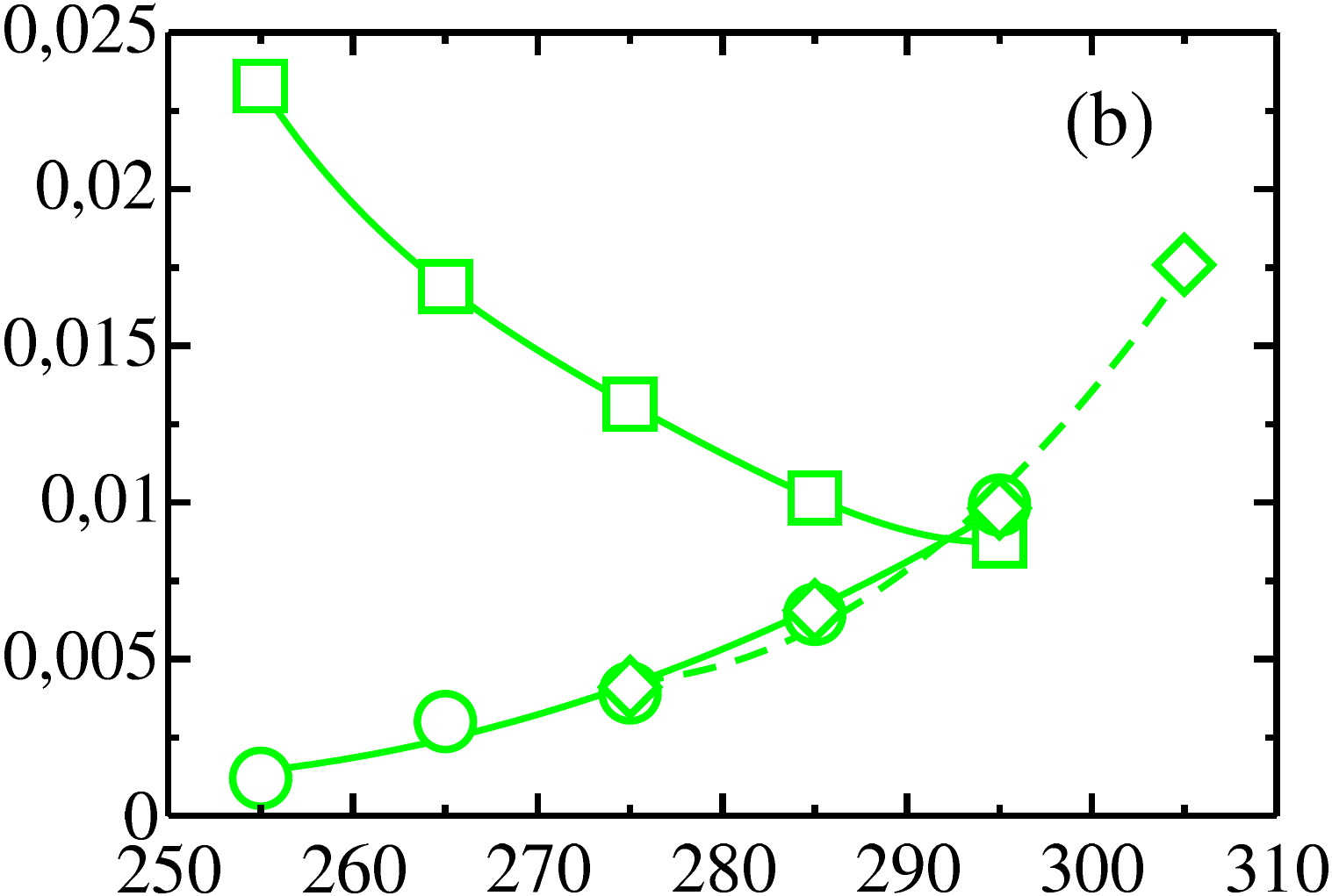}
         \includegraphics[width=0.28\textwidth]{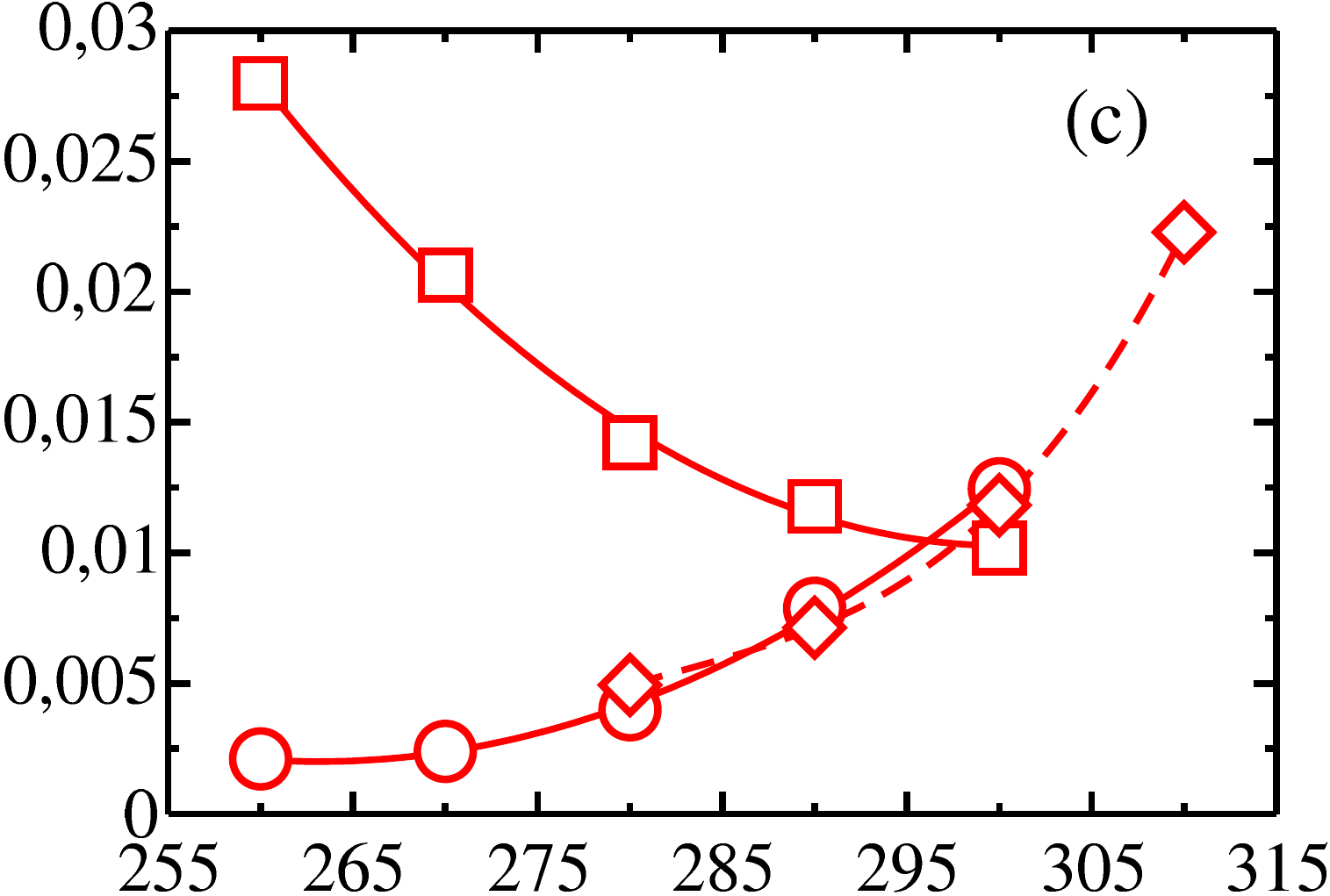}
   \\   
   \vspace{0.5cm}
\includegraphics[width=0.30\textwidth]{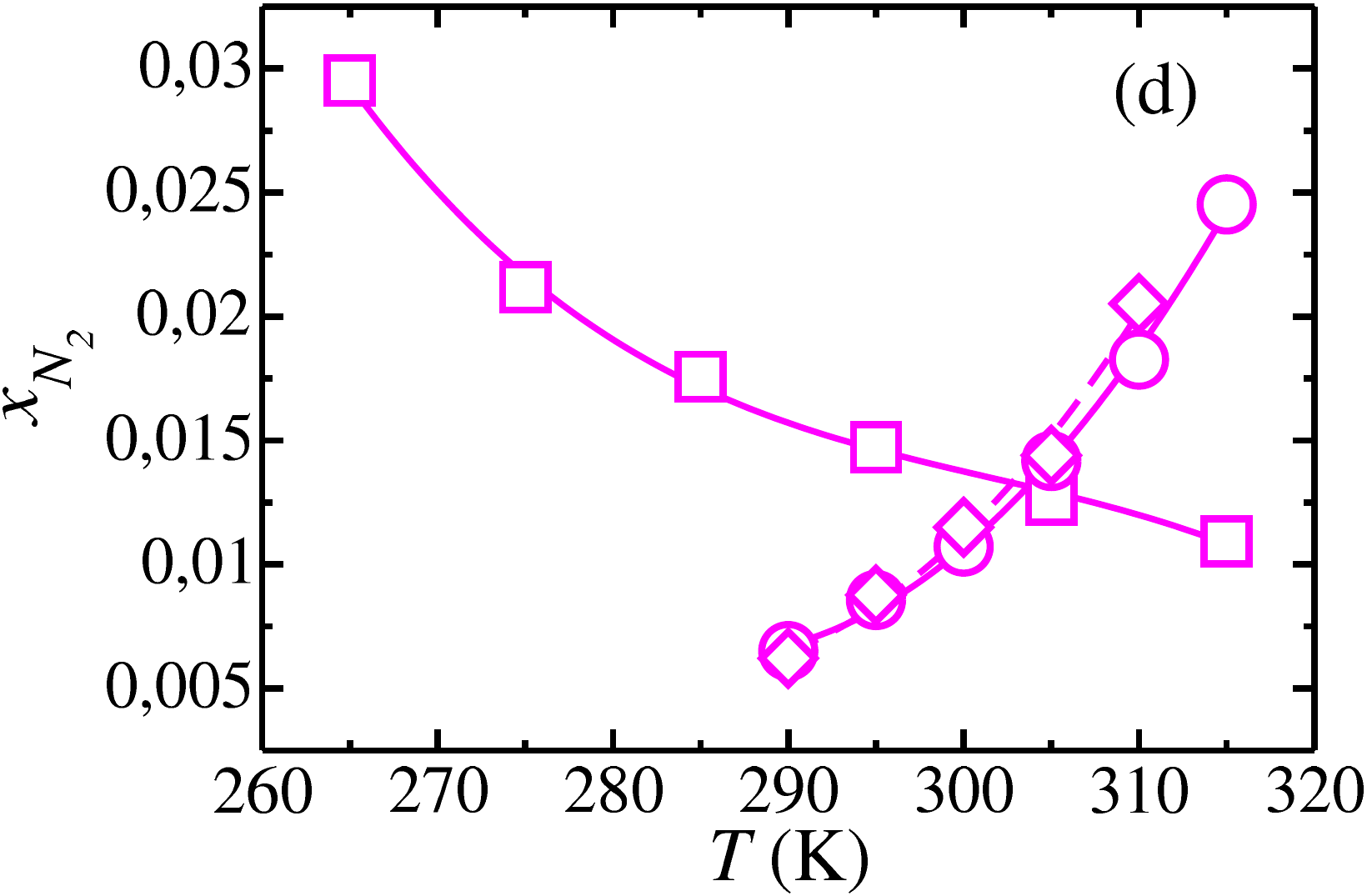}
         \includegraphics[width=0.28\textwidth]{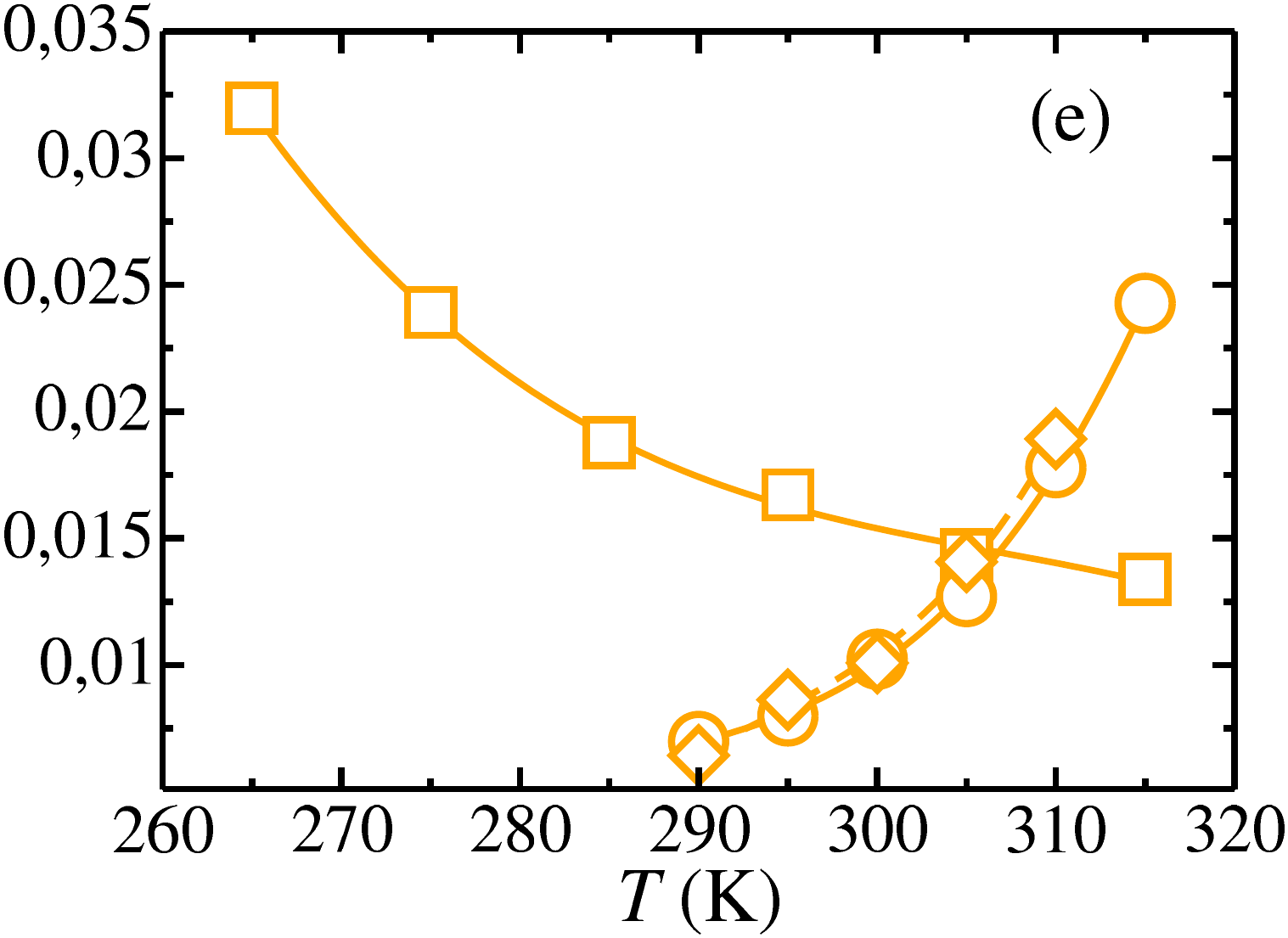}
          \includegraphics[width=0.28\textwidth]{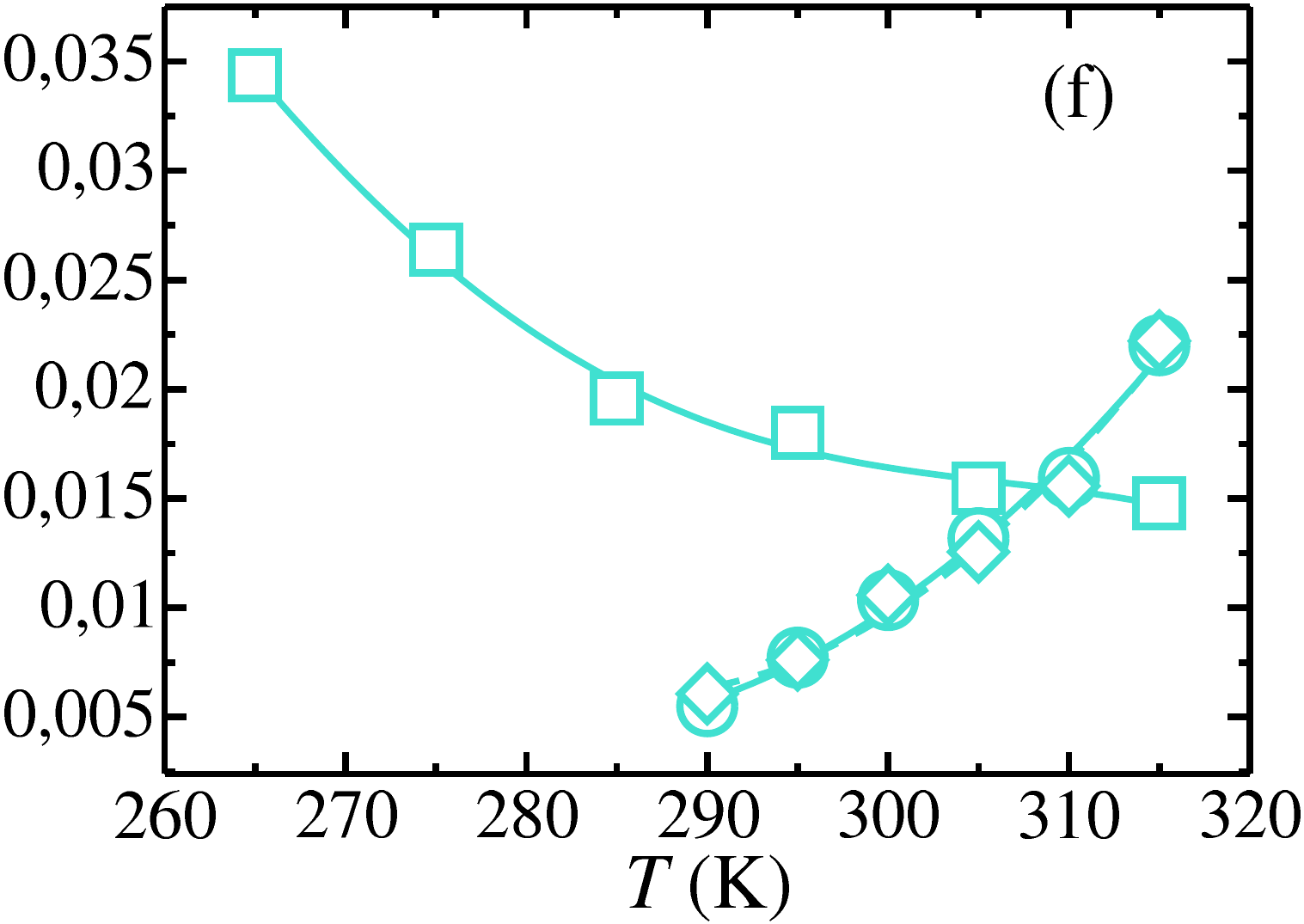}
\caption{Solubility of N$_2$ in the aqueous phase, as a function of temperature, at $500$ (blue), $1000$ (green), $1500$ (red), $2500$ (magenta), $3500$ (orange), and $4500\,\text{bar}$ (cyan) when the solution is in contact with the N$_{2}$ liquid and the hydrate phases via a planar interface. The squares represent the solubility of N$_{2}$ in the aqueous phase when it is in contact with the N$_{2}$ liquid. The diamonds and circles represent the solubility of N$_{2}$ in the aqueous phase when in contact with a single- and double-occupied hydrate phase, respectively. The crossing of each pair of curves (continuous and dashed) determines the dissociation temperature of the single- and double-occupied hydrate, $T_3$, at the corresponding pressure. In all cases, the curves are included as guide to the eyes.}
\label{figure8}
\end{figure*}

Fig.~\ref{figure8} shows the solubility values of N$_{2}$ in water obtained from the L$_{\text{w}}$~--~L$_{\text{N}_{2}}$ and the H~--~L$_{\text{w}}$ equilibria at different pressures and temperatures. The values obtained from the H~--~L$_{\text{w}}$ equilibria are represented taking into account the single and double occupancy of the N$_{2}$ hydrate. Hence, two $T_3$ values are reported for each pressure, one for a single-occupied hydrate and the other for a double-occupied hydrate. As discussed previously, the solubility of N$_{2}$ in the aqueous phase, when it is in contact with a hydrate phase, is almost the same independent of the occupancy of the hydrate. According to this, the occupancy has a negligible effect on the solubility of N$_{2}$ in water, and hence, on the dissociation temperature of the N$_2$ hydrate. Table~\ref{Table1} shows all the T$_{3}$ values obtained in this work, at different pressures, assuming single and double occupancy of the hydrate.

\begin{figure}
\includegraphics[width=\columnwidth]{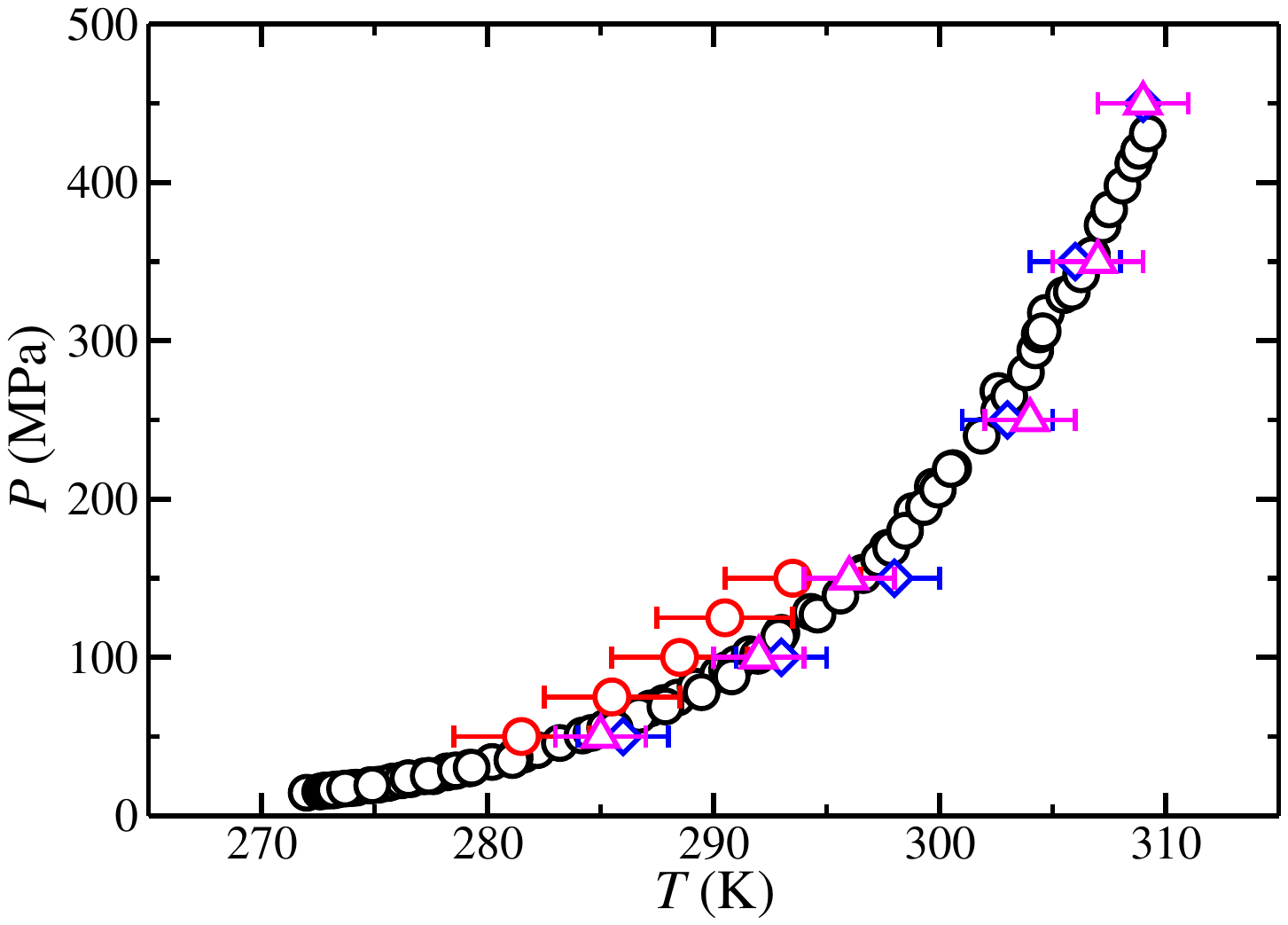}\\
\caption{Pressure-temperature or $PT$ projection of the dissociation line of the N$_{2}$ hydrate (single- and double-occupied). Blue diamonds (single-occupied) and magenta up triangles (double-occupied) are the results obtained in this work using the solubility method, the TIP4P/Ice model for water, and the TraPPE model for N$_{2}$ with the modified Bertherlot rule $\xi_{\text{ON}}=1.15$. Red circles are the results obtained by Yi \emph{et al.}~\cite{Yi2019a} using the direct coexistence technique, the TIP4P/2005 model for water, and the 2CLJQ model for N$_{2}$. Black circles correspond to experimental data taken from the literature.~\cite{vanCleeff1960a,Marshall1964a,Jhaveri1965a,Sugahara2002a,Mohammadi2003a}}
\label{figure9}
\end{figure}

Figure~\ref{figure9} shows the $PT$ projection of the dissociation line of the N$_{2}$ hydrate as obtained from computer simulation. We have also included experimental data~\cite{vanCleeff1960a,Marshall1964a,Jhaveri1965a,Sugahara2002a,Mohammadi2003a} and previous simulation results~\cite{Yi2019a,algaba2023b} taken from the literature. Note that the results obtained at $500$, $1000$, and $1500\,\text{bar}$ when the hydrate phase is single-occupied have already been reported in our previous work.~\cite{algaba2023b} The value of $\xi_{\text{ON}}$ was fitted to the experimental dissociation temperature of the N$_{2}$ hydrate at $500\,\text{bar}$ in our previous work~\cite{algaba2023b} and then transferred to other pressures to predict, without further fitting, the dissociation of the single- and double-occupied N$_{2}$ hydrate. Fig.~\ref{figure9} shows that 
the agreement between simulation results and experimental data from the literature is excellent in all cases. Note also that the effect of the occupancy on the dissociation temperatures is negligible in the whole range of pressures (see Table~\ref{Table1} for further details).

\begin{table}
\caption{Dissociation temperature, $T_3$, of the N$_{2}$ hydrate, at different pressures, as obtained in this work using the solubility method when the hydrate is single- and double-occupied by molecules of N$_2$. Numbers in parenthesis indicate the uncertainty of the results.}
\label{Table1}
\centering
\begin{tabular}{ccc}
\hline\hline
\multirow{2}{*}{$P$ (bar)} & $T_3$ (K) & $T_3$  (K)\\
& Single occupancy & Double occupancy \\
\hline
500 & 286(2) & 285(2)\\
1000 & 293(2) & 292(2) \\
1500 & 298(2) & 296(2) \\
2500 & 303(2) & 304(2) \\
3500 & 306(2) & 307(2) \\
4500 & 309(2) & 309(2)\\
\hline\hline
\end{tabular}
\end{table}

\subsection{Driving force for nucleation $\Delta\mu^{\text{EC}}_{N}$}

According to the Classical Nucleation Theory (CNT) of crystallization,~\cite{Debenedetti1996a} homogeneous nucleation process in a metastable liquid phase begins with the spontaneous formation of an initial cluster of the new solid phase (stable phase) from which the crystallization of the liquid phase take place. Forming the initial cluster is an activated process since the system has to overcome a free energy barrier. In the case of the N$_2$ hydrate, the free energy barrier that the system has to overcome to crystallize is reduced when the amount of N$_2$ in the aqueous solution phase is increased (supersaturated conditions)~\cite{Walsh2011a,Liang2011a,Yagasaki2014a,Bagherzadeh2015a,Grabowska2022a,Fang2023a}  and when the temperature is below the dissociation temperature (supercooling conditions).~\cite{Kashchiev2002a,Kashchiev2002b,Debenedetti1996a} Also, it is important to remark, that according to the CNT, the free energy barrier depends on the hydrate-aqueous solution interfacial free energy, $\gamma$, and the driving force for nucleation, $\Delta\mu_{\text{N}}$. The driving force for nucleation is the difference between the chemical potential of water and N$_2$ molecules in the hydrate, which is the stable phase, and those in the aqueous solution phase (metastable phase). When $\Delta\mu_{\text{N}}$ becomes negative, the hydrate phase becomes thermodynamically more stable than the liquid phase. Even so, the nucleation could not take place since there is still an energy barrier that the system has to overcome due to the formation of a (spherical) interface between the 
aqueous solution of N$_{2}$ and the hydrate. The more negative the value of the driving force, the smaller the free energy barrier. In our previous work,~\cite{algaba2023b} $\Delta\mu_{\text{N}}$ for the N$_2$ hydrate system was calculated at $500$, $1000$, and $1500\,\text{bar}$ and assuming single occupancy of the hydrate. Following the same procedure, we extend the study at higher pressures. In this work, we explicitly account for the effect of occupancy on $\Delta\mu_{\text{N}}$ assuming single and double occupancy of the hydrate.

According to Algaba \emph{et al.},~\cite{Algaba2023a} when comparing driving forces for nucleation of hydrates with different occupancies, it is more convenient to express $\Delta\mu_{\text{N}}$ per cage of hydrate formed from the aqueous solution than per guest molecule (N$_{2}$ in this case) used to form the hydrate from solution. Following this approach, we define the occupancy of the hydrate as the fraction of cages occupied by N$_{2}$, $x_{\text{occ}}=n_{\text{N}_{2}}/n_{\text{cg}}$, where $n_{\text{N}_{2}}$ and $n_{\text{cg}}$ are the number of N$_{2}$ molecules and cages per unit cell, respectively.~\cite{Algaba2023a} In this work, we follow the view of Kashchiev and Firoozabadi~\cite{Kashchiev2002a,Kashchiev2002b,Kashchiev2003a} and describe the formation of one cage of hydrate, with occupancy $x_{\text{occ}}$, from the aqueous solution phase as a chemical reaction at constant $P$ and $T$,~\cite{Grabowska2022a,Algaba2023a,algaba2023b}

\begin{equation}
x_{\text{occ}}\,\text{N}_{2} (\text{aq},x_{\text{N}_{2}}) +
5.67\,\text{H}_{2}\text{O} (\text{aq},x_{\text{N}_{2}}) 
\rightarrow [(\text{N}_{2})_{x_{\text{occ}}}(\text{H}_{2}\text{O})_{5.67}]_{\text{H}}
\label{reaction}
\end{equation}

\noindent
$\text{N}_{2}(\text{aq},x_{\text{N}_{2}})$ and $\text{H}_{2}\text{O} (\text{aq},x_{\text{N}_{2}})$ are the molecules of N$_{2}$ and water in the aqueous solution phase with composition $x_{\text{N}_{2}}$, respectively, while the $[(\text{N}_{2})_{x_{\text{occ}}}(\text{H}_{2}\text{O})_{5.67}]_{\text{H}}$ represents a ``molecule'' of the hydrate in the solid phase. The $5.67$ factor arises because the N$_{2}$ hydrate unit cell is formed by $136$ molecules of water and $24$ cages. Assuming single occupancy for each cage, for each molecule of N$_{2}$ there are $136/24\approx 5.67$ molecules of water. When the hydrate is single-occupied, there is only one molecule of N$_{2}$ per cage, and $x_{\text{occ}}=n_{\text{N}_{2}}/n_{\text{cg}}=24/24=1$. However, when the H (large) cages of the N$_{2}$ hydrate are double-occupied the stoichiometry of the N$_2$ hydrate changes, and $x_{\text{occ}}=n_{\text{N}_{2}}/n_{\text{cg}}=32/24\approx 1.33$.

To be consistent with Eq.~\eqref{reaction} and following the previous work of some of us,~\cite{Algaba2023a} we consider the driving force for nucleation of the hydrate expressed per cage of hydrate instead of per molecule of N$_{2}$. Taking the occupancy of the hydrate into account, the driving force for nucleation per cage of hydrates can be written as,~\cite{Kashchiev2002a,Grabowska2022a,Algaba2023a}

\begin{align}
\Delta\mu_{\text{N}}(P,T,x_{\text{N}_{2}},x_{\text{occ}})&=\tilde{\mu}^{\text{H}}_{\text{H}}(P,T,x_{\text{occ}}) \nonumber\\
& -x_{\text{occ}}\mu^{\text{aq}}_{\text{N}_{2}}(P,T,x_{\text{N}_{2}})-5.67\,\mu^{\text{aq}}_{\text{H}_{2}\text{O}}(P,T,x_{\text{N}_{2}})
\label{driving_force}
\end{align}

\noindent where $\tilde{\mu}^{\text{H}}_{\text{H}}(P,T,x_{\text{occ}})$ is the chemical potential of the hydrate in the solid stable phase (expressed as the Gibbs free energy of the hydrate per cage instead of per N$_{2}$ molecule) and $\mu_{\text{N}_{2}}^{\text{aq}}(P,T,x_{\text{N}_{2}})$ and $\mu_{\text{H}_{2}\text{O}}^{\text{aq}}(P,T,x_{\text{N}_{2}})$ are the chemical potentials of N$_2$ and water in the aqueous solution phase, respectively. Note that the occupancy of the hydrate is explicitly taken into account by the $x_{\text{occ}}$ factor and $\tilde{\mu}^{\text{H}}_{\text{H}}(P,T,x_{\text{occ}})$.

As we have pointed out in previous works,~\cite{Grabowska2022a,Algaba2023a,algaba2023b} it is possible to calculate $\Delta\mu_{\text{N}}$ at any $P$, $T$, and $x_{\text{N}_{2}}$. However, it is especially interesting to calculate $\Delta\mu_{\text{N}}$ at the conditions at which nucleation experiments are carried out, i.e., along the $\text{L}_{\text{w}}-\text{L}_{\text{N}_\text{2}}$ coexistence curve. Notice that the $\text{L}_{\text{w}}-\text{L}_{\text{N}_\text{2}}$ isobar curves were obtained by us in previous work at $500$, $1000$, and $1500\,\text{bar}$~\cite{algaba2023b} and we now extend the study at $2500$, $3500$, and $4500\,\text{bar}$. Following the notation introduced by some of us in previous works,~\cite{Grabowska2022a,Algaba2023a,algaba2023b} the driving force for nucleation per cage of hydrate at experimental conditions is expressed as,

\begin{align}
\Delta\mu^{\text{EC}}_{\text{N}}(P,T,x_{\text{N}_{2}}^{\text{eq}},x_{\text{occ}})&=\tilde{\mu}^{\text{H}}_{\text{H}}(P,T,x_{\text{occ}}) -x_{\text{occ}}\mu^{\text{aq}}_{\text{N}_{2}}(P,T,x_{\text{N}_{2}}^{\text{eq}}(P,T)) \nonumber\\
& -5.67\,\mu^{\text{aq}}_{\text{H}_{2}\text{O}}(P,T,x_{\text{N}_{2}}^{\text{eq}}(P,T))
\label{driving_force_EN}
\end{align}

\noindent For each pressure and at the temperature at which the three phases coexist, $T_{3}$, the chemical potentials are equal in each phase and, hence, $\Delta\mu^{\text{EC}}_{\text{N}}=0$. Note that since we are interested in the change of the chemical potential and not in its absolute value, we can set to zero the chemical potential of each component, including that of the ``hydrate molecule'' at $T_{3}$ for each pressure considered in this work.

\subsubsection{$\Delta\mu^{\text{EC}}_{\text{N}}$ via route 1}

In a previous work, Grabowska and collaborators~\cite{Grabowska2022a} proposed the use of the so-called route 1 to calculate the driving force for nucleation of the CH$_4$ hydrate. Later on, Algaba \emph{et al.}~\cite{Algaba2023a} applied the same route to estimate $\Delta\mu^{\text{EC}}_{\text{N}}$ of the CO$_{2}$ hydrate. In that work, they extended the route 1, originally proposed by Grabowska \emph{et al.},~\cite{Grabowska2022a} to deal with semi-occupied hydrates. In particular, they calculated the driving force for nucleation of the CO$_{2}$ hydrate considering an occupancy of $87.5\%$. This corresponds to a CO$_{2}$ hydrate in which half of the small or D cages are empty. In this work, we consider single and double occupancy of the N$_2$ hydrate. The expression introduced by some of us\cite{Algaba2023a} to deal with semi-occupied hydrate cages ($x_{\text{occ}}<1$), can be applied straightforwardly to take into account the double occupancy in the N$_{2}$ hydrate ($x_{\text{occ}}\approx 1.33$). According to this, the driving force for nucleation per cage of N$_{2}$ hydrate with occupancy $x_{\text{occ}}$ can be calculated as,~\cite{Algaba2023a,algaba2023b}

\begin{widetext}
\begin{align}
\dfrac{\Delta\mu^{\text{EC}}_{\text{N}}(P,T,x^{\text{eq}}_{\text{N}_{2}},x_{occ})}
{k_{B}T}=&
-\bigintss_{T_{3}}^{T} 
\dfrac{\tilde{h}_{\text{H}}^{\text{H}}(P,T',x_{occ})- \Big\{x_{occ}h_{\text{N}_{2}}(P,T')+5.67\,
h_{\text{H}_{2}\text{O}}(P,T')\Big\}}{k_{B}T'^{2}}dT'
\nonumber\\
& - 5.67\Big[k_{B}T\ln\{x^{\text{eq}}_{\text{H}_{2}\text{O}}(P,T)\}-k_{B}T_{3}\ln\{x^{\text{eq}}_{\text{H}_{2}\text{O}}(P,T_{3})\}\Big]
\label{driving_force_route1}
\end{align}
\end{widetext}

\noindent
Here $k_B$ is the Boltzmann constant and $\tilde{h}_{\text{H}}^{\text{H}}(P,T',x_{occ})$ represents the enthalpy of the hydrate at $P$ and $T'$ with occupancy $x_{\text{occ}}$. Note that $\tilde{h}_{\text{H}}$ is expressed as an enthalpy per cage of hydrate instead of a molar enthalpy or enthalpy per N$_{2}$ molecule, ${h}_{\text{H}}$. Both enthalpies are related by Eq.~(28) of the work of Algaba \emph{et al.},~\cite{Algaba2023a}

\begin{equation}
h_{\text{H}}^{\text{H}}=\dfrac{H}{N_{\text{N}_{2}}}=
\dfrac{H}{N_{\text{cg}}}\left(\frac{N_{\text{cg}}}{N_{\text{N}_{2}}}\right)=
\tilde{h}_{\text{H}}^{\text{H}}\left(\frac{N_{\text{cg}}}{N_{\text{N}_{2}}}\right)=
\dfrac{\tilde{h}_{\text{H}}^{\text{H}}}{x_{\text{occ}}}
\label{enthalpy_per_cage}    
\end{equation}

\noindent
Here $H$ is the enthalpy of the hydrate, and $N_{\text{N}_{2}}$ and $N_{\text{cg}}$ are the total number of N$_{2}$ molecules and cages used in the simulations, respectively. $h_{\text{N}_2}(P,T')$ and $h_{\text{H}_{2}\text{O}}(P,T')$ in Eq.~\eqref{driving_force_route1} are the enthalpies of pure N$_2$ and water at $P$ and $T'$. Finally, $x^{\text{eq}}_{\text{H}_{2}\text{O}}(P,T)$ and  $x^{\text{eq}}_{\text{H}_{2}\text{O}}(P,T_{3})$ represent the water composition in the aqueous solution phase (obtained from the $\text{L}_{\text{w}}-\text{L}_{\text{N}_\text{2}}$ solubility coexistence curves) at $T$ and $T_3$ (at pressure $P$), respectively.

Molar enthalpies of pure N$_{2}$ and water systems, $h_{\text{N}_2}(P,T')$ and $h_{\text{H}_{2}\text{O}}(P,T')$, are calculated by performing bulk simulations of pure phases of N$_2$ and water, respectively. Enthalpy of the hydrate per cage of hydrate, $\tilde{h}_{\text{H}}^{\text{H}}$, is obtained by dividing the total enthalpy of the N$_2$ hydrate by the number of hydrate cages. Notice that when single occupancy is assumed, it is equivalent to dividing the total enthalpy by the number of cages or the number of N$_{2}$ molecules as in our previous study\cite{algaba2023b}. For a detailed explanation of the simulation details such as the number of molecules, simulation times, as well as the barostat and thermostat, we refer the lector to our previous work.\cite{algaba2023b}

\subsubsection{$\Delta\mu^{\text{EC}}_{N}$ via dissociation route}

Following our previous works,~\cite{Grabowska2022a,Algaba2023a,algaba2023b} there is a simple and approximated route for $\Delta\mu^{\text{EC}}_{\text{N}}$ calculations based on the enthalpy of dissociation of the hydrate.~\cite{Kashchiev2002a} According to the chemical reaction given by Eq.~\eqref{reaction}, the dissociation enthalpy of a "hydrate molecule" is defined as the enthalpy change of the hydrate dissociation,

\begin{equation}
[(\text{N}_{2})_{x_{\text{occ}}}(\text{H}_{2}\text{O})_{5.67}]_{\text{H}}\rightarrow 
x_{\text{occ}}\text{N}_{2} (\text{liq}) +
5.67\,\text{H}_{2}\text{O} (\text{liq}) 
\label{dissoc}
\end{equation}

\noindent Notice that in this equation we are assuming that N$_2$ hydrate dissociates into pure N$_2$ and pure water. From the study of the $\text{L}_{\text{w}}-\text{L}_{\text{N}_\text{2}}$ equilibria we already know that this is not entirely true. However, it is still a good approximation due to the low solubility of N$_2$ in water. Additional approximations have to be undertaken in order to calculate $\Delta\mu^{\text{EC}}_{\text{N}}$. Following our previous works,~\cite{Grabowska2022a,Algaba2023a,algaba2023b} these approximations are: (1) the enthalpy of dissociation of the hydrate, $\tilde{h}_{\text{H}}^{\text{diss}}$ does not change with the temperature; (2) its value can be taken from its value at $T_{3}$; and (3) the change in the composition of the aqueous solution phase with the temperature is not taken into account. According to these approximations, Eq.~\eqref{driving_force_route1} reduces to:

\begin{equation}
\Delta\mu_{N}^{\text{EC}}=k_{B}T
\bigintsss_{T_{3}}^{T}\dfrac{\tilde{h}^{\text{diss}}_{\text{H}}}{k_{B}T'^{2}}\,dT'\approxeq -\tilde{h}^{\text{diss}}_{\text{H}} (T_{3}) \bigg(1-\dfrac{T}{T_{3}}\bigg)
\label{h_dissoc}
\end{equation}

\noindent Here the occupancy of the hydrate is explicitly taken into account by the $x_{occ}$ factor from Eq.~\eqref{dissoc}.

\subsubsection{$\Delta\mu^{\text{EC}}_{N}$ results}

\begin{figure}
\includegraphics[width=\columnwidth]{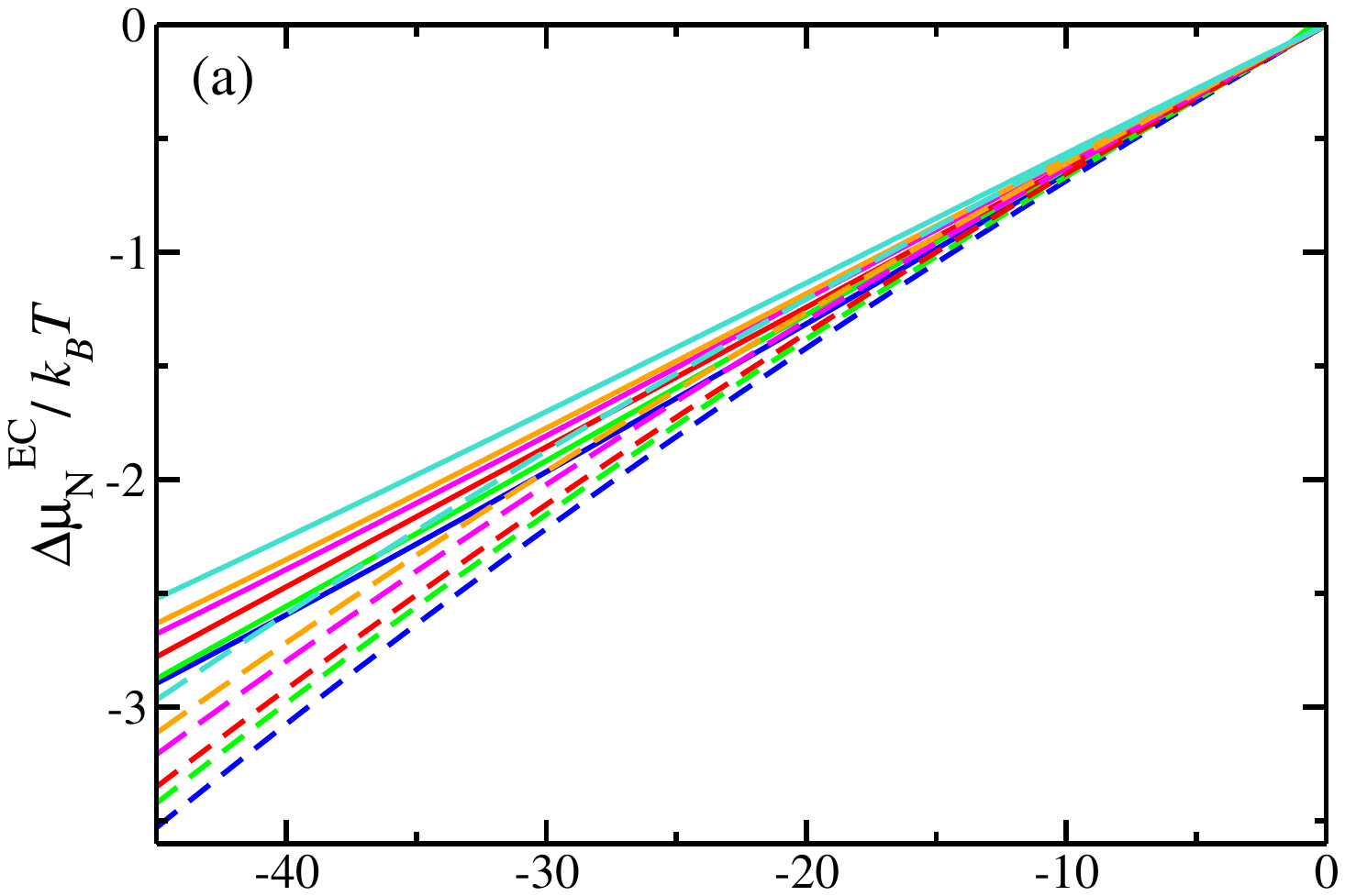}\\
\includegraphics[width=\columnwidth]{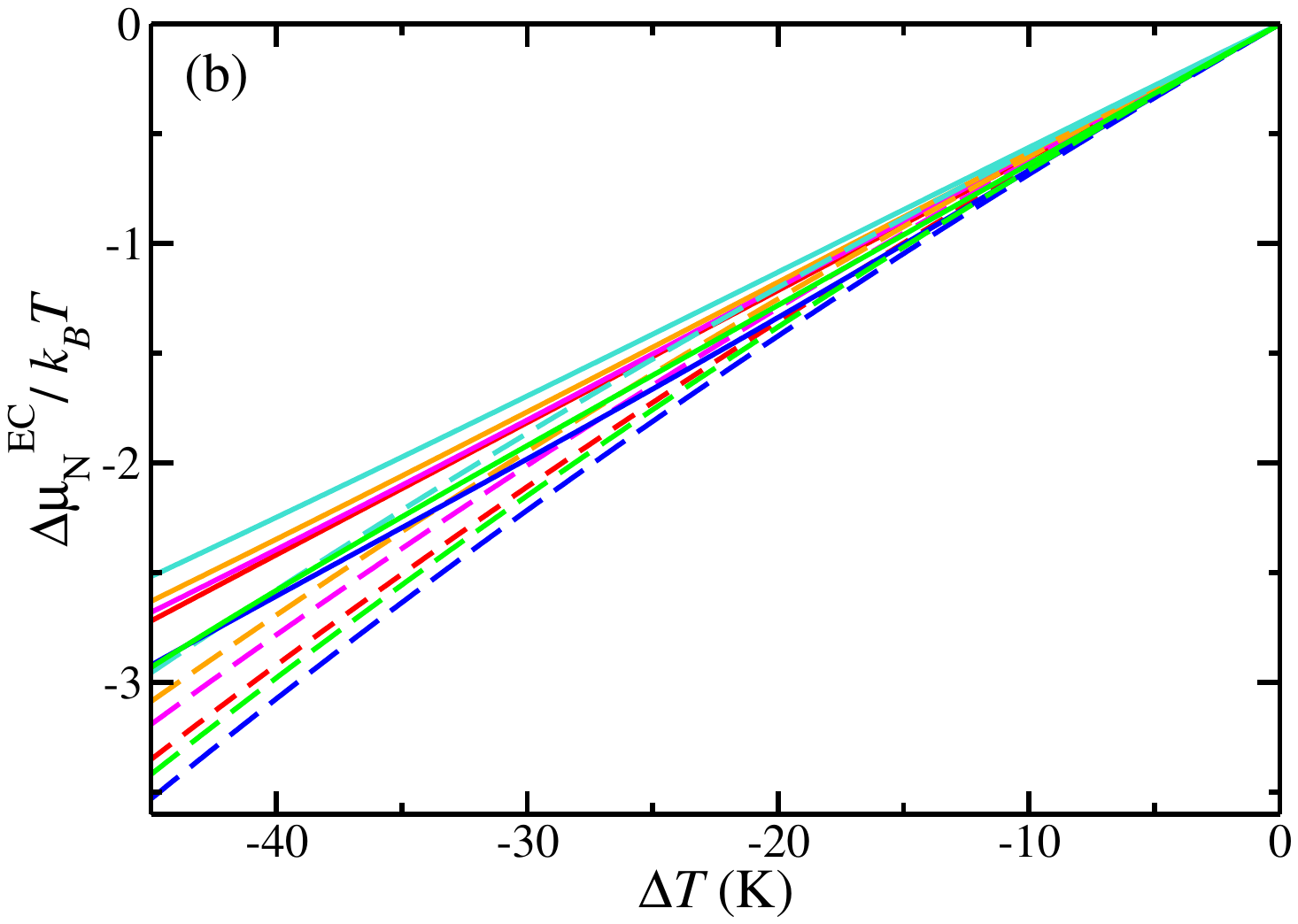}\\
\caption{Driving force for nucleation, $\Delta\mu^{\text{EC}}_{N}$, of N$_2$ hydrate at experimental conditions, as function of the supercooling, $\Delta T$, at $500$ (dark blue), $1000$ (green), $1500$ (red), $2500$ (magenta), $3500$ (orange), and $4500\,\text{bar}$ (cyan). The curves correspond to results obtained using route 1 (continuous curves) and dissociation route (dashed curves). Results have been obtained assuming single (a) and double (b) occupancy of the N$_2$ hydrate.}
\label{figure10}
\end{figure}

We have obtained the driving force for nucleation at experimental conditions of the N$_2$ hydrate via route 1 and the dissociation route. $\Delta\mu_{N}^{\text{EC}}$, as a function of the supercooling $\Delta T$, is shown in Fig.~\ref{figure10} at the different pressures and assuming single (a) and double (b) occupancy of the hydrate. We also include the results obtained in our previous work only assuming single occupancy of the hydrate at $500$, $1000$, and $1500\,\text{bar}$.~\cite{algaba2023b} 

Since the behavior of $\Delta\mu^{\text{EC}}_{N}$ assuming single and double occupancy is nearly identical,  we first concentrate on the differences between the results obtained using both routes. $\Delta\mu^{\text{EC}}_{N}$ becomes more negative as the supercooling degree is increased. This conclusion is valid for both routes. As some of us have previously discussed for methane~\cite{Grabowska2022a} and CO$_{2}$ hydrates,~\cite{Algaba2023a} agreement between the route 1 and the dissociation route is excellent at low supercooling degrees, from $|\Delta T|=0$ to $10-15\,\text{K}$, approximately. This is also true for the N$_{2}$ hydrate at low pressures and single occupancy.~\cite{algaba2023b} However, as the supercooling increases ($|\Delta T|\gtrapprox 15\,\text{K}$), differences between both routes become larger. The reasons are basically two: (1) $\tilde{h}^{\text{diss}}_{\text{H}}$ values at the $T_{3}$ differ from the real values evaluated at supercooling conditions. And (2), the dissociation route assumes that the hydrate dissociates into pure water and pure N$_{2}$. However, when the temperature decreases (supercooling increases) the solubility of N$_{2}$ in the water increases, making the predictions of the dissociation route less accurate since the assumptions on which they are based fail. To recap, the dissociation route provides a simple and fast method to estimate $\Delta\mu^{\text{EC}}_{N}$, but we do not generally recommend it except for temperatures close to the $T_3$ value.

We now turn our attention to the effect of the pressure on $\Delta\mu^{\text{EC}}_{N}$. As can be observed in Fig.~\ref{figure10}, $\Delta\mu^{\text{EC}}_{N}$ becomes smaller, in absolute value, when the pressure is increased. From $500$ to $4500\,\text{bar}$ and at the highest degree of supercooling studied in this work ($|\Delta T|=45\,\text{K}$), $\Delta\mu^{\text{EC}}_{N}$ decreases by approximately 10$\%$. We can conclude that this is a non-negligible effect. This is in agreement with our previous work\cite{algaba2023b} where we reported a maximum variation of $\Delta\mu_{\text{N}}^{\text{EC}}$, at fixed $\Delta T=40\,\text{K}$ and assuming single occupancy, below $5\%$ when pressure is increased from $500$ to $1500\,\text{bar}$. Note that the pressure range is now larger, from $500$ to $4500\,\text{bar}$.

\begin{figure}
\includegraphics[width=\columnwidth]{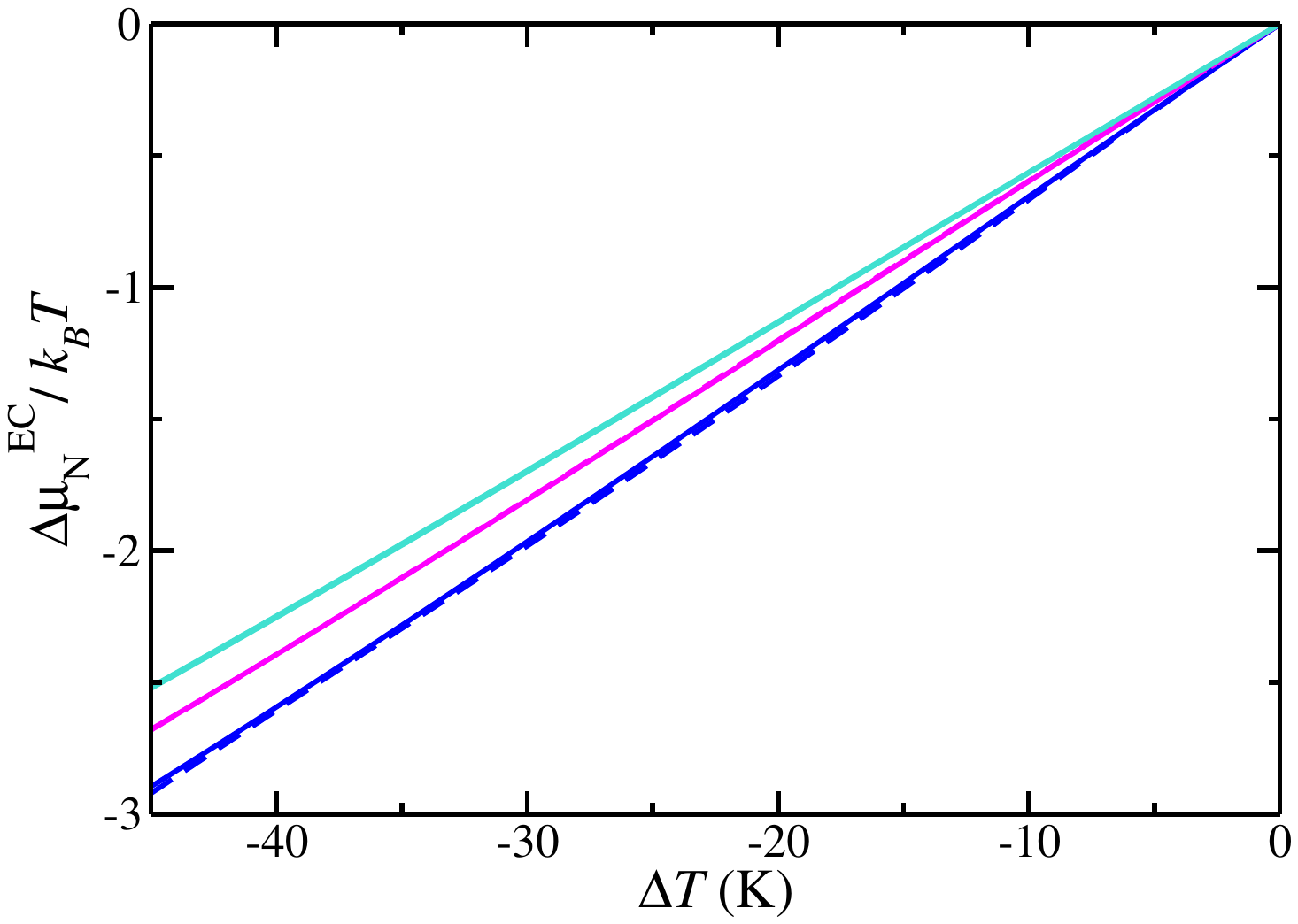}\\
\caption{Driving force for nucleation, $\Delta\mu^{\text{EC}}_{N}$, of N$_2$ hydrate at experimental conditions, as a function of the supercooling, $\Delta T$, at $500$ (dark blue), $2500$ (magenta), and $4500\,\text{bar}$ (cyan). The curves correspond to results obtained using route 1 assuming single (continuous curves) and double (dashed curves) occupancy of the N$_2$ hydrate.}
\label{figure11}
\end{figure}

To end Section D, we analyze the effect of the occupancy on the driving force for nucleation of the N$_{2}$ hydrate. Fig.~\ref{figure11} shows the result obtained at $500$, $2500$, and $4500\,\text{bar}$ via route 1 assuming single (continuous line) and double occupancy (dashed line) of the N$_{2}$ hydrate. According to Fig.~\ref{figure11}, the effect of the occupancy is negligible in the whole range of pressures and supercooling degrees studied in this work. This is an expected result since occupancy does not affect the dissociation temperature, $T_3$, of the N$_2$ hydrate.  As far as the authors know, this is the first time that the effect of double occupancy on the $T_3$ and driving force for nucleation of a hydrate that exhibits the sII crystallographic structure has been studied from computer simulation. This is a relevant result since hydrates with sII structure formed from small guest molecules, including N$_{2}$ and also H$_{2}$, have a potential interest in gas storage as occupancy plays a key role in the storage capabilities of these compounds.

\subsection{Effect of the pressure, temperature, and occupancy on unit-cell size.}

\begin{figure}
\includegraphics[width=\columnwidth]{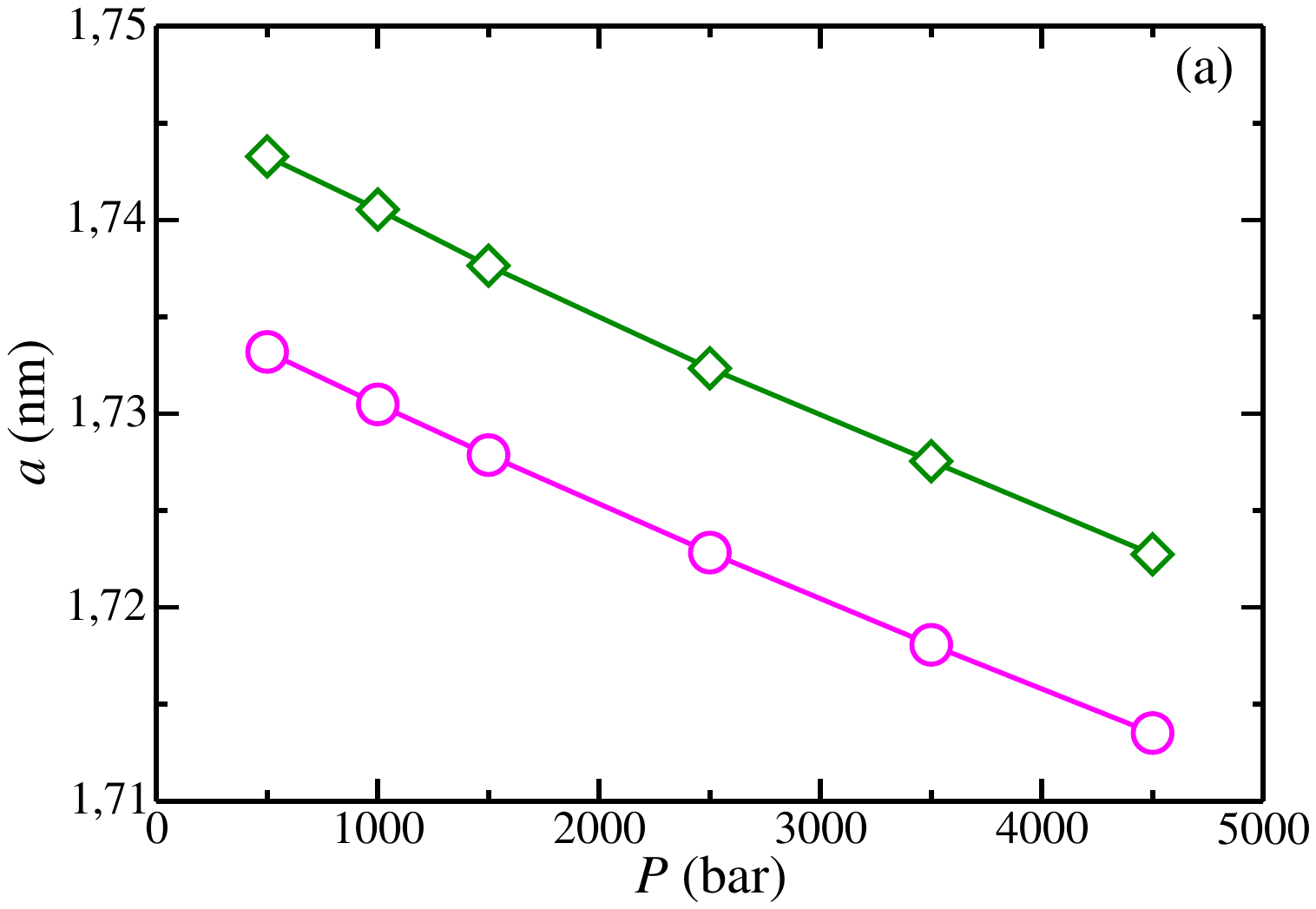}\\
\includegraphics[width=\columnwidth]{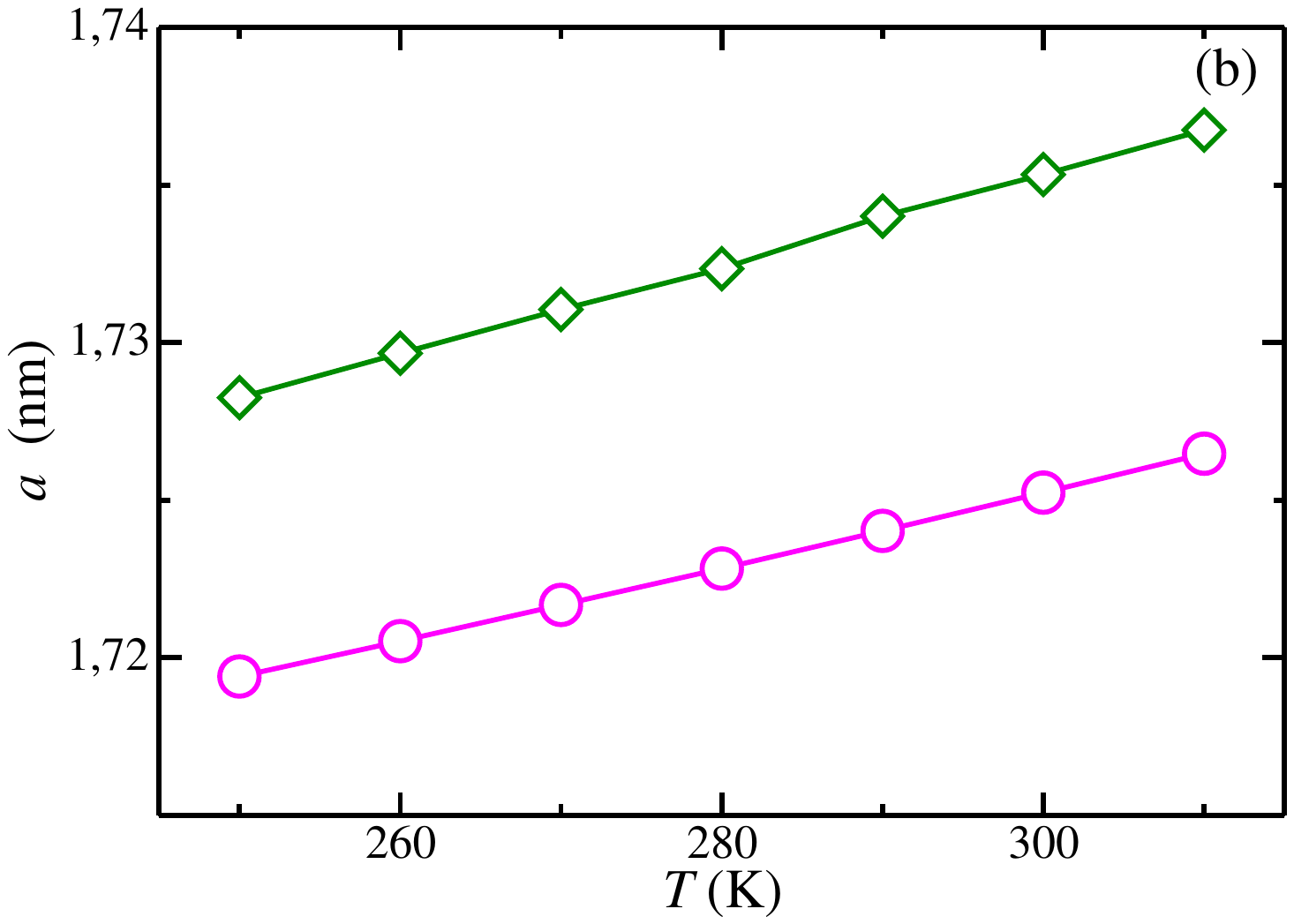}\\
\caption{Average unit-cell size or lattice constant obtained from MD $NPT$ simulations, as a function of the pressure at $280\,\text{K}$ (a), and as a function of the temperature at $2500\,\text{bar}$ (b), for single (magenta circles) and double occupancy (green diamonds).}
\label{figure12}
\end{figure}

Finally, we consider the effect of pressure, temperature, and occupancy on the unit-cell size of the N$_{2}$ hydrate. This information can be easily obtained from the analysis of the N$_2$ hydrate bulk simulations performed previously for the calculation of the N$_2$ hydrate enthalpy in Section D. Fig.~\ref{figure12} shows the average unit-cell size or lattice constant obtained from the N$_2$ hydrate bulk simulations. Since the sII crystallographic structure presents cubic symmetry, Fig.~\ref{figure12} shows the average length of the lattice constant, $a$, one of the sides of the unit-cell size of the sII structure. Notice that we can select arbitrarily any side of the N$_2$ hydrate structure due to the cubic symmetry of the sII crystallographic structure. As we can see in Fig.~\ref{figure12}a, when the temperature is fixed and the pressure is increased from $500$ to $4500\,\text{bar}$, the size of the unit cell is slightly reduced by $\approx1\%$ in both occupancies. On the other hand, when the pressure is fixed at 2500 bar and the temperature increases from $250$ to $310\,\text{K}$, the size of the unit cell is slightly increased by less than $\approx0.5\%$ in both occupancies. We have observed the same behavior in all the range of $T$ and $P$ conditions studied in this work. Hence, we can conclude that the pressure has a higher impact on the unit-cell size than the temperature, but in both cases, the effect is almost negligible. Finally, we analyze the effect of the occupancy on the lattice constant. As is shown in Fig.\ref{figure12}, the lattice constant with the hydrate with double occupancy is $\approx0.6\%$ larger than the lattice constant when single occupancy is assumed. As it happens with the effect of pressure and temperature on the lattice constant, the double occupancy has a measurable but almost negligible effect on the unit-cell size of the N$_{2}$ hydrate.

\section{Conclusions}

We have determined the three-phase hydrate -- aqueous solution -- N$_{2}$-rich liquid coexistence or dissociation line of the N$_{2}$ hydrate, in a wide range of pressures, using computer simulation and the solubility method recently proposed by some of us.~\cite{Grabowska2022a} In our previous work,\cite{algaba2023b} we considered lower pressures, from $500$ to $1500\,\text{bar}$ and single occupancy of N$_{2}$ molecules in the sII crystallographic structure of the hydrate. In this work, we go a step forward and extend the study to a wider range of pressures and consider explicitly the effect of double occupancy of N$_{2}$ molecules in the hydrate from $500$ to $4500\,\text{bar}$. As in our previous work,~\cite{Algaba2024b} water and N$_{2}$ are described using the TIP4P/Ice~\cite{Abascal2005b} and TraPPE models,~\cite{Potoff2001a} respectively. Unlike dispersive interaction parameters between the oxygen atom of water and the nitrogen atoms of N$_{2}$
are described by a modified Berthelot rule where $\xi_{\text{ON}}=1.15$. This value was optimized by us in our previous study\cite{algaba2023b} to match the experimental N$_2$ hydrate dissociation temperature at $500\,\text{bar}$. The same value of $\xi_{\text{ON}}$ is used in a transferable way to predict
the dissociation temperature of the hydrate along the whole three-phase coexistence line assuming single and double occupancy of the hydrate.

According to the solubility methodology, we have determined the dissociation temperature $T_3$ of the hydrate, at each pressure, by analyzing the solubility of N$_{2}$ in water when is in contact, via a planar interface, with a pure N$_2$ liquid phase (L$_{\text{w}}$--L$_{\text{N}_{2}}$ equilibrium), and with the N$_{2}$ hydrate phase (H--L$_{\text{N}_{2}}$ equilibrium). In this later case, we consider explicitly single and double occupancy. At fixed pressure, the temperature at which the L$_{\text{w}}$--L$_{\text{N}_{2}}$ and H--L$_{\text{N}_{2}}$ solubility lines cross determine the dissociation temperature of the hydrate. We have also compared the predictions obtained from computer simulations with experimental data taken from the literature.~\cite{vanCleeff1960a,Marshall1964a,Jhaveri1965a,Sugahara2002a,Mohammadi2003a} Agreement between both results, within the error bars of the simulation values, is excellent in the whole range of pressures considered. In addition, our results indicate that occupancy does not affect the dissociation temperature of the hydrate at all the pressures considered.

We have also estimated the driving force for nucleation, $\Delta\mu^{\text{EC}}_{N}$, of the N$_{2}$ hydrate as a function of the supercooling degree at several pressures. We have explicitly considered single and double occupancy of N$_{2}$ molecules in the hydrate. We have found that $\Delta\mu^{\text{EC}}_{N}$ becomes more positive when the pressure is increased, i.e., the hydrate phase becomes less stable as the pressure increases. Particularly, at $\Delta T=-40\,\text{K}$, the difference between $\Delta\mu^{\text{EC}}_{N}$ at $500$ and $4500\,\text{bar}$ is $10\%$, approximately. More importantly, the occupancy of N$_{2}$ in the hydrate does not affect the $\Delta\mu^{\text{EC}}_{N}$ values in the whole range of pressure and supercooling conditions studied in this work. To the best of our knowledge, this is the first time that the effect of the occupancy on the driving force for nucleation of a hydrate that exhibits sII crystallographic structure, along its dissociation line, is studied from computer simulation.

\section*{Author declarations}

\noindent
\textbf{Conflict of interests}

The authors declare no conflicts to disclose.

\section*{Author contributions}

\noindent
\textbf{Miguel J. Torrej\'on:} Methodology (equal); Investigation (equal); Writing – original draft (equal).
\textbf{Jesús Algaba:} Conceptualization (equal); Methodology (equal); Investigation (equal); Writing – original draft (equal); Writing – review \& editing (equal).
\textbf{Felipe J. Blas:} Conceptualization (equal); Funding acquisition (lead); Methodology (equal); Writing – original draft (equal); Writing – review \& editing (equal).

\section*{Acknowledgements}
This work was financed by Ministerio de Ciencia e Innovaci\'on (Grant No.~PID2021-125081NB-I00) and Universidad de Huelva (P.O. FEDER EPIT1282023), both co-financed by EU FEDER funds. MJT acknowledges the research contract (Ref.~01/2022/38143) of Programa Investigo (Plan de Recuperaci\'on, Transformaci\'on y Resiliencia, Fondos NextGeneration EU) from Junta de Andaluc\'{\i}a (HU/INV/0004/2022). We greatly acknowledge RES resources provided by Barcelona Supercomputing Center in Mare Nostrum to FI-2023-3-0011 and by The Supercomputing and Bioinnovation Center of the University of Malaga in Picasso to FI-2024-1-0017.

\section*{Data availability}

The data that supports the findings of this study are available within the article.

\bibliography{bibfjblas}

\begin{thebibliography}{96}%
\makeatletter
\providecommand \@ifxundefined [1]{%
 \@ifx{#1\undefined}
}%
\providecommand \@ifnum [1]{%
 \ifnum #1\expandafter \@firstoftwo
 \else \expandafter \@secondoftwo
 \fi
}%
\providecommand \@ifx [1]{%
 \ifx #1\expandafter \@firstoftwo
 \else \expandafter \@secondoftwo
 \fi
}%
\providecommand \natexlab [1]{#1}%
\providecommand \enquote  [1]{``#1''}%
\providecommand \bibnamefont  [1]{#1}%
\providecommand \bibfnamefont [1]{#1}%
\providecommand \citenamefont [1]{#1}%
\providecommand \href@noop [0]{\@secondoftwo}%
\providecommand \href [0]{\begingroup \@sanitize@url \@href}%
\providecommand \@href[1]{\@@startlink{#1}\@@href}%
\providecommand \@@href[1]{\endgroup#1\@@endlink}%
\providecommand \@sanitize@url [0]{\catcode `\\12\catcode `\$12\catcode
  `\&12\catcode `\#12\catcode `\^12\catcode `\_12\catcode `\%12\relax}%
\providecommand \@@startlink[1]{}%
\providecommand \@@endlink[0]{}%
\providecommand \url  [0]{\begingroup\@sanitize@url \@url }%
\providecommand \@url [1]{\endgroup\@href {#1}{\urlprefix }}%
\providecommand \urlprefix  [0]{URL }%
\providecommand \Eprint [0]{\href }%
\providecommand \doibase [0]{http://dx.doi.org/}%
\providecommand \selectlanguage [0]{\@gobble}%
\providecommand \bibinfo  [0]{\@secondoftwo}%
\providecommand \bibfield  [0]{\@secondoftwo}%
\providecommand \translation [1]{[#1]}%
\providecommand \BibitemOpen [0]{}%
\providecommand \bibitemStop [0]{}%
\providecommand \bibitemNoStop [0]{.\EOS\space}%
\providecommand \EOS [0]{\spacefactor3000\relax}%
\providecommand \BibitemShut  [1]{\csname bibitem#1\endcsname}%
\let\auto@bib@innerbib\@empty
\bibitem [{\citenamefont {Sloan}\ and\ \citenamefont {Koh}(2008)}]{Sloan2008a}%
  \BibitemOpen
  \bibfield  {author} {\bibinfo {author} {\bibfnamefont {E.~D.}\ \bibnamefont
  {Sloan}}\ and\ \bibinfo {author} {\bibfnamefont {C.}~\bibnamefont {Koh}},\
  }\href@noop {} {\emph {\bibinfo {title} {{C}lathrate {H}ydrates of {N}atural
  {G}ases}}},\ \bibinfo {edition} {3rd}\ ed.\ (\bibinfo  {publisher} {CRC
  Press},\ \bibinfo {address} {New York},\ \bibinfo {year} {2008})\BibitemShut
  {NoStop}%
\bibitem [{\citenamefont {Ripmeester}\ and\ \citenamefont
  {Alavi}(2022)}]{Ripmeester2022a}%
  \BibitemOpen
  \bibfield  {author} {\bibinfo {author} {\bibfnamefont {J.~A.}\ \bibnamefont
  {Ripmeester}}\ and\ \bibinfo {author} {\bibfnamefont {S.}~\bibnamefont
  {Alavi}},\ }\href@noop {} {\emph {\bibinfo {title} {Clathrate Hydrates:
  Molecular Science and Characterization}}}\ (\bibinfo  {publisher} {Wiley-VCH:
  Weinheim, Germany},\ \bibinfo {year} {2022})\BibitemShut {NoStop}%
\bibitem [{\citenamefont {Ripmeester}\ and\ \citenamefont
  {Alavi}(2016)}]{Ripmeester2016a}%
  \BibitemOpen
  \bibfield  {author} {\bibinfo {author} {\bibfnamefont {J.~A.}\ \bibnamefont
  {Ripmeester}}\ and\ \bibinfo {author} {\bibfnamefont {S.}~\bibnamefont
  {Alavi}},\ }\bibfield  {title} {\enquote {\bibinfo {title} {Some current
  challenges in clathrate hydrate science: Nucleation, decomposition and the
  memory effect},}\ }\href@noop {} {\bibfield  {journal} {\bibinfo  {journal}
  {Curr. Opin. Solid State Mater Sci.}\ }\textbf {\bibinfo {volume} {20}},\
  \bibinfo {pages} {344--351} (\bibinfo {year} {2016})}\BibitemShut {NoStop}%
\bibitem [{\citenamefont {Ratcliffe}(2022)}]{Ratcliffe2022a}%
  \BibitemOpen
  \bibfield  {author} {\bibinfo {author} {\bibfnamefont {C.~I.}\ \bibnamefont
  {Ratcliffe}},\ }\bibfield  {title} {\enquote {\bibinfo {title} {The
  development of clathrate hydrate science},}\ }\href@noop {} {\bibfield
  {journal} {\bibinfo  {journal} {Energy Fuels}\ }\textbf {\bibinfo {volume}
  {36}},\ \bibinfo {pages} {10412--10429} (\bibinfo {year} {2022})}\BibitemShut
  {NoStop}%
\bibitem [{\citenamefont {Ma}\ \emph {et~al.}(2016)\citenamefont {Ma},
  \citenamefont {Zhang}, \citenamefont {Bao},\ and\ \citenamefont
  {Deng}}]{ma2016review}%
  \BibitemOpen
  \bibfield  {author} {\bibinfo {author} {\bibfnamefont {Z.}~\bibnamefont
  {Ma}}, \bibinfo {author} {\bibfnamefont {P.}~\bibnamefont {Zhang}}, \bibinfo
  {author} {\bibfnamefont {H.}~\bibnamefont {Bao}}, \ and\ \bibinfo {author}
  {\bibfnamefont {S.}~\bibnamefont {Deng}},\ }\bibfield  {title} {\enquote
  {\bibinfo {title} {Review of fundamental properties of {CO$_{2}$} hydrates
  and {CO$_{2}$} capture and separation using hydration method},}\ }\href@noop
  {} {\bibfield  {journal} {\bibinfo  {journal} {Renew. Sustain. Energy Rev.}\
  }\textbf {\bibinfo {volume} {53}},\ \bibinfo {pages} {1273--1302} (\bibinfo
  {year} {2016})}\BibitemShut {NoStop}%
\bibitem [{\citenamefont {Dashti}, \citenamefont {Yew},\ and\ \citenamefont
  {Lou}(2015)}]{dashti2015recent}%
  \BibitemOpen
  \bibfield  {author} {\bibinfo {author} {\bibfnamefont {H.}~\bibnamefont
  {Dashti}}, \bibinfo {author} {\bibfnamefont {L.~Z.}\ \bibnamefont {Yew}}, \
  and\ \bibinfo {author} {\bibfnamefont {X.}~\bibnamefont {Lou}},\ }\bibfield
  {title} {\enquote {\bibinfo {title} {Recent advances in gas hydrate-based
  {CO$_{2}$} capture},}\ }\href@noop {} {\bibfield  {journal} {\bibinfo
  {journal} {J. Natural Gas Sci. Eng.}\ }\textbf {\bibinfo {volume} {23}},\
  \bibinfo {pages} {195--207} (\bibinfo {year} {2015})}\BibitemShut {NoStop}%
\bibitem [{\citenamefont {Cannone}, \citenamefont {Lanzini},\ and\
  \citenamefont {Santarelli}(2021)}]{cannone2021review}%
  \BibitemOpen
  \bibfield  {author} {\bibinfo {author} {\bibfnamefont {S.~F.}\ \bibnamefont
  {Cannone}}, \bibinfo {author} {\bibfnamefont {A.}~\bibnamefont {Lanzini}}, \
  and\ \bibinfo {author} {\bibfnamefont {M.}~\bibnamefont {Santarelli}},\
  }\bibfield  {title} {\enquote {\bibinfo {title} {A review on {CO$_{2}$}
  capture technologies with focus on {CO$_{2}$}-enhanced methane recovery from
  hydrates},}\ }\href@noop {} {\bibfield  {journal} {\bibinfo  {journal}
  {Energies}\ }\textbf {\bibinfo {volume} {14}},\ \bibinfo {pages} {387}
  (\bibinfo {year} {2021})}\BibitemShut {NoStop}%
\bibitem [{\citenamefont {Duc}, \citenamefont {Chauvy},\ and\ \citenamefont
  {Herri}(2007)}]{duc2007co2}%
  \BibitemOpen
  \bibfield  {author} {\bibinfo {author} {\bibfnamefont {N.~H.}\ \bibnamefont
  {Duc}}, \bibinfo {author} {\bibfnamefont {F.}~\bibnamefont {Chauvy}}, \ and\
  \bibinfo {author} {\bibfnamefont {J.-M.}\ \bibnamefont {Herri}},\ }\bibfield
  {title} {\enquote {\bibinfo {title} {{CO$_{2}$} capture by hydrate
  crystallization--{A} potential solution for gas emission of steelmaking
  industry},}\ }\href@noop {} {\bibfield  {journal} {\bibinfo  {journal}
  {Energy Convers. Manag.}\ }\textbf {\bibinfo {volume} {48}},\ \bibinfo
  {pages} {1313--1322} (\bibinfo {year} {2007})}\BibitemShut {NoStop}%
\bibitem [{\citenamefont {Choi}\ \emph {et~al.}(2022)\citenamefont {Choi},
  \citenamefont {Mok}, \citenamefont {Lee}, \citenamefont {Lee}, \citenamefont
  {Lee}, \citenamefont {Sum},\ and\ \citenamefont {Seo}}]{choi2022effective}%
  \BibitemOpen
  \bibfield  {author} {\bibinfo {author} {\bibfnamefont {W.}~\bibnamefont
  {Choi}}, \bibinfo {author} {\bibfnamefont {J.}~\bibnamefont {Mok}}, \bibinfo
  {author} {\bibfnamefont {J.}~\bibnamefont {Lee}}, \bibinfo {author}
  {\bibfnamefont {Y.}~\bibnamefont {Lee}}, \bibinfo {author} {\bibfnamefont
  {J.}~\bibnamefont {Lee}}, \bibinfo {author} {\bibfnamefont {A.~K.}\
  \bibnamefont {Sum}}, \ and\ \bibinfo {author} {\bibfnamefont
  {Y.}~\bibnamefont {Seo}},\ }\bibfield  {title} {\enquote {\bibinfo {title}
  {Effective {CH$_{4}$} production and novel {CO$_{2}$} storage through
  depressurization-assisted replacement in natural gas hydrate-bearing
  sediment},}\ }\href@noop {} {\bibfield  {journal} {\bibinfo  {journal} {Appl.
  Energy}\ }\textbf {\bibinfo {volume} {326}},\ \bibinfo {pages} {119971}
  (\bibinfo {year} {2022})}\BibitemShut {NoStop}%
\bibitem [{\citenamefont {Lee}, \citenamefont {Koh},\ and\ \citenamefont
  {Sum}(2014)}]{lee2014quantitative}%
  \BibitemOpen
  \bibfield  {author} {\bibinfo {author} {\bibfnamefont {B.~R.}\ \bibnamefont
  {Lee}}, \bibinfo {author} {\bibfnamefont {C.~A.}\ \bibnamefont {Koh}}, \ and\
  \bibinfo {author} {\bibfnamefont {A.~K.}\ \bibnamefont {Sum}},\ }\bibfield
  {title} {\enquote {\bibinfo {title} {Quantitative measurement and mechanisms
  for {CH$_{4}$} production from hydrates with the injection of liquid
  {CO$_{2}$}},}\ }\href@noop {} {\bibfield  {journal} {\bibinfo  {journal}
  {Phys. Chem. Chem. Phys.}\ }\textbf {\bibinfo {volume} {16}},\ \bibinfo
  {pages} {14922--14927} (\bibinfo {year} {2014})}\BibitemShut {NoStop}%
\bibitem [{\citenamefont {Veluswamy}, \citenamefont {Kumar},\ and\
  \citenamefont {Linga}(2014)}]{veluswamy2014hydrogen}%
  \BibitemOpen
  \bibfield  {author} {\bibinfo {author} {\bibfnamefont {H.~P.}\ \bibnamefont
  {Veluswamy}}, \bibinfo {author} {\bibfnamefont {R.}~\bibnamefont {Kumar}}, \
  and\ \bibinfo {author} {\bibfnamefont {P.}~\bibnamefont {Linga}},\ }\bibfield
   {title} {\enquote {\bibinfo {title} {Hydrogen storage in clathrate hydrates:
  Current state of the art and future directions},}\ }\href@noop {} {\bibfield
  {journal} {\bibinfo  {journal} {Appl. Energy}\ }\textbf {\bibinfo {volume}
  {122}},\ \bibinfo {pages} {112--132} (\bibinfo {year} {2014})}\BibitemShut
  {NoStop}%
\bibitem [{\citenamefont {Lee}\ \emph {et~al.}(2005)\citenamefont {Lee},
  \citenamefont {Lee}, \citenamefont {Kim}, \citenamefont {Park}, \citenamefont
  {Seo}, \citenamefont {Zeng}, \citenamefont {Moudrakovski}, \citenamefont
  {Ratcliffe},\ and\ \citenamefont {Ripmeester}}]{lee2005tuning}%
  \BibitemOpen
  \bibfield  {author} {\bibinfo {author} {\bibfnamefont {H.}~\bibnamefont
  {Lee}}, \bibinfo {author} {\bibfnamefont {J.-W.}\ \bibnamefont {Lee}},
  \bibinfo {author} {\bibfnamefont {D.~Y.}\ \bibnamefont {Kim}}, \bibinfo
  {author} {\bibfnamefont {J.}~\bibnamefont {Park}}, \bibinfo {author}
  {\bibfnamefont {Y.-T.}\ \bibnamefont {Seo}}, \bibinfo {author} {\bibfnamefont
  {H.}~\bibnamefont {Zeng}}, \bibinfo {author} {\bibfnamefont {I.~L.}\
  \bibnamefont {Moudrakovski}}, \bibinfo {author} {\bibfnamefont {C.~I.}\
  \bibnamefont {Ratcliffe}}, \ and\ \bibinfo {author} {\bibfnamefont {J.~A.}\
  \bibnamefont {Ripmeester}},\ }\bibfield  {title} {\enquote {\bibinfo {title}
  {Tuning clathrate hydrates for hydrogen storage},}\ }\href@noop {} {\bibfield
   {journal} {\bibinfo  {journal} {Nature}\ }\textbf {\bibinfo {volume}
  {434}},\ \bibinfo {pages} {743--746} (\bibinfo {year} {2005})}\BibitemShut
  {NoStop}%
\bibitem [{\citenamefont {Yi}\ \emph {et~al.}(2019)\citenamefont {Yi},
  \citenamefont {Zhou}, \citenamefont {He}, \citenamefont {Cai}, \citenamefont
  {Zhao}, \citenamefont {Zhang},\ and\ \citenamefont {Shao}}]{Yi2019a}%
  \BibitemOpen
  \bibfield  {author} {\bibinfo {author} {\bibfnamefont {L.}~\bibnamefont
  {Yi}}, \bibinfo {author} {\bibfnamefont {X.}~\bibnamefont {Zhou}}, \bibinfo
  {author} {\bibfnamefont {Y.}~\bibnamefont {He}}, \bibinfo {author}
  {\bibfnamefont {Z.}~\bibnamefont {Cai}}, \bibinfo {author} {\bibfnamefont
  {L.}~\bibnamefont {Zhao}}, \bibinfo {author} {\bibfnamefont {W.}~\bibnamefont
  {Zhang}}, \ and\ \bibinfo {author} {\bibfnamefont {Y.}~\bibnamefont {Shao}},\
  }\bibfield  {title} {\enquote {\bibinfo {title} {Molecular dynamics
  simulation study on the growth of structure {II} nitrogen hydrate},}\
  }\href@noop {} {\bibfield  {journal} {\bibinfo  {journal} {J. Phys. Chem. B}\
  }\textbf {\bibinfo {volume} {123}},\ \bibinfo {pages} {9180--9186} (\bibinfo
  {year} {2019})}\BibitemShut {NoStop}%
\bibitem [{\citenamefont {Hassanpouryouzband}\ \emph
  {et~al.}(2018)\citenamefont {Hassanpouryouzband}, \citenamefont {Yang},
  \citenamefont {Tohidi}, \citenamefont {Chuvilin}, \citenamefont {Istomin},
  \citenamefont {Bukhanov},\ and\ \citenamefont
  {Cheremisin}}]{hassanpouryouzband2018co2}%
  \BibitemOpen
  \bibfield  {author} {\bibinfo {author} {\bibfnamefont {A.}~\bibnamefont
  {Hassanpouryouzband}}, \bibinfo {author} {\bibfnamefont {J.}~\bibnamefont
  {Yang}}, \bibinfo {author} {\bibfnamefont {B.}~\bibnamefont {Tohidi}},
  \bibinfo {author} {\bibfnamefont {E.}~\bibnamefont {Chuvilin}}, \bibinfo
  {author} {\bibfnamefont {V.}~\bibnamefont {Istomin}}, \bibinfo {author}
  {\bibfnamefont {B.}~\bibnamefont {Bukhanov}}, \ and\ \bibinfo {author}
  {\bibfnamefont {A.}~\bibnamefont {Cheremisin}},\ }\bibfield  {title}
  {\enquote {\bibinfo {title} {{CO$_{2}$} capture by injection of flue gas or
  {CO$_{2}$}--{N$_{2}$} mixtures into hydrate reservoirs: Dependence of
  {CO$_{2}$} capture efficiency on gas hydrate reservoir conditions},}\
  }\href@noop {} {\bibfield  {journal} {\bibinfo  {journal} {Environ. Sci.
  Technol.}\ }\textbf {\bibinfo {volume} {52}},\ \bibinfo {pages} {4324--4330}
  (\bibinfo {year} {2018})}\BibitemShut {NoStop}%
\bibitem [{\citenamefont {Lee}\ and\ \citenamefont
  {Holder}(2001)}]{lee2001methane}%
  \BibitemOpen
  \bibfield  {author} {\bibinfo {author} {\bibfnamefont {S.-Y.}\ \bibnamefont
  {Lee}}\ and\ \bibinfo {author} {\bibfnamefont {G.~D.}\ \bibnamefont
  {Holder}},\ }\bibfield  {title} {\enquote {\bibinfo {title} {Methane hydrates
  potential as a future energy source},}\ }\href@noop {} {\bibfield  {journal}
  {\bibinfo  {journal} {Fuel Process. Technol.}\ }\textbf {\bibinfo {volume}
  {71}},\ \bibinfo {pages} {181--186} (\bibinfo {year} {2001})}\BibitemShut
  {NoStop}%
\bibitem [{\citenamefont {Ruppel}\ and\ \citenamefont
  {Kessler}(2017)}]{ruppel2017interaction}%
  \BibitemOpen
  \bibfield  {author} {\bibinfo {author} {\bibfnamefont {C.~D.}\ \bibnamefont
  {Ruppel}}\ and\ \bibinfo {author} {\bibfnamefont {J.~D.}\ \bibnamefont
  {Kessler}},\ }\bibfield  {title} {\enquote {\bibinfo {title} {The interaction
  of climate change and methane hydrates},}\ }\href@noop {} {\bibfield
  {journal} {\bibinfo  {journal} {Rev. Geophys.}\ }\textbf {\bibinfo {volume}
  {55}},\ \bibinfo {pages} {126--168} (\bibinfo {year} {2017})}\BibitemShut
  {NoStop}%
\bibitem [{\citenamefont {Tsimpanogiannis}\ and\ \citenamefont
  {Economou}(2017)}]{Tsimpanogiannis2017a}%
  \BibitemOpen
  \bibfield  {author} {\bibinfo {author} {\bibfnamefont {I.~N.}\ \bibnamefont
  {Tsimpanogiannis}}\ and\ \bibinfo {author} {\bibfnamefont {I.~G.}\
  \bibnamefont {Economou}},\ }\bibfield  {title} {\enquote {\bibinfo {title}
  {Monte carlo simulation studies of clathrate hydrates: A review},}\
  }\href@noop {} {\bibfield  {journal} {\bibinfo  {journal} {J. Supercrit.
  Fluids}\ }\textbf {\bibinfo {volume} {134}},\ \bibinfo {pages} {51--60}
  (\bibinfo {year} {2017})}\BibitemShut {NoStop}%
\bibitem [{\citenamefont {Brumby}\ \emph {et~al.}(2019)\citenamefont {Brumby},
  \citenamefont {Yuhara}, \citenamefont {Hasegawa}, \citenamefont {Wu},
  \citenamefont {Sum},\ and\ \citenamefont {Yasuoka}}]{Brumby2019a}%
  \BibitemOpen
  \bibfield  {author} {\bibinfo {author} {\bibfnamefont {P.~E.}\ \bibnamefont
  {Brumby}}, \bibinfo {author} {\bibfnamefont {D.}~\bibnamefont {Yuhara}},
  \bibinfo {author} {\bibfnamefont {T.}~\bibnamefont {Hasegawa}}, \bibinfo
  {author} {\bibfnamefont {D.~T.}\ \bibnamefont {Wu}}, \bibinfo {author}
  {\bibfnamefont {A.~K.}\ \bibnamefont {Sum}}, \ and\ \bibinfo {author}
  {\bibfnamefont {K.}~\bibnamefont {Yasuoka}},\ }\bibfield  {title} {\enquote
  {\bibinfo {title} {Cage occupancies, lattice constants, and guest chemical
  potentials for structure {II} hydrogen clathrate hydrate from {G}ibbs
  ensemble {M}onte {C}arlo simulations},}\ }\href@noop {} {\bibfield  {journal}
  {\bibinfo  {journal} {J. Chem. Phys.}\ }\textbf {\bibinfo {volume} {150}},\
  \bibinfo {pages} {134503} (\bibinfo {year} {2019})}\BibitemShut {NoStop}%
\bibitem [{\citenamefont {Michalis}\ \emph {et~al.}(2022)\citenamefont
  {Michalis}, \citenamefont {Economou}, \citenamefont {Stubos},\ and\
  \citenamefont {Tsimpanogiannis}}]{Michalis2022a}%
  \BibitemOpen
  \bibfield  {author} {\bibinfo {author} {\bibfnamefont {V.~K.}\ \bibnamefont
  {Michalis}}, \bibinfo {author} {\bibfnamefont {I.~G.}\ \bibnamefont
  {Economou}}, \bibinfo {author} {\bibfnamefont {A.~K.}\ \bibnamefont
  {Stubos}}, \ and\ \bibinfo {author} {\bibfnamefont {I.~N.}\ \bibnamefont
  {Tsimpanogiannis}},\ }\bibfield  {title} {\enquote {\bibinfo {title} {Phase
  equilibria molecular simulations of hydrogen hydrates via the direct phase
  coexistence approach},}\ }\href@noop {} {\bibfield  {journal} {\bibinfo
  {journal} {J. Chem. Phys.}\ }\textbf {\bibinfo {volume} {157}},\ \bibinfo
  {pages} {154501} (\bibinfo {year} {2022})}\BibitemShut {NoStop}%
\bibitem [{\citenamefont {Kuhs}\ \emph {et~al.}(1997)\citenamefont {Kuhs},
  \citenamefont {Chazallon}, \citenamefont {Radaelli},\ and\ \citenamefont
  {Pauer}}]{Kuhs1997a}%
  \BibitemOpen
  \bibfield  {author} {\bibinfo {author} {\bibfnamefont {W.~F.}\ \bibnamefont
  {Kuhs}}, \bibinfo {author} {\bibfnamefont {B.}~\bibnamefont {Chazallon}},
  \bibinfo {author} {\bibfnamefont {P.~G.}\ \bibnamefont {Radaelli}}, \ and\
  \bibinfo {author} {\bibfnamefont {F.}~\bibnamefont {Pauer}},\ }\bibfield
  {title} {\enquote {\bibinfo {title} {Cage occupancy and compressibility of
  deuterated {N$_{2}$}--clathrate hydrate by neutron diffraction},}\
  }\href@noop {} {\bibfield  {journal} {\bibinfo  {journal} {J. Inclusion
  Phenom.}\ }\textbf {\bibinfo {volume} {29}},\ \bibinfo {pages} {65--77}
  (\bibinfo {year} {1997})}\BibitemShut {NoStop}%
\bibitem [{\citenamefont {Chazallon}\ and\ \citenamefont
  {Kuhs}(2002)}]{Chazallon2002a}%
  \BibitemOpen
  \bibfield  {author} {\bibinfo {author} {\bibfnamefont {B.}~\bibnamefont
  {Chazallon}}\ and\ \bibinfo {author} {\bibfnamefont {W.~F.}\ \bibnamefont
  {Kuhs}},\ }\bibfield  {title} {\enquote {\bibinfo {title} {In situ structural
  properties of {N$_{2}$-}, {O$_{2}$-}, and air-clathrates by neutron
  diffraction},}\ }\href@noop {} {\bibfield  {journal} {\bibinfo  {journal} {J.
  Chem. Phys.}\ }\textbf {\bibinfo {volume} {117}},\ \bibinfo {pages}
  {308--320} (\bibinfo {year} {2002})}\BibitemShut {NoStop}%
\bibitem [{\citenamefont {Sasaki}\ \emph {et~al.}(2003)\citenamefont {Sasaki},
  \citenamefont {Hori}, \citenamefont {Kume},\ and\ \citenamefont
  {Shimizu}}]{Sasaki2003a}%
  \BibitemOpen
  \bibfield  {author} {\bibinfo {author} {\bibfnamefont {S.}~\bibnamefont
  {Sasaki}}, \bibinfo {author} {\bibfnamefont {S.}~\bibnamefont {Hori}},
  \bibinfo {author} {\bibfnamefont {T.}~\bibnamefont {Kume}}, \ and\ \bibinfo
  {author} {\bibfnamefont {H.}~\bibnamefont {Shimizu}},\ }\bibfield  {title}
  {\enquote {\bibinfo {title} {Microscopic observation and in situ raman
  scattering studies on high-pressure phase transformations of a synthetic
  nitrogen hydrate},}\ }\href@noop {} {\bibfield  {journal} {\bibinfo
  {journal} {J. Chem. Phys.}\ }\textbf {\bibinfo {volume} {118}},\ \bibinfo
  {pages} {7892--7897} (\bibinfo {year} {2003})}\BibitemShut {NoStop}%
\bibitem [{\citenamefont {van Klaveren}\ \emph
  {et~al.}(2001{\natexlab{a}})\citenamefont {van Klaveren}, \citenamefont
  {Michels}, \citenamefont {Schouten}, \citenamefont {Klug},\ and\
  \citenamefont {Tse}}]{vanKlaveren2001a}%
  \BibitemOpen
  \bibfield  {author} {\bibinfo {author} {\bibfnamefont {E.~P.}\ \bibnamefont
  {van Klaveren}}, \bibinfo {author} {\bibfnamefont {J.~P.~J.}\ \bibnamefont
  {Michels}}, \bibinfo {author} {\bibfnamefont {J.~A.}\ \bibnamefont
  {Schouten}}, \bibinfo {author} {\bibfnamefont {D.~D.}\ \bibnamefont {Klug}},
  \ and\ \bibinfo {author} {\bibfnamefont {J.~S.}\ \bibnamefont {Tse}},\
  }\bibfield  {title} {\enquote {\bibinfo {title} {Stability of doubly occupied
  {N$_{2}$} clathrate hydrates investigated by molecular dynamics
  simulations},}\ }\href@noop {} {\bibfield  {journal} {\bibinfo  {journal} {J.
  Chem. Phys.}\ }\textbf {\bibinfo {volume} {114}},\ \bibinfo {pages}
  {5745--5754} (\bibinfo {year} {2001}{\natexlab{a}})}\BibitemShut {NoStop}%
\bibitem [{\citenamefont {van Klaveren}\ \emph
  {et~al.}(2001{\natexlab{b}})\citenamefont {van Klaveren}, \citenamefont
  {Michels}, \citenamefont {Schouten}, \citenamefont {Klug},\ and\
  \citenamefont {Tse}}]{vanKlaveren2001b}%
  \BibitemOpen
  \bibfield  {author} {\bibinfo {author} {\bibfnamefont {E.~P.}\ \bibnamefont
  {van Klaveren}}, \bibinfo {author} {\bibfnamefont {J.~P.~J.}\ \bibnamefont
  {Michels}}, \bibinfo {author} {\bibfnamefont {J.~A.}\ \bibnamefont
  {Schouten}}, \bibinfo {author} {\bibfnamefont {D.~D.}\ \bibnamefont {Klug}},
  \ and\ \bibinfo {author} {\bibfnamefont {J.~S.}\ \bibnamefont {Tse}},\
  }\bibfield  {title} {\enquote {\bibinfo {title} {Molecular dynamics
  simulation study of the properties of doubly occupied {N$_{2}$} clathrate
  hydrates},}\ }\href@noop {} {\bibfield  {journal} {\bibinfo  {journal} {J.
  Chem. Phys.}\ }\textbf {\bibinfo {volume} {115}},\ \bibinfo {pages}
  {10500--10508} (\bibinfo {year} {2001}{\natexlab{b}})}\BibitemShut {NoStop}%
\bibitem [{\citenamefont {van Klaveren}\ \emph {et~al.}(2002)\citenamefont {van
  Klaveren}, \citenamefont {Michels}, \citenamefont {Schouten}, \citenamefont
  {Klug},\ and\ \citenamefont {Tse}}]{VanKlaveren2002a}%
  \BibitemOpen
  \bibfield  {author} {\bibinfo {author} {\bibfnamefont {E.~P.}\ \bibnamefont
  {van Klaveren}}, \bibinfo {author} {\bibfnamefont {J.~P.~J.}\ \bibnamefont
  {Michels}}, \bibinfo {author} {\bibfnamefont {J.~A.}\ \bibnamefont
  {Schouten}}, \bibinfo {author} {\bibfnamefont {D.~D.}\ \bibnamefont {Klug}},
  \ and\ \bibinfo {author} {\bibfnamefont {J.~S.}\ \bibnamefont {Tse}},\
  }\bibfield  {title} {\enquote {\bibinfo {title} {Computer simulations of the
  dynamics of doubly occupied {N$_{2}$} clathrate hydrates},}\ }\href@noop {}
  {\bibfield  {journal} {\bibinfo  {journal} {J. Chem. Phys.}\ }\textbf
  {\bibinfo {volume} {117}},\ \bibinfo {pages} {6637--6645} (\bibinfo {year}
  {2002})}\BibitemShut {NoStop}%
\bibitem [{\citenamefont {Algaba}, \citenamefont {Torrej{\'o}n},\ and\
  \citenamefont {Blas}(2023)}]{algaba2023b}%
  \BibitemOpen
  \bibfield  {author} {\bibinfo {author} {\bibfnamefont {J.}~\bibnamefont
  {Algaba}}, \bibinfo {author} {\bibfnamefont {M.~J.}\ \bibnamefont
  {Torrej{\'o}n}}, \ and\ \bibinfo {author} {\bibfnamefont {F.~J.}\
  \bibnamefont {Blas}},\ }\bibfield  {title} {\enquote {\bibinfo {title}
  {Dissociation line and driving force for nucleation of the nitrogen hydrate
  from computer simulation},}\ }\href@noop {} {\bibfield  {journal} {\bibinfo
  {journal} {J. Chem. Phys.}\ }\textbf {\bibinfo {volume} {159}},\ \bibinfo
  {pages} {224707} (\bibinfo {year} {2023})}\BibitemShut {NoStop}%
\bibitem [{\citenamefont {Grabowska}\ \emph {et~al.}(2022)\citenamefont
  {Grabowska}, \citenamefont {Bl{\'a}zquez}, \citenamefont {Sanz},
  \citenamefont {Zer{\'o}n}, \citenamefont {Algaba}, \citenamefont
  {M{\'{\i}}guez}, \citenamefont {Blas},\ and\ \citenamefont
  {Vega}}]{Grabowska2022a}%
  \BibitemOpen
  \bibfield  {author} {\bibinfo {author} {\bibfnamefont {J.}~\bibnamefont
  {Grabowska}}, \bibinfo {author} {\bibfnamefont {S.}~\bibnamefont
  {Bl{\'a}zquez}}, \bibinfo {author} {\bibfnamefont {E.}~\bibnamefont {Sanz}},
  \bibinfo {author} {\bibfnamefont {I.~M.}\ \bibnamefont {Zer{\'o}n}}, \bibinfo
  {author} {\bibfnamefont {J.}~\bibnamefont {Algaba}}, \bibinfo {author}
  {\bibfnamefont {J.~M.}\ \bibnamefont {M{\'{\i}}guez}}, \bibinfo {author}
  {\bibfnamefont {F.~J.}\ \bibnamefont {Blas}}, \ and\ \bibinfo {author}
  {\bibfnamefont {C.}~\bibnamefont {Vega}},\ }\bibfield  {title} {\enquote
  {\bibinfo {title} {Solubility of methane in water: some useful results for
  hydrate nucleation},}\ }\href@noop {} {\bibfield  {journal} {\bibinfo
  {journal} {J. Phys. Chem. B}\ }\textbf {\bibinfo {volume} {126}},\ \bibinfo
  {pages} {8553--8570} (\bibinfo {year} {2022})}\BibitemShut {NoStop}%
\bibitem [{\citenamefont {Algaba}\ \emph {et~al.}(2023)\citenamefont {Algaba},
  \citenamefont {Zer{\'o}n}, \citenamefont {M{\'\i}guez}, \citenamefont
  {Grabowska}, \citenamefont {Blazquez}, \citenamefont {Sanz}, \citenamefont
  {Vega},\ and\ \citenamefont {Blas}}]{Algaba2023a}%
  \BibitemOpen
  \bibfield  {author} {\bibinfo {author} {\bibfnamefont {J.}~\bibnamefont
  {Algaba}}, \bibinfo {author} {\bibfnamefont {I.~M.}\ \bibnamefont
  {Zer{\'o}n}}, \bibinfo {author} {\bibfnamefont {J.~M.}\ \bibnamefont
  {M{\'\i}guez}}, \bibinfo {author} {\bibfnamefont {J.}~\bibnamefont
  {Grabowska}}, \bibinfo {author} {\bibfnamefont {S.}~\bibnamefont {Blazquez}},
  \bibinfo {author} {\bibfnamefont {E.}~\bibnamefont {Sanz}}, \bibinfo {author}
  {\bibfnamefont {C.}~\bibnamefont {Vega}}, \ and\ \bibinfo {author}
  {\bibfnamefont {F.~J.}\ \bibnamefont {Blas}},\ }\bibfield  {title} {\enquote
  {\bibinfo {title} {Solubility of carbon dioxide in water: Some useful results
  for hydrate nucleation},}\ }\href@noop {} {\bibfield  {journal} {\bibinfo
  {journal} {J. Chem. Phys.}\ }\textbf {\bibinfo {volume} {158}},\ \bibinfo
  {pages} {054505} (\bibinfo {year} {2023})}\BibitemShut {NoStop}%
\bibitem [{\citenamefont {Debenedetti}(1997)}]{Debenedetti1996a}%
  \BibitemOpen
  \bibfield  {author} {\bibinfo {author} {\bibfnamefont {P.~G.}\ \bibnamefont
  {Debenedetti}},\ }\href@noop {} {\emph {\bibinfo {title} {Metastable Liquids:
  Concepts and Principles}}}\ (\bibinfo  {publisher} {Princeton University
  Press},\ \bibinfo {year} {1997})\BibitemShut {NoStop}%
\bibitem [{\citenamefont {Kashchiev}(2000)}]{Kashchiev2000a}%
  \BibitemOpen
  \bibfield  {author} {\bibinfo {author} {\bibfnamefont {D.}~\bibnamefont
  {Kashchiev}},\ }\href@noop {} {\emph {\bibinfo {title} {Nucleation}}}\
  (\bibinfo  {publisher} {Butterworth-Heinemann: Oxford, UK},\ \bibinfo {year}
  {2000})\BibitemShut {NoStop}%
\bibitem [{\citenamefont {Maeda}(2020)}]{Maeda2020a}%
  \BibitemOpen
  \bibfield  {author} {\bibinfo {author} {\bibfnamefont {N.}~\bibnamefont
  {Maeda}},\ }\href@noop {} {\emph {\bibinfo {title} {Nucleation of Gas
  Hydrates}}}\ (\bibinfo  {publisher} {Springer Nature Switzerland,
  Switzerland},\ \bibinfo {year} {2020})\BibitemShut {NoStop}%
\bibitem [{\citenamefont {Kashchiev}\ and\ \citenamefont
  {Firoozabadi}(2002{\natexlab{a}})}]{Kashchiev2002a}%
  \BibitemOpen
  \bibfield  {author} {\bibinfo {author} {\bibfnamefont {D.}~\bibnamefont
  {Kashchiev}}\ and\ \bibinfo {author} {\bibfnamefont {A.}~\bibnamefont
  {Firoozabadi}},\ }\bibfield  {title} {\enquote {\bibinfo {title} {Driving
  force for crystallization of gas hydrates},}\ }\href@noop {} {\bibfield
  {journal} {\bibinfo  {journal} {J. Cryst. Growth}\ }\textbf {\bibinfo
  {volume} {241}},\ \bibinfo {pages} {220--230} (\bibinfo {year}
  {2002}{\natexlab{a}})}\BibitemShut {NoStop}%
\bibitem [{\citenamefont {Kashchiev}\ and\ \citenamefont
  {Firoozabadi}(2002{\natexlab{b}})}]{Kashchiev2002b}%
  \BibitemOpen
  \bibfield  {author} {\bibinfo {author} {\bibfnamefont {D.}~\bibnamefont
  {Kashchiev}}\ and\ \bibinfo {author} {\bibfnamefont {A.}~\bibnamefont
  {Firoozabadi}},\ }\bibfield  {title} {\enquote {\bibinfo {title} {Nucleation
  of gas hydrates},}\ }\href@noop {} {\bibfield  {journal} {\bibinfo  {journal}
  {J. Cryst. Growth}\ }\textbf {\bibinfo {volume} {243}},\ \bibinfo {pages}
  {476--489} (\bibinfo {year} {2002}{\natexlab{b}})}\BibitemShut {NoStop}%
\bibitem [{\citenamefont {Jacobson}, \citenamefont {Hujo},\ and\ \citenamefont
  {Molinero}(2010{\natexlab{a}})}]{Jacobson2010a}%
  \BibitemOpen
  \bibfield  {author} {\bibinfo {author} {\bibfnamefont {L.~C.}\ \bibnamefont
  {Jacobson}}, \bibinfo {author} {\bibfnamefont {W.}~\bibnamefont {Hujo}}, \
  and\ \bibinfo {author} {\bibfnamefont {V.}~\bibnamefont {Molinero}},\
  }\bibfield  {title} {\enquote {\bibinfo {title} {Amorphous precursors in the
  nucleation of clathrate hydrates},}\ }\href@noop {} {\bibfield  {journal}
  {\bibinfo  {journal} {J. Am. Chem. Soc.}\ }\textbf {\bibinfo {volume}
  {132}},\ \bibinfo {pages} {11806--11811} (\bibinfo {year}
  {2010}{\natexlab{a}})}\BibitemShut {NoStop}%
\bibitem [{\citenamefont {Jacobson}, \citenamefont {Hujo},\ and\ \citenamefont
  {Molinero}(2010{\natexlab{b}})}]{Jacobson2010b}%
  \BibitemOpen
  \bibfield  {author} {\bibinfo {author} {\bibfnamefont {L.~C.}\ \bibnamefont
  {Jacobson}}, \bibinfo {author} {\bibfnamefont {W.}~\bibnamefont {Hujo}}, \
  and\ \bibinfo {author} {\bibfnamefont {V.}~\bibnamefont {Molinero}},\
  }\bibfield  {title} {\enquote {\bibinfo {title} {Nucleation pathways of
  clathrate hydrates: Effect of guest size and solubility},}\ }\href@noop {}
  {\bibfield  {journal} {\bibinfo  {journal} {J. Phys. Chem. B}\ }\textbf
  {\bibinfo {volume} {114}},\ \bibinfo {pages} {13796--13807} (\bibinfo {year}
  {2010}{\natexlab{b}})}\BibitemShut {NoStop}%
\bibitem [{\citenamefont {Walsh}\ \emph {et~al.}(2011)\citenamefont {Walsh},
  \citenamefont {Beckham}, \citenamefont {Koh}, \citenamefont {Sloan},
  \citenamefont {Wu},\ and\ \citenamefont {Sum}}]{Walsh2011a}%
  \BibitemOpen
  \bibfield  {author} {\bibinfo {author} {\bibfnamefont {M.~R.}\ \bibnamefont
  {Walsh}}, \bibinfo {author} {\bibfnamefont {G.~T.}\ \bibnamefont {Beckham}},
  \bibinfo {author} {\bibfnamefont {C.~A.}\ \bibnamefont {Koh}}, \bibinfo
  {author} {\bibfnamefont {E.~D.}\ \bibnamefont {Sloan}}, \bibinfo {author}
  {\bibfnamefont {D.~T.}\ \bibnamefont {Wu}}, \ and\ \bibinfo {author}
  {\bibfnamefont {A.~K.}\ \bibnamefont {Sum}},\ }\bibfield  {title} {\enquote
  {\bibinfo {title} {Methane hydrate nucleation rates from molecular dynamics
  simulations: Effects of aqueous methane concentration, interfacial curvature,
  and system size},}\ }\href {\doibase https://doi.org/10.1021/jp206483q}
  {\bibfield  {journal} {\bibinfo  {journal} {J. Phys. Chem. C}\ }\textbf
  {\bibinfo {volume} {115}},\ \bibinfo {pages} {21241} (\bibinfo {year}
  {2011})}\BibitemShut {NoStop}%
\bibitem [{\citenamefont {Jacobson}\ and\ \citenamefont
  {Molinero}(2011)}]{Jacobson2011a}%
  \BibitemOpen
  \bibfield  {author} {\bibinfo {author} {\bibfnamefont {L.~C.}\ \bibnamefont
  {Jacobson}}\ and\ \bibinfo {author} {\bibfnamefont {V.}~\bibnamefont
  {Molinero}},\ }\bibfield  {title} {\enquote {\bibinfo {title} {Can amorphous
  nuclei grow crystalline clathrates? the size and crystallinity of critical
  clathrate nuclei},}\ }\href@noop {} {\bibfield  {journal} {\bibinfo
  {journal} {J. Am. Chem. Soc.}\ }\textbf {\bibinfo {volume} {133}},\ \bibinfo
  {pages} {6458--6463} (\bibinfo {year} {2011})}\BibitemShut {NoStop}%
\bibitem [{\citenamefont {Sarupria}\ and\ \citenamefont
  {Debenedetti}(2012)}]{Sarupria2011a}%
  \BibitemOpen
  \bibfield  {author} {\bibinfo {author} {\bibfnamefont {S.}~\bibnamefont
  {Sarupria}}\ and\ \bibinfo {author} {\bibfnamefont {P.~G.}\ \bibnamefont
  {Debenedetti}},\ }\bibfield  {title} {\enquote {\bibinfo {title} {Molecular
  dynamics study of carbon dioxide hydrate dissociation},}\ }\href@noop {}
  {\bibfield  {journal} {\bibinfo  {journal} {J. Phys. Chem. Lett.}\ }\textbf
  {\bibinfo {volume} {3}},\ \bibinfo {pages} {2942--2947} (\bibinfo {year}
  {2012})}\BibitemShut {NoStop}%
\bibitem [{\citenamefont {Knott}\ \emph {et~al.}(2012)\citenamefont {Knott},
  \citenamefont {Molinero}, \citenamefont {Doherty},\ and\ \citenamefont
  {Peters}}]{Knott2012a}%
  \BibitemOpen
  \bibfield  {author} {\bibinfo {author} {\bibfnamefont {B.~C.}\ \bibnamefont
  {Knott}}, \bibinfo {author} {\bibfnamefont {V.}~\bibnamefont {Molinero}},
  \bibinfo {author} {\bibfnamefont {M.~F.}\ \bibnamefont {Doherty}}, \ and\
  \bibinfo {author} {\bibfnamefont {B.}~\bibnamefont {Peters}},\ }\bibfield
  {title} {\enquote {\bibinfo {title} {Homogeneous nucleation of methane
  hydrates: Unrealistic under realistic conditions},}\ }\href@noop {}
  {\bibfield  {journal} {\bibinfo  {journal} {J. Am. Chem. Soc.}\ }\textbf
  {\bibinfo {volume} {134}},\ \bibinfo {pages} {19544--19547} (\bibinfo {year}
  {2012})}\BibitemShut {NoStop}%
\bibitem [{\citenamefont {Sarupria}\ and\ \citenamefont
  {Debenedetti}(2011)}]{Sarupria2012a}%
  \BibitemOpen
  \bibfield  {author} {\bibinfo {author} {\bibfnamefont {S.}~\bibnamefont
  {Sarupria}}\ and\ \bibinfo {author} {\bibfnamefont {P.~G.}\ \bibnamefont
  {Debenedetti}},\ }\bibfield  {title} {\enquote {\bibinfo {title} {Homogeneous
  nucleation of methane hydrate in microsecond molecular dynamics
  simulations},}\ }\href@noop {} {\bibfield  {journal} {\bibinfo  {journal} {J.
  Phys. Chem. A}\ }\textbf {\bibinfo {volume} {115}},\ \bibinfo {pages}
  {6102--6111} (\bibinfo {year} {2011})}\BibitemShut {NoStop}%
\bibitem [{\citenamefont {Liang}\ and\ \citenamefont
  {Kusalik}(2013)}]{Liang2013a}%
  \BibitemOpen
  \bibfield  {author} {\bibinfo {author} {\bibfnamefont {S.}~\bibnamefont
  {Liang}}\ and\ \bibinfo {author} {\bibfnamefont {P.~G.}\ \bibnamefont
  {Kusalik}},\ }\bibfield  {title} {\enquote {\bibinfo {title} {Nucleation of
  gas hydrates within constant energy systems},}\ }\href {\doibase
  https://dx.doi.org/10.1021/jp308395x} {\bibfield  {journal} {\bibinfo
  {journal} {J. Phys. Chem. B}\ }\textbf {\bibinfo {volume} {117}},\ \bibinfo
  {pages} {1403} (\bibinfo {year} {2013})}\BibitemShut {NoStop}%
\bibitem [{\citenamefont {Barnes}\ \emph {et~al.}(2014)\citenamefont {Barnes},
  \citenamefont {Knott}, \citenamefont {Beckham}, \citenamefont {Wu},\ and\
  \citenamefont {Sum}}]{Barnes2014a}%
  \BibitemOpen
  \bibfield  {author} {\bibinfo {author} {\bibfnamefont {B.~C.}\ \bibnamefont
  {Barnes}}, \bibinfo {author} {\bibfnamefont {B.~C.}\ \bibnamefont {Knott}},
  \bibinfo {author} {\bibfnamefont {G.~T.}\ \bibnamefont {Beckham}}, \bibinfo
  {author} {\bibfnamefont {D.}~\bibnamefont {Wu}}, \ and\ \bibinfo {author}
  {\bibfnamefont {A.~K.}\ \bibnamefont {Sum}},\ }\bibfield  {title} {\enquote
  {\bibinfo {title} {Molecular dynamics study of carbon dioxide hydrate
  dissociation},}\ }\href {\doibase https://doi.org/10.1021/jp507959q}
  {\bibfield  {journal} {\bibinfo  {journal} {J. Phys. Chem. B}\ }\textbf
  {\bibinfo {volume} {118}},\ \bibinfo {pages} {13236--13243} (\bibinfo {year}
  {2014})}\BibitemShut {NoStop}%
\bibitem [{\citenamefont {Yuhara}\ \emph {et~al.}(2015)\citenamefont {Yuhara},
  \citenamefont {Barnes}, \citenamefont {Suh}, \citenamefont {knott},
  \citenamefont {Beckham}, \citenamefont {Yasuoka}, \citenamefont {Wu},\ and\
  \citenamefont {Sum}}]{Yuhara2015a}%
  \BibitemOpen
  \bibfield  {author} {\bibinfo {author} {\bibfnamefont {D.}~\bibnamefont
  {Yuhara}}, \bibinfo {author} {\bibfnamefont {B.~C.}\ \bibnamefont {Barnes}},
  \bibinfo {author} {\bibfnamefont {D.}~\bibnamefont {Suh}}, \bibinfo {author}
  {\bibfnamefont {B.~C.}\ \bibnamefont {knott}}, \bibinfo {author}
  {\bibfnamefont {G.~T.}\ \bibnamefont {Beckham}}, \bibinfo {author}
  {\bibfnamefont {K.}~\bibnamefont {Yasuoka}}, \bibinfo {author} {\bibfnamefont
  {D.}~\bibnamefont {Wu}}, \ and\ \bibinfo {author} {\bibfnamefont {A.~K.}\
  \bibnamefont {Sum}},\ }\bibfield  {title} {\enquote {\bibinfo {title}
  {Nucleation rate analysis of methane hydrate from molecular dynamics
  simualtions},}\ }\href {\doibase https://doi.org/10.1039/c4fd00219a}
  {\bibfield  {journal} {\bibinfo  {journal} {Faraday Discuss.}\ }\textbf
  {\bibinfo {volume} {179}},\ \bibinfo {pages} {463--474} (\bibinfo {year}
  {2015})}\BibitemShut {NoStop}%
\bibitem [{\citenamefont {Warrier}\ \emph {et~al.}(2016)\citenamefont
  {Warrier}, \citenamefont {Khan}, \citenamefont {Srivastava}, \citenamefont
  {Maupin},\ and\ \citenamefont {Koh}}]{Warrier2016a}%
  \BibitemOpen
  \bibfield  {author} {\bibinfo {author} {\bibfnamefont {P.}~\bibnamefont
  {Warrier}}, \bibinfo {author} {\bibfnamefont {M.~N.}\ \bibnamefont {Khan}},
  \bibinfo {author} {\bibfnamefont {V.}~\bibnamefont {Srivastava}}, \bibinfo
  {author} {\bibfnamefont {C.~M.}\ \bibnamefont {Maupin}}, \ and\ \bibinfo
  {author} {\bibfnamefont {C.~A.}\ \bibnamefont {Koh}},\ }\bibfield  {title}
  {\enquote {\bibinfo {title} {Overview: Nucleation of clathrate hydrates},}\
  }\href {\doibase https://doi.org/10.1063/1.4968590} {\bibfield  {journal}
  {\bibinfo  {journal} {J. Chem. Phys.}\ }\textbf {\bibinfo {volume} {145}},\
  \bibinfo {pages} {211705} (\bibinfo {year} {2016})}\BibitemShut {NoStop}%
\bibitem [{\citenamefont {Zhang}, \citenamefont {Kusalik},\ and\ \citenamefont
  {Guo}(2018)}]{Zhang2018a}%
  \BibitemOpen
  \bibfield  {author} {\bibinfo {author} {\bibfnamefont {Z.}~\bibnamefont
  {Zhang}}, \bibinfo {author} {\bibfnamefont {P.~G.}\ \bibnamefont {Kusalik}},
  \ and\ \bibinfo {author} {\bibfnamefont {G.-J.}\ \bibnamefont {Guo}},\
  }\bibfield  {title} {\enquote {\bibinfo {title} {Molecular insight into the
  growth of hydrogen and methane binary hydrates},}\ }\href {\doibase
  10.1021/acs.jpcc.8b00842} {\bibfield  {journal} {\bibinfo  {journal} {J.
  Phys. Chem. C}\ }\textbf {\bibinfo {volume} {122}},\ \bibinfo {pages}
  {7771--7778} (\bibinfo {year} {2018})}\BibitemShut {NoStop}%
\bibitem [{\citenamefont {Arjun}, \citenamefont {Berendsen},\ and\
  \citenamefont {Bolhuis}(2019)}]{Arjun2019a}%
  \BibitemOpen
  \bibfield  {author} {\bibinfo {author} {\bibfnamefont {A.}~\bibnamefont
  {Arjun}}, \bibinfo {author} {\bibfnamefont {T.~A.}\ \bibnamefont
  {Berendsen}}, \ and\ \bibinfo {author} {\bibfnamefont {P.~G.}\ \bibnamefont
  {Bolhuis}},\ }\bibfield  {title} {\enquote {\bibinfo {title} {Unbiased
  atomistic insight in the competing nucleation mechanisms of methane
  hydrates},}\ }\href {\doibase https://doi.org/10.1073/pnas.1906502116}
  {\bibfield  {journal} {\bibinfo  {journal} {Proc. Natl. Acad. Sci.}\ }\textbf
  {\bibinfo {volume} {116}},\ \bibinfo {pages} {19305} (\bibinfo {year}
  {2019})}\BibitemShut {NoStop}%
\bibitem [{\citenamefont {Zhang}\ \emph {et~al.}(2020)\citenamefont {Zhang},
  \citenamefont {Guo}, \citenamefont {Wu},\ and\ \citenamefont
  {Kusalik}}]{Zhang2020a}%
  \BibitemOpen
  \bibfield  {author} {\bibinfo {author} {\bibfnamefont {Z.}~\bibnamefont
  {Zhang}}, \bibinfo {author} {\bibfnamefont {G.-J.}\ \bibnamefont {Guo}},
  \bibinfo {author} {\bibfnamefont {N.}~\bibnamefont {Wu}}, \ and\ \bibinfo
  {author} {\bibfnamefont {P.~G.}\ \bibnamefont {Kusalik}},\ }\bibfield
  {title} {\enquote {\bibinfo {title} {Molecular insights into guest and
  composition dependence of mixed hydrate nucleation},}\ }\href {\doibase
  10.1021/acs.jpcc.0c07375} {\bibfield  {journal} {\bibinfo  {journal} {J.
  Phys. Chem. C}\ }\textbf {\bibinfo {volume} {124}},\ \bibinfo {pages}
  {25078--25086} (\bibinfo {year} {2020})}\BibitemShut {NoStop}%
\bibitem [{\citenamefont {Arjun}\ and\ \citenamefont
  {Bolhuis}(2020)}]{Arjun2020a}%
  \BibitemOpen
  \bibfield  {author} {\bibinfo {author} {\bibfnamefont {A.}~\bibnamefont
  {Arjun}}\ and\ \bibinfo {author} {\bibfnamefont {P.~G.}\ \bibnamefont
  {Bolhuis}},\ }\bibfield  {title} {\enquote {\bibinfo {title} {Rate prediction
  for homogeneous nucleation of methane hydrate at moderate supersaturation
  using transition interface sampling},}\ }\href {\doibase
  https://dx.doi.org/10.1021/acs.jpcb.0c04582} {\bibfield  {journal} {\bibinfo
  {journal} {J. Phys. Chem. B}\ }\textbf {\bibinfo {volume} {124}},\ \bibinfo
  {pages} {8099} (\bibinfo {year} {2020})}\BibitemShut {NoStop}%
\bibitem [{\citenamefont {Arjun}\ and\ \citenamefont
  {Bolhuis}(2021)}]{Arjun2021a}%
  \BibitemOpen
  \bibfield  {author} {\bibinfo {author} {\bibfnamefont {A.}~\bibnamefont
  {Arjun}}\ and\ \bibinfo {author} {\bibfnamefont {P.~G.}\ \bibnamefont
  {Bolhuis}},\ }\bibfield  {title} {\enquote {\bibinfo {title} {Homogenous
  nucleation rate of {CO$_{2}$} hydrates using transition interface
  sampling},}\ }\href {\doibase https://dx.doi.org/10.1063/5.0044883}
  {\bibfield  {journal} {\bibinfo  {journal} {J. Chem. Phys.}\ }\textbf
  {\bibinfo {volume} {154}},\ \bibinfo {pages} {164507} (\bibinfo {year}
  {2021})}\BibitemShut {NoStop}%
\bibitem [{\citenamefont {Wang}\ \emph {et~al.}(2022)\citenamefont {Wang},
  \citenamefont {Hall}, \citenamefont {Chang},\ and\ \citenamefont
  {kusalik}}]{Wang2022a}%
  \BibitemOpen
  \bibfield  {author} {\bibinfo {author} {\bibfnamefont {L.}~\bibnamefont
  {Wang}}, \bibinfo {author} {\bibfnamefont {K.}~\bibnamefont {Hall}}, \bibinfo
  {author} {\bibfnamefont {Z.}~\bibnamefont {Chang}}, \ and\ \bibinfo {author}
  {\bibfnamefont {P.~G.}\ \bibnamefont {kusalik}},\ }\bibfield  {title}
  {\enquote {\bibinfo {title} {Mixed hydrate nucleation: molecular mechanics
  and cage strcutures},}\ }\href {\doibase 10.1021/acs.jpcb.2c03223} {\bibfield
   {journal} {\bibinfo  {journal} {J. Phys. Chem. B}\ }\textbf {\bibinfo
  {volume} {126}},\ \bibinfo {pages} {7015--7026} (\bibinfo {year}
  {2022})}\BibitemShut {NoStop}%
\bibitem [{\citenamefont {Grabowska}\ \emph {et~al.}(2023)\citenamefont
  {Grabowska}, \citenamefont {Bl{\'a}zquez}, \citenamefont {Sanz},
  \citenamefont {Noya}, \citenamefont {Zer{\'o}n}, \citenamefont {Algaba},
  \citenamefont {M{\'{\i}}guez}, \citenamefont {Blas},\ and\ \citenamefont
  {Vega}}]{Grabowska2023a}%
  \BibitemOpen
  \bibfield  {author} {\bibinfo {author} {\bibfnamefont {J.}~\bibnamefont
  {Grabowska}}, \bibinfo {author} {\bibfnamefont {S.}~\bibnamefont
  {Bl{\'a}zquez}}, \bibinfo {author} {\bibfnamefont {E.}~\bibnamefont {Sanz}},
  \bibinfo {author} {\bibfnamefont {E.~G.}\ \bibnamefont {Noya}}, \bibinfo
  {author} {\bibfnamefont {I.~M.}\ \bibnamefont {Zer{\'o}n}}, \bibinfo {author}
  {\bibfnamefont {J.}~\bibnamefont {Algaba}}, \bibinfo {author} {\bibfnamefont
  {J.~M.}\ \bibnamefont {M{\'{\i}}guez}}, \bibinfo {author} {\bibfnamefont
  {F.~J.}\ \bibnamefont {Blas}}, \ and\ \bibinfo {author} {\bibfnamefont
  {C.}~\bibnamefont {Vega}},\ }\bibfield  {title} {\enquote {\bibinfo {title}
  {Homogeneous nucleation rate of methane hydrate formation under experimental
  conditions from seeding simulations},}\ }\href@noop {} {\bibfield  {journal}
  {\bibinfo  {journal} {J. Chem. Phys.}\ }\textbf {\bibinfo {volume} {158}},\
  \bibinfo {pages} {114505} (\bibinfo {year} {2023})}\BibitemShut {NoStop}%
\bibitem [{\citenamefont {Wang}\ and\ \citenamefont
  {Kusalik}(2023)}]{Wang2023a}%
  \BibitemOpen
  \bibfield  {author} {\bibinfo {author} {\bibfnamefont {L.}~\bibnamefont
  {Wang}}\ and\ \bibinfo {author} {\bibfnamefont {P.~G.}\ \bibnamefont
  {Kusalik}},\ }\bibfield  {title} {\enquote {\bibinfo {title} {Understanding
  why constant energy or constant temperature may affect nucleation behavior in
  {MD} simulations: A study of gas hydrate nucleation},}\ }\href {\doibase
  10.1063/5.0169669} {\bibfield  {journal} {\bibinfo  {journal} {J. Chem.
  Phys.}\ }\textbf {\bibinfo {volume} {159}},\ \bibinfo {pages} {184501}
  (\bibinfo {year} {2023})}\BibitemShut {NoStop}%
\bibitem [{\citenamefont {Abascal}\ \emph {et~al.}(2005)\citenamefont
  {Abascal}, \citenamefont {Sanz}, \citenamefont {Fern\'andez},\ and\
  \citenamefont {Vega}}]{Abascal2005b}%
  \BibitemOpen
  \bibfield  {author} {\bibinfo {author} {\bibfnamefont {J.~L.~F.}\
  \bibnamefont {Abascal}}, \bibinfo {author} {\bibfnamefont {E.}~\bibnamefont
  {Sanz}}, \bibinfo {author} {\bibfnamefont {R.~G.}\ \bibnamefont
  {Fern\'andez}}, \ and\ \bibinfo {author} {\bibfnamefont {C.}~\bibnamefont
  {Vega}},\ }\bibfield  {title} {\enquote {\bibinfo {title} {A potential model
  for the study of ices and amorphous water: {TIP4P/Ice}},}\ }\href@noop {}
  {\bibfield  {journal} {\bibinfo  {journal} {J. Chem. Phys.}\ }\textbf
  {\bibinfo {volume} {122}},\ \bibinfo {pages} {234511} (\bibinfo {year}
  {2005})}\BibitemShut {NoStop}%
\bibitem [{\citenamefont {Potoff}\ and\ \citenamefont
  {Siepmann}(2001)}]{Potoff2001a}%
  \BibitemOpen
  \bibfield  {author} {\bibinfo {author} {\bibfnamefont {J.~J.}\ \bibnamefont
  {Potoff}}\ and\ \bibinfo {author} {\bibfnamefont {J.~I.}\ \bibnamefont
  {Siepmann}},\ }\bibfield  {title} {\enquote {\bibinfo {title} {Vapor-liquid
  equilibria of mixtures containing alkanes, carbon dioxide, and nitrogen},}\
  }\href@noop {} {\bibfield  {journal} {\bibinfo  {journal} {AIChE J.}\
  }\textbf {\bibinfo {volume} {47}},\ \bibinfo {pages} {1676--1682} (\bibinfo
  {year} {2001})}\BibitemShut {NoStop}%
\bibitem [{\citenamefont {{van der Spoel}}\ \emph {et~al.}(2005)\citenamefont
  {{van der Spoel}}, \citenamefont {Lindahl}, \citenamefont {Hess},
  \citenamefont {Groenhof}, \citenamefont {Mark},\ and\ \citenamefont
  {Berendsen}}]{VanDerSpoel2005a}%
  \BibitemOpen
  \bibfield  {author} {\bibinfo {author} {\bibfnamefont {D.}~\bibnamefont {{van
  der Spoel}}}, \bibinfo {author} {\bibfnamefont {E.}~\bibnamefont {Lindahl}},
  \bibinfo {author} {\bibfnamefont {B.}~\bibnamefont {Hess}}, \bibinfo {author}
  {\bibfnamefont {G.}~\bibnamefont {Groenhof}}, \bibinfo {author}
  {\bibfnamefont {A.~E.}\ \bibnamefont {Mark}}, \ and\ \bibinfo {author}
  {\bibfnamefont {H.~J.}\ \bibnamefont {Berendsen}},\ }\bibfield  {title}
  {\enquote {\bibinfo {title} {Gromacs: Fast, flexible, and free},}\
  }\href@noop {} {\bibfield  {journal} {\bibinfo  {journal} {J. Comput. Chem.}\
  }\textbf {\bibinfo {volume} {26}},\ \bibinfo {pages} {1701--1718} (\bibinfo
  {year} {2005})}\BibitemShut {NoStop}%
\bibitem [{\citenamefont {Cuendet}\ and\ \citenamefont
  {Gunsteren}(2007)}]{Cuendet2007a}%
  \BibitemOpen
  \bibfield  {author} {\bibinfo {author} {\bibfnamefont {M.~A.}\ \bibnamefont
  {Cuendet}}\ and\ \bibinfo {author} {\bibfnamefont {W.~F.~V.}\ \bibnamefont
  {Gunsteren}},\ }\bibfield  {title} {\enquote {\bibinfo {title} {On the
  calculation of velocity-dependent properties in molecular dynamics
  simulations using the leapfrog integration algorithm},}\ }\href@noop {}
  {\bibfield  {journal} {\bibinfo  {journal} {J. Chem. Phys.}\ }\textbf
  {\bibinfo {volume} {127}},\ \bibinfo {pages} {184102} (\bibinfo {year}
  {2007})}\BibitemShut {NoStop}%
\bibitem [{\citenamefont {Nos{\'e}}(1984)}]{Nose1984a}%
  \BibitemOpen
  \bibfield  {author} {\bibinfo {author} {\bibfnamefont {S.}~\bibnamefont
  {Nos{\'e}}},\ }\bibfield  {title} {\enquote {\bibinfo {title} {A molecular
  dynamics method for simulations in the canonical ensemble},}\ }\href@noop {}
  {\bibfield  {journal} {\bibinfo  {journal} {Mol. Phys.}\ }\textbf {\bibinfo
  {volume} {52}},\ \bibinfo {pages} {255--268} (\bibinfo {year}
  {1984})}\BibitemShut {NoStop}%
\bibitem [{\citenamefont {Parrinello}\ and\ \citenamefont
  {Rahman}(1981)}]{Parrinello1981a}%
  \BibitemOpen
  \bibfield  {author} {\bibinfo {author} {\bibfnamefont {M.}~\bibnamefont
  {Parrinello}}\ and\ \bibinfo {author} {\bibfnamefont {A.}~\bibnamefont
  {Rahman}},\ }\bibfield  {title} {\enquote {\bibinfo {title} {Polymorphic
  transitions in single crystals: A new molecular dynamics method},}\
  }\href@noop {} {\bibfield  {journal} {\bibinfo  {journal} {J. Appl. Phys.}\
  }\textbf {\bibinfo {volume} {52}},\ \bibinfo {pages} {7182--7190} (\bibinfo
  {year} {1981})}\BibitemShut {NoStop}%
\bibitem [{\citenamefont {Essmann}\ \emph {et~al.}(1995)\citenamefont
  {Essmann}, \citenamefont {Perera}, \citenamefont {Berkowitz}, \citenamefont
  {Darden}, \citenamefont {Lee},\ and\ \citenamefont
  {Pedersen}}]{Essmann1995a}%
  \BibitemOpen
  \bibfield  {author} {\bibinfo {author} {\bibfnamefont {U.}~\bibnamefont
  {Essmann}}, \bibinfo {author} {\bibfnamefont {L.}~\bibnamefont {Perera}},
  \bibinfo {author} {\bibfnamefont {M.~L.}\ \bibnamefont {Berkowitz}}, \bibinfo
  {author} {\bibfnamefont {T.}~\bibnamefont {Darden}}, \bibinfo {author}
  {\bibfnamefont {H.}~\bibnamefont {Lee}}, \ and\ \bibinfo {author}
  {\bibfnamefont {L.~G.}\ \bibnamefont {Pedersen}},\ }\bibfield  {title}
  {\enquote {\bibinfo {title} {A smooth particle mesh {Ewald} method},}\
  }\href@noop {} {\bibfield  {journal} {\bibinfo  {journal} {J. Chem. Phys.}\
  }\textbf {\bibinfo {volume} {103}},\ \bibinfo {pages} {8577--8593} (\bibinfo
  {year} {1995})}\BibitemShut {NoStop}%
\bibitem [{\citenamefont {Ladd}\ and\ \citenamefont
  {Woodcock}(1977)}]{Ladd1977a}%
  \BibitemOpen
  \bibfield  {author} {\bibinfo {author} {\bibfnamefont {J.~C.}\ \bibnamefont
  {Ladd}}\ and\ \bibinfo {author} {\bibfnamefont {L.~V.}\ \bibnamefont
  {Woodcock}},\ }\bibfield  {title} {\enquote {\bibinfo {title} {Triple-point
  coexistence properties of the lennard-jones system},}\ }\href@noop {}
  {\bibfield  {journal} {\bibinfo  {journal} {Chem. Phys. Lett.}\ }\textbf
  {\bibinfo {volume} {51}},\ \bibinfo {pages} {159155} (\bibinfo {year}
  {1977})}\BibitemShut {NoStop}%
\bibitem [{\citenamefont {Vega}\ \emph {et~al.}(2008)\citenamefont {Vega},
  \citenamefont {Sanz}, \citenamefont {Abacal},\ and\ \citenamefont
  {Noya}}]{Vega2008a}%
  \BibitemOpen
  \bibfield  {author} {\bibinfo {author} {\bibfnamefont {C.}~\bibnamefont
  {Vega}}, \bibinfo {author} {\bibfnamefont {E.}~\bibnamefont {Sanz}}, \bibinfo
  {author} {\bibfnamefont {J.~L.~F.}\ \bibnamefont {Abacal}}, \ and\ \bibinfo
  {author} {\bibfnamefont {E.~G.}\ \bibnamefont {Noya}},\ }\bibfield  {title}
  {\enquote {\bibinfo {title} {Determination of phase diagrams via computer
  simulation: Methodology and applications to water, electrolytes and
  proteins},}\ }\href@noop {} {\bibfield  {journal} {\bibinfo  {journal} {J.
  Phys.: Condensed Matter}\ }\textbf {\bibinfo {volume} {20}},\ \bibinfo
  {pages} {153101} (\bibinfo {year} {2008})}\BibitemShut {NoStop}%
\bibitem [{\citenamefont {Conde}\ and\ \citenamefont
  {Vega}(2010)}]{Conde2010a}%
  \BibitemOpen
  \bibfield  {author} {\bibinfo {author} {\bibfnamefont {M.~M.}\ \bibnamefont
  {Conde}}\ and\ \bibinfo {author} {\bibfnamefont {C.}~\bibnamefont {Vega}},\
  }\bibfield  {title} {\enquote {\bibinfo {title} {Determining the three-phase
  coexistence line in methane hydrates using computer simulations},}\ }\href
  {\doibase https://doi.org/10.1063/1.3466751} {\bibfield  {journal} {\bibinfo
  {journal} {J. Chem. Phys.}\ }\textbf {\bibinfo {volume} {133}},\ \bibinfo
  {pages} {064507} (\bibinfo {year} {2010})}\BibitemShut {NoStop}%
\bibitem [{\citenamefont {Conde}\ and\ \citenamefont
  {Vega}(2013)}]{Conde2013a}%
  \BibitemOpen
  \bibfield  {author} {\bibinfo {author} {\bibfnamefont {M.~M.}\ \bibnamefont
  {Conde}}\ and\ \bibinfo {author} {\bibfnamefont {C.}~\bibnamefont {Vega}},\
  }\bibfield  {title} {\enquote {\bibinfo {title} {Note: A simple correlation
  to locate the three phase coexistence line in methane-hydrate simulations},}\
  }\href {\doibase https://doi.org/10.1063/1.4790647} {\bibfield  {journal}
  {\bibinfo  {journal} {J. Chem. Phys.}\ }\textbf {\bibinfo {volume} {138}},\
  \bibinfo {pages} {056101} (\bibinfo {year} {2013})}\BibitemShut {NoStop}%
\bibitem [{\citenamefont {Michalis}\ \emph {et~al.}(2015)\citenamefont
  {Michalis}, \citenamefont {Costandy}, \citenamefont {Tsimpanogiannis},
  \citenamefont {Stubos},\ and\ \citenamefont {Economou}}]{Michalis2015a}%
  \BibitemOpen
  \bibfield  {author} {\bibinfo {author} {\bibfnamefont {V.~K.}\ \bibnamefont
  {Michalis}}, \bibinfo {author} {\bibfnamefont {J.}~\bibnamefont {Costandy}},
  \bibinfo {author} {\bibfnamefont {I.~N.}\ \bibnamefont {Tsimpanogiannis}},
  \bibinfo {author} {\bibfnamefont {A.~K.}\ \bibnamefont {Stubos}}, \ and\
  \bibinfo {author} {\bibfnamefont {I.~G.}\ \bibnamefont {Economou}},\
  }\bibfield  {title} {\enquote {\bibinfo {title} {Prediction of the phase
  equilibria of methane hydrates using the direct phase coexistence
  methodology},}\ }\href {\doibase https://doi.org/10.1063/1.4905572}
  {\bibfield  {journal} {\bibinfo  {journal} {J. Chem. Phys.}\ }\textbf
  {\bibinfo {volume} {142}},\ \bibinfo {pages} {044501} (\bibinfo {year}
  {2015})}\BibitemShut {NoStop}%
\bibitem [{\citenamefont {M{\'\i}guez}\ \emph {et~al.}(2015)\citenamefont
  {M{\'\i}guez}, \citenamefont {Conde}, \citenamefont {Torr{\'e}},
  \citenamefont {Blas}, \citenamefont {Pi{\~n}eiro},\ and\ \citenamefont
  {Vega}}]{Miguez2015a}%
  \BibitemOpen
  \bibfield  {author} {\bibinfo {author} {\bibfnamefont {J.~M.}\ \bibnamefont
  {M{\'\i}guez}}, \bibinfo {author} {\bibfnamefont {M.~M.}\ \bibnamefont
  {Conde}}, \bibinfo {author} {\bibfnamefont {J.-P.}\ \bibnamefont
  {Torr{\'e}}}, \bibinfo {author} {\bibfnamefont {F.~J.}\ \bibnamefont {Blas}},
  \bibinfo {author} {\bibfnamefont {M.~M.}\ \bibnamefont {Pi{\~n}eiro}}, \ and\
  \bibinfo {author} {\bibfnamefont {C.}~\bibnamefont {Vega}},\ }\bibfield
  {title} {\enquote {\bibinfo {title} {Molecular dynamics simulation of
  {CO$_2$} hydrates: Prediction of three phase coexistence line},}\ }\href
  {\doibase https://doi.org/10.1063/1.4916119} {\bibfield  {journal} {\bibinfo
  {journal} {J. Chem. Phys.}\ }\textbf {\bibinfo {volume} {142}},\ \bibinfo
  {pages} {124505} (\bibinfo {year} {2015})}\BibitemShut {NoStop}%
\bibitem [{\citenamefont {Costandy}\ \emph {et~al.}(2015)\citenamefont
  {Costandy}, \citenamefont {Michalis}, \citenamefont {Tsimpanogiannis},
  \citenamefont {Stubos},\ and\ \citenamefont {Economou}}]{Costandy2015a}%
  \BibitemOpen
  \bibfield  {author} {\bibinfo {author} {\bibfnamefont {J.}~\bibnamefont
  {Costandy}}, \bibinfo {author} {\bibfnamefont {V.~K.}\ \bibnamefont
  {Michalis}}, \bibinfo {author} {\bibfnamefont {I.~N.}\ \bibnamefont
  {Tsimpanogiannis}}, \bibinfo {author} {\bibfnamefont {A.~K.}\ \bibnamefont
  {Stubos}}, \ and\ \bibinfo {author} {\bibfnamefont {I.~G.}\ \bibnamefont
  {Economou}},\ }\bibfield  {title} {\enquote {\bibinfo {title} {The role of
  intermolecular interactions in the prediction of the phase equilibria of
  carbon dioxide hydrates},}\ }\href {\doibase
  https://doi.org/10.1063/1.4929805} {\bibfield  {journal} {\bibinfo  {journal}
  {J. Chem. Phys.}\ }\textbf {\bibinfo {volume} {143}},\ \bibinfo {pages}
  {094506} (\bibinfo {year} {2015})}\BibitemShut {NoStop}%
\bibitem [{\citenamefont {P{\'e}rez-Rodr{\'\i}guez}\ \emph
  {et~al.}(2017)\citenamefont {P{\'e}rez-Rodr{\'\i}guez}, \citenamefont
  {Vidal-Vidal}, \citenamefont {M{\'\i}guez}, \citenamefont {Blas},
  \citenamefont {Torr{\'e}},\ and\ \citenamefont
  {Pi{\~n}eiro}}]{Perez-Rodriguez2017a}%
  \BibitemOpen
  \bibfield  {author} {\bibinfo {author} {\bibfnamefont {M.}~\bibnamefont
  {P{\'e}rez-Rodr{\'\i}guez}}, \bibinfo {author} {\bibfnamefont
  {A.}~\bibnamefont {Vidal-Vidal}}, \bibinfo {author} {\bibfnamefont
  {J.}~\bibnamefont {M{\'\i}guez}}, \bibinfo {author} {\bibfnamefont {F.~J.}\
  \bibnamefont {Blas}}, \bibinfo {author} {\bibfnamefont {J.-P.}\ \bibnamefont
  {Torr{\'e}}}, \ and\ \bibinfo {author} {\bibfnamefont {M.~M.}\ \bibnamefont
  {Pi{\~n}eiro}},\ }\bibfield  {title} {\enquote {\bibinfo {title}
  {Computational study of the interplay between intermolecular interactions and
  {CO$_{2}$} orientations in type {I} hydrates},}\ }\href {\doibase
  https://doi.org/10.1039/C6CP07097C} {\bibfield  {journal} {\bibinfo
  {journal} {Phys. Chem. Chem. Phys.}\ }\textbf {\bibinfo {volume} {19}},\
  \bibinfo {pages} {3384--3393} (\bibinfo {year} {2017})}\BibitemShut {NoStop}%
\bibitem [{\citenamefont {Algaba}\ \emph
  {et~al.}(2024{\natexlab{a}})\citenamefont {Algaba}, \citenamefont
  {Romero-Guzm{\'a}n}, \citenamefont {Torrej{\'o}n},\ and\ \citenamefont
  {Blas}}]{Algaba2024c}%
  \BibitemOpen
  \bibfield  {author} {\bibinfo {author} {\bibfnamefont {J.}~\bibnamefont
  {Algaba}}, \bibinfo {author} {\bibfnamefont {C.}~\bibnamefont
  {Romero-Guzm{\'a}n}}, \bibinfo {author} {\bibfnamefont {M.~J.}\ \bibnamefont
  {Torrej{\'o}n}}, \ and\ \bibinfo {author} {\bibfnamefont {F.~J.}\
  \bibnamefont {Blas}},\ }\bibfield  {title} {\enquote {\bibinfo {title}
  {Prediction of the univariant two-phase coexistence line of the
  tetrahydrofuran hydrate from computer simulation},}\ }\href@noop {}
  {\bibfield  {journal} {\bibinfo  {journal} {J. Phys. Chem.}\ }\textbf
  {\bibinfo {volume} {160}},\ \bibinfo {pages} {164718} (\bibinfo {year}
  {2024}{\natexlab{a}})}\BibitemShut {NoStop}%
\bibitem [{\citenamefont {Fern{\'a}ndez-Fern{\'a}ndez}\ \emph
  {et~al.}(2019)\citenamefont {Fern{\'a}ndez-Fern{\'a}ndez}, \citenamefont
  {P{\'e}rez-Rodr{\'\i}guez}, \citenamefont {Comesa{\~n}a},\ and\ \citenamefont
  {Pi{\~n}eiro}}]{Fernandez-Fernandez2019a}%
  \BibitemOpen
  \bibfield  {author} {\bibinfo {author} {\bibfnamefont {A.~M.}\ \bibnamefont
  {Fern{\'a}ndez-Fern{\'a}ndez}}, \bibinfo {author} {\bibfnamefont
  {M.}~\bibnamefont {P{\'e}rez-Rodr{\'\i}guez}}, \bibinfo {author}
  {\bibfnamefont {A.}~\bibnamefont {Comesa{\~n}a}}, \ and\ \bibinfo {author}
  {\bibfnamefont {M.~M.}\ \bibnamefont {Pi{\~n}eiro}},\ }\bibfield  {title}
  {\enquote {\bibinfo {title} {Three-phase equilibrium curve shift for methane
  hydrate in oceanic conditions calculated from molecular dynamics
  simulations},}\ }\href {\doibase
  https://doi.org/10.1016/j.molliq.2018.10.146} {\bibfield  {journal} {\bibinfo
   {journal} {J. Mol. Liq.}\ }\textbf {\bibinfo {volume} {274}},\ \bibinfo
  {pages} {426--433} (\bibinfo {year} {2019})}\BibitemShut {NoStop}%
\bibitem [{\citenamefont {Blazquez}, \citenamefont {Vega},\ and\ \citenamefont
  {Conde}(2023)}]{Blazquez2023b}%
  \BibitemOpen
  \bibfield  {author} {\bibinfo {author} {\bibfnamefont {S.}~\bibnamefont
  {Blazquez}}, \bibinfo {author} {\bibfnamefont {C.}~\bibnamefont {Vega}}, \
  and\ \bibinfo {author} {\bibfnamefont {M.~M.}\ \bibnamefont {Conde}},\
  }\bibfield  {title} {\enquote {\bibinfo {title} {Three phase equilibria of
  the methane hydrate in {NaCl} solutions: A simulation study},}\ }\href@noop
  {} {\bibfield  {journal} {\bibinfo  {journal} {J. Mol. Liq.}\ }\textbf
  {\bibinfo {volume} {383}},\ \bibinfo {pages} {122031} (\bibinfo {year}
  {2023})}\BibitemShut {NoStop}%
\bibitem [{\citenamefont {Zhang}\ \emph {et~al.}(2022)\citenamefont {Zhang},
  \citenamefont {Kusalik}, \citenamefont {Wu}, \citenamefont {Liu},\ and\
  \citenamefont {Zhang}}]{Zhang2022a}%
  \BibitemOpen
  \bibfield  {author} {\bibinfo {author} {\bibfnamefont {Z.}~\bibnamefont
  {Zhang}}, \bibinfo {author} {\bibfnamefont {P.~G.}\ \bibnamefont {Kusalik}},
  \bibinfo {author} {\bibfnamefont {N.}~\bibnamefont {Wu}}, \bibinfo {author}
  {\bibfnamefont {C.}~\bibnamefont {Liu}}, \ and\ \bibinfo {author}
  {\bibfnamefont {Y.}~\bibnamefont {Zhang}},\ }\bibfield  {title} {\enquote
  {\bibinfo {title} {Molecular simulation study on the stability of methane
  hydrate confined in slit-shaped pores},}\ }\href {\doibase
  https://doi.org/10.1016/j.energy.2022.124738} {\bibfield  {journal} {\bibinfo
   {journal} {Energy}\ }\textbf {\bibinfo {volume} {257}},\ \bibinfo {pages}
  {124738} (\bibinfo {year} {2022})}\BibitemShut {NoStop}%
\bibitem [{\citenamefont {Zhang}\ \emph {et~al.}(2023)\citenamefont {Zhang},
  \citenamefont {Kusalik}, \citenamefont {Liu},\ and\ \citenamefont
  {Wu}}]{Zhang2023a}%
  \BibitemOpen
  \bibfield  {author} {\bibinfo {author} {\bibfnamefont {Z.}~\bibnamefont
  {Zhang}}, \bibinfo {author} {\bibfnamefont {P.~G.}\ \bibnamefont {Kusalik}},
  \bibinfo {author} {\bibfnamefont {C.}~\bibnamefont {Liu}}, \ and\ \bibinfo
  {author} {\bibfnamefont {N.}~\bibnamefont {Wu}},\ }\bibfield  {title}
  {\enquote {\bibinfo {title} {Methane hydrate formation in slit-shaped pores:
  Impacts of surface hydrophilicity},}\ }\href {\doibase
  https://doi.org/10.1016/j.energy.2023.129414} {\bibfield  {journal} {\bibinfo
   {journal} {Energy}\ }\textbf {\bibinfo {volume} {285}},\ \bibinfo {pages}
  {129414} (\bibinfo {year} {2023})}\BibitemShut {NoStop}%
\bibitem [{\citenamefont {Fern{\'a}ndez-Fern{\'a}ndez}\ \emph
  {et~al.}(2024)\citenamefont {Fern{\'a}ndez-Fern{\'a}ndez}, \citenamefont
  {B{\'a}rcena}, \citenamefont {Conde}, \citenamefont {P{\'e}rez-S{\'a}nchez},
  \citenamefont {P{\'e}rez-Rodr{\'{\i}}guez},\ and\ \citenamefont
  {Pi{\~n}eiro}}]{Fernandez-Fernandez2024a}%
  \BibitemOpen
  \bibfield  {author} {\bibinfo {author} {\bibfnamefont {A.~M.}\ \bibnamefont
  {Fern{\'a}ndez-Fern{\'a}ndez}}, \bibinfo {author} {\bibfnamefont
  {A.}~\bibnamefont {B{\'a}rcena}}, \bibinfo {author} {\bibfnamefont {M.~M.}\
  \bibnamefont {Conde}}, \bibinfo {author} {\bibfnamefont {G.}~\bibnamefont
  {P{\'e}rez-S{\'a}nchez}}, \bibinfo {author} {\bibfnamefont {M.}~\bibnamefont
  {P{\'e}rez-Rodr{\'{\i}}guez}}, \ and\ \bibinfo {author} {\bibfnamefont
  {M.~M.}\ \bibnamefont {Pi{\~n}eiro}},\ }\bibfield  {title} {\enquote
  {\bibinfo {title} {Modeling oceanic sedimentary methane hydrate growth
  through molecular dynamics simulation},}\ }\href {\doibase
  https://doi.org/10.1063/5.0203116} {\bibfield  {journal} {\bibinfo  {journal}
  {J. Chem. Phys.}\ }\textbf {\bibinfo {volume} {160}},\ \bibinfo {pages}
  {144107} (\bibinfo {year} {2024})}\BibitemShut {NoStop}%
\bibitem [{\citenamefont {Waage}, \citenamefont {Vlugt},\ and\ \citenamefont
  {Kjelstrup}(2017)}]{Waage2017a}%
  \BibitemOpen
  \bibfield  {author} {\bibinfo {author} {\bibfnamefont {M.~H.}\ \bibnamefont
  {Waage}}, \bibinfo {author} {\bibfnamefont {T.~J.~H.}\ \bibnamefont {Vlugt}},
  \ and\ \bibinfo {author} {\bibfnamefont {S.}~\bibnamefont {Kjelstrup}},\
  }\bibfield  {title} {\enquote {\bibinfo {title} {Phase diagram of methane and
  carbon dioxide hydrates computed by {Monte Carlo} simulations},}\ }\href
  {\doibase https://doi.org/10.1021/acs.jpcb.7b03071} {\bibfield  {journal}
  {\bibinfo  {journal} {J. Phys. Chem. B}\ }\textbf {\bibinfo {volume} {121}},\
  \bibinfo {pages} {7336--7350} (\bibinfo {year} {2017})}\BibitemShut {NoStop}%
\bibitem [{\citenamefont {Jensen}\ \emph {et~al.}(2010)\citenamefont {Jensen},
  \citenamefont {Thomsen}, \citenamefont {von Solms}, \citenamefont
  {Wierzchowski}, \citenamefont {Walsh}, \citenamefont {Koh}, \citenamefont
  {Sloan}, \citenamefont {Wu},\ and\ \citenamefont {Sum}}]{Jensen2010a}%
  \BibitemOpen
  \bibfield  {author} {\bibinfo {author} {\bibfnamefont {L.}~\bibnamefont
  {Jensen}}, \bibinfo {author} {\bibfnamefont {K.}~\bibnamefont {Thomsen}},
  \bibinfo {author} {\bibfnamefont {N.}~\bibnamefont {von Solms}}, \bibinfo
  {author} {\bibfnamefont {S.}~\bibnamefont {Wierzchowski}}, \bibinfo {author}
  {\bibfnamefont {M.~R.}\ \bibnamefont {Walsh}}, \bibinfo {author}
  {\bibfnamefont {C.~A.}\ \bibnamefont {Koh}}, \bibinfo {author} {\bibfnamefont
  {E.~D.}\ \bibnamefont {Sloan}}, \bibinfo {author} {\bibfnamefont {D.~T.}\
  \bibnamefont {Wu}}, \ and\ \bibinfo {author} {\bibfnamefont {A.~K.}\
  \bibnamefont {Sum}},\ }\bibfield  {title} {\enquote {\bibinfo {title}
  {Calculation of liquid water--hydrate--methane vapor phase equilibria from
  molecular simulations},}\ }\href@noop {} {\bibfield  {journal} {\bibinfo
  {journal} {J. Phys. Chem. B}\ }\textbf {\bibinfo {volume} {114}},\ \bibinfo
  {pages} {5775--5782} (\bibinfo {year} {2010})}\BibitemShut {NoStop}%
\bibitem [{\citenamefont {Wiebe}, \citenamefont {Gaddy},\ and\ \citenamefont
  {Heins~Jr}(1933)}]{Wiebe1933}%
  \BibitemOpen
  \bibfield  {author} {\bibinfo {author} {\bibfnamefont {R.}~\bibnamefont
  {Wiebe}}, \bibinfo {author} {\bibfnamefont {V.}~\bibnamefont {Gaddy}}, \ and\
  \bibinfo {author} {\bibfnamefont {C.}~\bibnamefont {Heins~Jr}},\ }\bibfield
  {title} {\enquote {\bibinfo {title} {The solubility of nitrogen in water at
  50, 75 and 100 from 25 to 1000 atmospheres},}\ }\href@noop {} {\bibfield
  {journal} {\bibinfo  {journal} {J. Am. Chem. Soc.}\ }\textbf {\bibinfo
  {volume} {55}},\ \bibinfo {pages} {947--953} (\bibinfo {year}
  {1933})}\BibitemShut {NoStop}%
\bibitem [{\citenamefont {Rowlinson}\ and\ \citenamefont
  {Widom}(1982)}]{Rowlinson1982b}%
  \BibitemOpen
  \bibfield  {author} {\bibinfo {author} {\bibfnamefont {J.~S.}\ \bibnamefont
  {Rowlinson}}\ and\ \bibinfo {author} {\bibfnamefont {B.}~\bibnamefont
  {Widom}},\ }\href@noop {} {\emph {\bibinfo {title} {Molecular Theory of
  Capillarity}}}\ (\bibinfo  {publisher} {Claredon Press},\ \bibinfo {year}
  {1982})\BibitemShut {NoStop}%
\bibitem [{\citenamefont {Miguel}\ and\ \citenamefont
  {Jackson}(2006{\natexlab{a}})}]{deMiguel2006c}%
  \BibitemOpen
  \bibfield  {author} {\bibinfo {author} {\bibfnamefont {E.~D.}\ \bibnamefont
  {Miguel}}\ and\ \bibinfo {author} {\bibfnamefont {G.}~\bibnamefont
  {Jackson}},\ }\bibfield  {title} {\enquote {\bibinfo {title} {Detailed
  examination of the calculation of the pressure in simulations of systems with
  discontinuous interactions from the mechanical and thermodynamic
  perspectives},}\ }\href@noop {} {\bibfield  {journal} {\bibinfo  {journal}
  {Mol. Phys.}\ }\textbf {\bibinfo {volume} {104}},\ \bibinfo {pages}
  {3717--3734} (\bibinfo {year} {2006}{\natexlab{a}})}\BibitemShut {NoStop}%
\bibitem [{\citenamefont {Miguel}\ and\ \citenamefont
  {Jackson}(2006{\natexlab{b}})}]{deMiguel2006b}%
  \BibitemOpen
  \bibfield  {author} {\bibinfo {author} {\bibfnamefont {E.~D.}\ \bibnamefont
  {Miguel}}\ and\ \bibinfo {author} {\bibfnamefont {G.}~\bibnamefont
  {Jackson}},\ }\bibfield  {title} {\enquote {\bibinfo {title} {The nature of
  the calculation of the pressure in molecular simulations of continuous models
  from volume perturbations},}\ }\href@noop {} {\bibfield  {journal} {\bibinfo
  {journal} {J. Chem. Phys.}\ }\textbf {\bibinfo {volume} {125}},\ \bibinfo
  {pages} {164109} (\bibinfo {year} {2006}{\natexlab{b}})}\BibitemShut
  {NoStop}%
\bibitem [{\citenamefont {Flyvbjerg}\ and\ \citenamefont
  {Petersen}(1989)}]{Flyvbjerg1989a}%
  \BibitemOpen
  \bibfield  {author} {\bibinfo {author} {\bibfnamefont {H.}~\bibnamefont
  {Flyvbjerg}}\ and\ \bibinfo {author} {\bibfnamefont {H.}~\bibnamefont
  {Petersen}},\ }\bibfield  {title} {\enquote {\bibinfo {title} {Error
  estimates on averages of correlated data},}\ }\href@noop {} {\bibfield
  {journal} {\bibinfo  {journal} {J. Chem. Phys.}\ }\textbf {\bibinfo {volume}
  {91}},\ \bibinfo {pages} {461--466} (\bibinfo {year} {1989})}\BibitemShut
  {NoStop}%
\bibitem [{\citenamefont {Wiegand}\ and\ \citenamefont
  {Franck}(1994)}]{Wiegand1994}%
  \BibitemOpen
  \bibfield  {author} {\bibinfo {author} {\bibfnamefont {G.}~\bibnamefont
  {Wiegand}}\ and\ \bibinfo {author} {\bibfnamefont {E.}~\bibnamefont
  {Franck}},\ }\bibfield  {title} {\enquote {\bibinfo {title} {Interfacial
  tension between water and non-polar fluids up to 473 {K} and 2800 bar},}\
  }\href@noop {} {\bibfield  {journal} {\bibinfo  {journal} {Ber.
  Bunsengesellschaft Phys. Chem.}\ }\textbf {\bibinfo {volume} {98}},\ \bibinfo
  {pages} {809--817} (\bibinfo {year} {1994})}\BibitemShut {NoStop}%
\bibitem [{\citenamefont {Buch}, \citenamefont {Sandler},\ and\ \citenamefont
  {Sadlej}(1998)}]{Buch1998a}%
  \BibitemOpen
  \bibfield  {author} {\bibinfo {author} {\bibfnamefont {V.}~\bibnamefont
  {Buch}}, \bibinfo {author} {\bibfnamefont {P.}~\bibnamefont {Sandler}}, \
  and\ \bibinfo {author} {\bibfnamefont {J.}~\bibnamefont {Sadlej}},\
  }\bibfield  {title} {\enquote {\bibinfo {title} {Simulations of {H$_{2}$O}
  solid, liquid, and clusters, with an emphasis on ferroelectric ordering
  transition in hexagonal ice},}\ }\href@noop {} {\bibfield  {journal}
  {\bibinfo  {journal} {J. Phys. Chem. B}\ }\textbf {\bibinfo {volume} {102}},\
  \bibinfo {pages} {8641--8653} (\bibinfo {year} {1998})}\BibitemShut {NoStop}%
\bibitem [{\citenamefont {Bernal}\ and\ \citenamefont
  {Fowler}(1933)}]{Bernal1933a}%
  \BibitemOpen
  \bibfield  {author} {\bibinfo {author} {\bibfnamefont {J.~D.}\ \bibnamefont
  {Bernal}}\ and\ \bibinfo {author} {\bibfnamefont {R.~H.}\ \bibnamefont
  {Fowler}},\ }\bibfield  {title} {\enquote {\bibinfo {title} {Simulations of
  {H$_{2}$O} solid, liquid, and clusters, with an emphasis on ferroelectric
  ordering transition in hexagonal ice},}\ }\href@noop {} {\bibfield  {journal}
  {\bibinfo  {journal} {J. Chem. Phys.}\ }\textbf {\bibinfo {volume} {1}},\
  \bibinfo {pages} {515--548} (\bibinfo {year} {1933})}\BibitemShut {NoStop}%
\bibitem [{\citenamefont {Blazquez}\ \emph {et~al.}(2024)\citenamefont
  {Blazquez}, \citenamefont {Algaba}, \citenamefont {Míguez}, \citenamefont
  {Vega}, \citenamefont {Blas},\ and\ \citenamefont {Conde}}]{Blazquez2024a}%
  \BibitemOpen
  \bibfield  {author} {\bibinfo {author} {\bibfnamefont {S.}~\bibnamefont
  {Blazquez}}, \bibinfo {author} {\bibfnamefont {J.}~\bibnamefont {Algaba}},
  \bibinfo {author} {\bibfnamefont {J.~M.}\ \bibnamefont {Míguez}}, \bibinfo
  {author} {\bibfnamefont {C.}~\bibnamefont {Vega}}, \bibinfo {author}
  {\bibfnamefont {F.~J.}\ \bibnamefont {Blas}}, \ and\ \bibinfo {author}
  {\bibfnamefont {M.~M.}\ \bibnamefont {Conde}},\ }\bibfield  {title} {\enquote
  {\bibinfo {title} {Three-phase equilibria of hydrates from computer
  simulation. {I}: {F}inite-size effects in the methane hydrate},}\ }\href@noop
  {} {\bibfield  {journal} {\bibinfo  {journal} {J. Chem. Phys.}\ }\textbf
  {\bibinfo {volume} {160}},\ \bibinfo {pages} {164721} (\bibinfo {year}
  {2024})}\BibitemShut {NoStop}%
\bibitem [{\citenamefont {Algaba}\ \emph
  {et~al.}(2024{\natexlab{b}})\citenamefont {Algaba}, \citenamefont {Blazquez},
  \citenamefont {Feria}, \citenamefont {Míguez}, \citenamefont {Conde},\ and\
  \citenamefont {Blas}}]{Algaba2024a}%
  \BibitemOpen
  \bibfield  {author} {\bibinfo {author} {\bibfnamefont {J.}~\bibnamefont
  {Algaba}}, \bibinfo {author} {\bibfnamefont {S.}~\bibnamefont {Blazquez}},
  \bibinfo {author} {\bibfnamefont {E.}~\bibnamefont {Feria}}, \bibinfo
  {author} {\bibfnamefont {J.~M.}\ \bibnamefont {Míguez}}, \bibinfo {author}
  {\bibfnamefont {M.~M.}\ \bibnamefont {Conde}}, \ and\ \bibinfo {author}
  {\bibfnamefont {F.~J.}\ \bibnamefont {Blas}},\ }\bibfield  {title} {\enquote
  {\bibinfo {title} {Three-phase equilibria of hydrates from computer
  simulation. {II}: {F}inite-size effects in the carbon dioxide hydrate},}\
  }\href@noop {} {\bibfield  {journal} {\bibinfo  {journal} {J. Chem. Phys.}\
  }\textbf {\bibinfo {volume} {160}},\ \bibinfo {pages} {164722} (\bibinfo
  {year} {2024}{\natexlab{b}})}\BibitemShut {NoStop}%
\bibitem [{\citenamefont {Algaba}\ \emph
  {et~al.}(2024{\natexlab{c}})\citenamefont {Algaba}, \citenamefont {Blazquez},
  \citenamefont {Míguez}, \citenamefont {Conde},\ and\ \citenamefont
  {Blas}}]{Algaba2024b}%
  \BibitemOpen
  \bibfield  {author} {\bibinfo {author} {\bibfnamefont {J.}~\bibnamefont
  {Algaba}}, \bibinfo {author} {\bibfnamefont {S.}~\bibnamefont {Blazquez}},
  \bibinfo {author} {\bibfnamefont {J.~M.}\ \bibnamefont {Míguez}}, \bibinfo
  {author} {\bibfnamefont {M.~M.}\ \bibnamefont {Conde}}, \ and\ \bibinfo
  {author} {\bibfnamefont {F.~J.}\ \bibnamefont {Blas}},\ }\bibfield  {title}
  {\enquote {\bibinfo {title} {Three-phase equilibria of hydrates from computer
  simulation. {III}: {E}ffect of dispersive interactions in methane and carbon
  dioxide hydrates},}\ }\href@noop {} {\bibfield  {journal} {\bibinfo
  {journal} {J. Chem. Phys.}\ }\textbf {\bibinfo {volume} {160}},\ \bibinfo
  {pages} {164723} (\bibinfo {year} {2024}{\natexlab{c}})}\BibitemShut
  {NoStop}%
\bibitem [{\citenamefont {van Cleeff}\ and\ \citenamefont
  {Diepen}(1960)}]{vanCleeff1960a}%
  \BibitemOpen
  \bibfield  {author} {\bibinfo {author} {\bibfnamefont {A.}~\bibnamefont {van
  Cleeff}}\ and\ \bibinfo {author} {\bibfnamefont {G.~A.~M.}\ \bibnamefont
  {Diepen}},\ }\bibfield  {title} {\enquote {\bibinfo {title} {Gas hydrates of
  nitrogen and oxygen},}\ }\href@noop {} {\bibfield  {journal} {\bibinfo
  {journal} {Rec. Trav. Chim.}\ }\textbf {\bibinfo {volume} {79}},\ \bibinfo
  {pages} {582--586} (\bibinfo {year} {1960})}\BibitemShut {NoStop}%
\bibitem [{\citenamefont {Marshall}, \citenamefont {Saito},\ and\ \citenamefont
  {Kobayashi}(1964)}]{Marshall1964a}%
  \BibitemOpen
  \bibfield  {author} {\bibinfo {author} {\bibfnamefont {D.~R.}\ \bibnamefont
  {Marshall}}, \bibinfo {author} {\bibfnamefont {S.}~\bibnamefont {Saito}}, \
  and\ \bibinfo {author} {\bibfnamefont {R.}~\bibnamefont {Kobayashi}},\
  }\bibfield  {title} {\enquote {\bibinfo {title} {Hydrates at high pressures:
  Part {I}. {M}ethane-water, argon-water, and nitrogen-water systems},}\
  }\href@noop {} {\bibfield  {journal} {\bibinfo  {journal} {AIChE J.}\
  }\textbf {\bibinfo {volume} {10}},\ \bibinfo {pages} {202--205} (\bibinfo
  {year} {1964})}\BibitemShut {NoStop}%
\bibitem [{\citenamefont {Jhaveri}\ and\ \citenamefont
  {Robinson}(1965)}]{Jhaveri1965a}%
  \BibitemOpen
  \bibfield  {author} {\bibinfo {author} {\bibfnamefont {J.}~\bibnamefont
  {Jhaveri}}\ and\ \bibinfo {author} {\bibfnamefont {D.~B.}\ \bibnamefont
  {Robinson}},\ }\bibfield  {title} {\enquote {\bibinfo {title} {Hydrates in
  the methane-nitrogen system},}\ }\href@noop {} {\bibfield  {journal}
  {\bibinfo  {journal} {Can. J. Chem Eng.}\ }\textbf {\bibinfo {volume} {43}},\
  \bibinfo {pages} {75--78} (\bibinfo {year} {1965})}\BibitemShut {NoStop}%
\bibitem [{\citenamefont {Sugahara}\ \emph {et~al.}(2002)\citenamefont
  {Sugahara}, \citenamefont {Tanaka}, \citenamefont {Sugahara},\ and\
  \citenamefont {Ohgaki}}]{Sugahara2002a}%
  \BibitemOpen
  \bibfield  {author} {\bibinfo {author} {\bibfnamefont {K.}~\bibnamefont
  {Sugahara}}, \bibinfo {author} {\bibfnamefont {Y.}~\bibnamefont {Tanaka}},
  \bibinfo {author} {\bibfnamefont {T.}~\bibnamefont {Sugahara}}, \ and\
  \bibinfo {author} {\bibfnamefont {K.}~\bibnamefont {Ohgaki}},\ }\bibfield
  {title} {\enquote {\bibinfo {title} {Thermodynamic stability and structure of
  nitrogen hydrate crystal},}\ }\href@noop {} {\bibfield  {journal} {\bibinfo
  {journal} {J. Supramol. Chem.}\ }\textbf {\bibinfo {volume} {2}},\ \bibinfo
  {pages} {365--368} (\bibinfo {year} {2002})}\BibitemShut {NoStop}%
\bibitem [{\citenamefont {Mohammadi}, \citenamefont {Tohidi},\ and\
  \citenamefont {Burgass}(2003)}]{Mohammadi2003a}%
  \BibitemOpen
  \bibfield  {author} {\bibinfo {author} {\bibfnamefont {A.~H.}\ \bibnamefont
  {Mohammadi}}, \bibinfo {author} {\bibfnamefont {B.}~\bibnamefont {Tohidi}}, \
  and\ \bibinfo {author} {\bibfnamefont {R.~W.}\ \bibnamefont {Burgass}},\
  }\bibfield  {title} {\enquote {\bibinfo {title} {Equilibrium data and
  thermodynamic modeling of nitrogen, oxygen, and air clathrate hydrates},}\
  }\href@noop {} {\bibfield  {journal} {\bibinfo  {journal} {J. Chem. Eng.
  Data}\ }\textbf {\bibinfo {volume} {48}},\ \bibinfo {pages} {612--616}
  (\bibinfo {year} {2003})}\BibitemShut {NoStop}%
\bibitem [{\citenamefont {Liang}\ and\ \citenamefont
  {Kusalik}(2011)}]{Liang2011a}%
  \BibitemOpen
  \bibfield  {author} {\bibinfo {author} {\bibfnamefont {S.}~\bibnamefont
  {Liang}}\ and\ \bibinfo {author} {\bibfnamefont {P.~G.}\ \bibnamefont
  {Kusalik}},\ }\bibfield  {title} {\enquote {\bibinfo {title} {Exploring
  nucleation of {H$_{2}$S} hydrates},}\ }\href@noop {} {\bibfield  {journal}
  {\bibinfo  {journal} {Chem. Sci.}\ }\textbf {\bibinfo {volume} {2}},\
  \bibinfo {pages} {1286--1292} (\bibinfo {year} {2011})}\BibitemShut {NoStop}%
\bibitem [{\citenamefont {Yagasaki}\ \emph {et~al.}(2014)\citenamefont
  {Yagasaki}, \citenamefont {Matsumoto}, \citenamefont {Andoh}, \citenamefont
  {Okazaki},\ and\ \citenamefont {Tanaka}}]{Yagasaki2014a}%
  \BibitemOpen
  \bibfield  {author} {\bibinfo {author} {\bibfnamefont {T.}~\bibnamefont
  {Yagasaki}}, \bibinfo {author} {\bibfnamefont {M.}~\bibnamefont {Matsumoto}},
  \bibinfo {author} {\bibfnamefont {Y.}~\bibnamefont {Andoh}}, \bibinfo
  {author} {\bibfnamefont {S.}~\bibnamefont {Okazaki}}, \ and\ \bibinfo
  {author} {\bibfnamefont {H.}~\bibnamefont {Tanaka}},\ }\bibfield  {title}
  {\enquote {\bibinfo {title} {Effect of bubble formation on the dissociation
  of methane hydrate in water: A molecular dynamics study},}\ }\href@noop {}
  {\bibfield  {journal} {\bibinfo  {journal} {J. Phys. Chem. B}\ }\textbf
  {\bibinfo {volume} {118}},\ \bibinfo {pages} {1900} (\bibinfo {year}
  {2014})}\BibitemShut {NoStop}%
\bibitem [{\citenamefont {Bagherzadeh}\ \emph {et~al.}(2015)\citenamefont
  {Bagherzadeh}, \citenamefont {Alavi}, \citenamefont {Ripmeester},\ and\
  \citenamefont {Englezos}}]{Bagherzadeh2015a}%
  \BibitemOpen
  \bibfield  {author} {\bibinfo {author} {\bibfnamefont {S.~A.}\ \bibnamefont
  {Bagherzadeh}}, \bibinfo {author} {\bibfnamefont {S.}~\bibnamefont {Alavi}},
  \bibinfo {author} {\bibfnamefont {J.}~\bibnamefont {Ripmeester}}, \ and\
  \bibinfo {author} {\bibfnamefont {P.}~\bibnamefont {Englezos}},\ }\bibfield
  {title} {\enquote {\bibinfo {title} {Formation of methane nano-bubbles during
  hydrate decomposition and their effect on hydrate growth},}\ }\href@noop {}
  {\bibfield  {journal} {\bibinfo  {journal} {J. Chem. Phys.}\ }\textbf
  {\bibinfo {volume} {142}},\ \bibinfo {pages} {214701} (\bibinfo {year}
  {2015})}\BibitemShut {NoStop}%
\bibitem [{\citenamefont {Fang}\ \emph {et~al.}(2023)\citenamefont {Fang},
  \citenamefont {Moultos}, \citenamefont {Sun}, \citenamefont {Liu},
  \citenamefont {Ning},\ and\ \citenamefont {Vlugt}}]{Fang2023a}%
  \BibitemOpen
  \bibfield  {author} {\bibinfo {author} {\bibfnamefont {B.}~\bibnamefont
  {Fang}}, \bibinfo {author} {\bibfnamefont {O.}~\bibnamefont {Moultos}},
  \bibinfo {author} {\bibfnamefont {T.~L.~J.}\ \bibnamefont {Sun}}, \bibinfo
  {author} {\bibfnamefont {Z.}~\bibnamefont {Liu}}, \bibinfo {author}
  {\bibfnamefont {F.}~\bibnamefont {Ning}}, \ and\ \bibinfo {author}
  {\bibfnamefont {T.~J.~H.}\ \bibnamefont {Vlugt}},\ }\bibfield  {title}
  {\enquote {\bibinfo {title} {Effects of nanobubbles on methane hydrate
  dissociation: A molecular simulation study},}\ }\href@noop {} {\bibfield
  {journal} {\bibinfo  {journal} {Fuel}\ }\textbf {\bibinfo {volume} {345}},\
  \bibinfo {pages} {128230} (\bibinfo {year} {2023})}\BibitemShut {NoStop}%
\bibitem [{\citenamefont {Kashchiev}\ and\ \citenamefont {van
  Rosmalen}(2003)}]{Kashchiev2003a}%
  \BibitemOpen
  \bibfield  {author} {\bibinfo {author} {\bibfnamefont {D.}~\bibnamefont
  {Kashchiev}}\ and\ \bibinfo {author} {\bibfnamefont {G.~M.}\ \bibnamefont
  {van Rosmalen}},\ }\bibfield  {title} {\enquote {\bibinfo {title} {Review:
  Nucleation in solutions revisited},}\ }\href {\doibase
  https://doi.org/10.1002/crat.200310070} {\bibfield  {journal} {\bibinfo
  {journal} {Cryst. Res. Technol.}\ }\textbf {\bibinfo {volume} {38}},\
  \bibinfo {pages} {555--574} (\bibinfo {year} {2003})}\BibitemShut {NoStop}%
\end{thebibliography}%

\end{document}